\documentclass[twocolumn,prd,floatfix,preprintnumbers,a4paper,nofootinbib,superscriptaddress]{revtex4-1}
\usepackage{epsfig}
\usepackage{graphics}
\usepackage{graphicx}
\usepackage{amsmath,amssymb}
\usepackage{amsfonts}
\usepackage{bm}
\usepackage[usenames,dvipsnames]{color}
\usepackage{amssymb}
\usepackage{psfrag}
\usepackage{times}
\usepackage[varg]{txfonts}
\usepackage{xspace}
\usepackage{float}
\usepackage{placeins} 
\usepackage{capt-of} 

\usepackage[colorlinks, pdfborder={0 0 0}]{hyperref} 
\definecolor{LinkColor}{rgb}{0.75, 0, 0}
\definecolor{CiteColor}{rgb}{0, 0.5, 0.5}
\definecolor{UrlColor}{rgb}{0, 0, 0.75}
\hypersetup{linkcolor=LinkColor}
\hypersetup{citecolor=CiteColor}
\hypersetup{urlcolor=UrlColor}

\let\oldtheequation\theequation
\makeatletter
\def\tagform@#1{\maketag@@@{\ignorespaces#1\unskip\@@italiccorr}}
\renewcommand{\theequation}{(\oldtheequation)}
\makeatother

\usepackage[utf8]{inputenc}
\usepackage{ulem}
\newcommand{\macro}[1]{#1}                      
\newcommand{\NRcount}{\macro{427}\xspace}       
\newcommand{\NRcountNS}{\macro{92}\xspace}      
\newcommand{\NRcountEqSqone}{\macro{37}\xspace}	
\newcommand{\NRcountEqSrest}{\macro{60}\xspace}	
\newcommand{\NRcountUneqS}{\macro{238}\xspace}	
\newcommand{\NRcountNZSUM}{\macro{298}\xspace}	
\newcommand{\NRcountSXS}{\macro{161}\xspace}
\newcommand{\NRcountBAM}{\macro{45}\xspace}
\newcommand{\NRcountGaT}{\macro{114}\xspace}
\newcommand{\NRcountRIT}{\macro{107}\xspace}
\newcommand{\NRcountBAMnew}{\macro{27}\xspace}	
\newcommand{\NRcountOutliers}{\macro{16}\xspace}
\newcommand{\NRduplicated}{\macro{41}\xspace}
\newcommand{\NRReltolerance}{\macro{0.001}\xspace}
\newcommand{\NRStricttolerance}{\macro{0.0002}\xspace}

\newcommand{\dcc}{LIGO-P1600270-v5}

\newcommand{\pade}{Pad{\'e}\xspace}
\newcommand{\ansaetze}{ans{\"a}tze\xspace}
\newcommand{\emr}{extreme-mass-ratio\xspace}
\newcommand{\Emr}{Extreme-mass-ratio\xspace}
\newcommand{\emrl}{extreme-mass-ratio limit\xspace}
\newcommand{\eqmeqS}{equal-mass-equal-spin\xspace}
\newcommand{\EqmeqS}{Equal-mass-equal-spin\xspace}
\newcommand{\BAM}{\texttt{BAM}\xspace}

\newcommand{\chieff}{\chi_{\mathrm{eff}}}
\newcommand{\Seff}{\widehat{S}}
\newcommand{\Stot}{S}
\newcommand{\chidiff}{\Delta\chi}
\newcommand{\af}{\chi_{\mathrm{f}}}
\newcommand{\Mf}{M_{\mathrm{f}}}
\newcommand{\isco}{\mathrm{ISCO}}
\newcommand{\Eisco}{E_\isco}
\newcommand{\rhoisco}{\rho_\isco}
\newcommand{\Lisco}{L_{\mathrm{orb},\isco}}
\newcommand{\Lorb}{L_{\mathrm{orb}}}
\newcommand{\trueLorb}{\Lorb}
\newcommand{\ourLorb}{\Lorb^\prime}
\newcommand{\ourLorbNR}{L^\prime_{\mathrm{orb,NR}}}
\newcommand{\oneDeta}{\left(\eta,\Seff=0\right)}
\newcommand{\oneDS}{\left(\eta=0.25,\Seff\right)}
\newcommand{\twoDparams}{\left(\eta,\Seff\right)}
\newcommand{\threeDparams}{\left(\eta,\Seff,\chidiff\right)}
\newcommand{\LorboneDeta}{\ourLorb\oneDeta}
\newcommand{\LorboneDS}{\ourLorb\oneDS}
\newcommand{\LorbtwoD}{\ourLorb\twoDparams}
\newcommand{\LorbthreeD}{\ourLorb\threeDparams}
\newcommand{\DLorb}{\Delta\ourLorb}
\newcommand{\DLorbthreeD}{\DLorb\,\threeDparams}
\newcommand{\Erad}{E_{\mathrm{rad}}}
\newcommand{\EradoneDeta}{\Erad\oneDeta}
\newcommand{\EradoneDS}{\Erad\oneDS}
\newcommand{\EradtwoD}{\Erad\twoDparams}
\newcommand{\EradthreeD}{\Erad\threeDparams}
\newcommand{\DErad}{\Delta\Erad}
\newcommand{\DEradthreeD}{\DErad\,\threeDparams}
\newcommand{\Ndata}{N_{\mathrm{data}}}
\newcommand{\Ncoef}{N_{\mathrm{coeffs}}}
\newcommand{\RMSE}{\mathrm{RMSE}}
\newcommand{\AIC}{\mathrm{AIC}}
\newcommand{\AICc}{\mathrm{AICc}}
\newcommand{\BIC}{\mathrm{BIC}}
\newcommand{\maxlnL}{\mathcal{L}_{\max}}
\newcommand{\covfit}{\mathcal{C}_{\mathrm{fit}}}
\newcommand{\estVar}{\widehat{\sigma}^2}

\begin{document}


\title{Hierarchical data-driven approach to fitting numerical relativity data\\ for nonprecessing binary black holes with an application to final spin and radiated energy}

\newcommand{\cardiff}{\affiliation{School of Physics and Astronomy, Cardiff University, The Parade, Cardiff, CF24 3AA, United Kingdom}}
\newcommand{\uib}{\affiliation{Universitat de les Illes Balears, IAC3---IEEC, 07122 Palma de Mallorca, Spain}}
\newcommand{\icts}{\affiliation{International Centre for Theoretical Sciences, Tata Institute of Fundamental Research, Bangalore 560012, India}}
\newcommand{\birm}{\affiliation{School of Physics and Astronomy, University of Birmingham, Edgbaston, Birmingham, B15 2TT, United Kingdom}}
\newcommand{\golm}{\affiliation{Albert-Einstein-Institut, Am M{\"u}hlenberg 1, D-14476 Potsdam-Golm, Germany}}
\newcommand{\glasgow}{\affiliation{School of Physics and Astronomy, University of Glasgow, Glasgow G12 8QQ, United Kingdom}}

\author{Xisco Jim{\'e}nez-Forteza}\email{francisco.forteza@ligo.org}\uib
\author{David~Keitel}\email{david.keitel@ligo.org}\uib\glasgow
\author{Sascha~Husa}\email{sascha.husa@ligo.org}\uib
\author{Mark Hannam}\cardiff
\author{Sebastian Khan}\cardiff
\author{Michael P{\"u}rrer}\cardiff\golm

\begin{abstract}
Numerical relativity is an essential tool in studying the coalescence of binary black holes (BBHs).
It is still computationally prohibitive to cover the BBH parameter space exhaustively,
making phenomenological fitting formulas for BBH waveforms and final-state properties important for practical applications.
We describe a general hierarchical bottom-up fitting methodology to design and calibrate fits to numerical relativity simulations
for the three-dimensional parameter space of quasicircular nonprecessing merging BBHs,
spanned by mass ratio and by the individual spin components orthogonal to the orbital plane.
Particular attention is paid to incorporating the extreme-mass-ratio limit
and to the subdominant unequal-spin effects.
As an illustration of the method, we provide two applications, to the final spin and final mass (or equivalently: radiated energy) of the remnant black hole.
Fitting to \NRcount numerical relativity simulations,
we obtain results broadly consistent with previously published fits,
but improving in overall accuracy and particularly in the approach to extremal limits and for unequal-spin configurations.
We also discuss the importance of data quality studies when combining simulations from diverse sources,
how detailed error budgets will be necessary for further improvements of these already highly accurate fits,
and how this first detailed study of unequal-spin effects helps in choosing the most informative parameters for future numerical relativity runs.
\end{abstract}

 \date{27 March 2017 \\
 ..LIGO document number: \dcc}

\maketitle

\section{Introduction}

According to general relativity, compact binary black holes (BBHs) coalesce
through the emission of gravitational waves (GWs),
as already observed by LIGO in at least two cases~\cite{Abbott:2016blz,TheLIGOScientific:2016wfe,Abbott:2016nmj,TheLIGOScientific:2016pea}. 
The merger remnant is a single Kerr BH~\cite{Kerr:1963ud} characterized only by its final spin and mass.
An essential and robust tool for predicting BBH evolution,
since the 2005 breakthrough~\cite{Pretorius:2005gq,Campanelli:2005dd,Baker:2005vv},
is numerical relativity (NR).
Due to the large computational cost of each NR simulation,
it is still computationally prohibitive to cover the BBH parameter space exhaustively.
Hence it is natural to develop simple, yet accurate model fits to the existing set of NR simulations.
One then has to study their quality of interpolation between NR simulations,
and of extrapolation to regions of parameter space beyond the calibration range.
For example, the phenomenological inspiral-merger-ringdown waveform models of~\cite{Husa:2015iqa,Khan:2015jqa,Hannam:2013oca,T1500602,PhenomPv2Paper},
though used very successfully as one of two waveform families
for LIGO O1 data analysis~\cite{Abbott:2016blz,TheLIGOScientific:2016wfe,Abbott:2016nmj,TheLIGOScientific:2016pea}, 
still include only a limited set of physical effects and were calibrated to small sets of NR simulations with relatively simple fitting methods.

In this paper we develop a more general, hierarchical fitting approach for BBH properties and waveforms.
The fit ansatz functions  are developed through studying hierarchical structures present in the NR data set itself,
and their complexity tailored to the actual predictive power of the data set by the use of information criteria.

We first apply this method to the spin and mass of the remnant black hole,
which determine the frequencies of the quasinormal-mode ringdown~\cite{Press:1971wr,Chandrasekhar:1975zza,Detweiler:1980gk,Kokkotas:1999bd} to a Kerr black hole.
The ringdown is a crucial part in modeling full inspiral-merger-ringdown
waveforms~\cite{Ajith:2009bn,Husa:2015iqa,Khan:2015jqa,Hannam:2013oca,T1500602,PhenomPv2Paper,Taracchini:2013rva,Pan:2013rra};
and from parameter estimation with the full waveforms, the final mass and spin can be estimated
with accuracy similar to other BBH parameters~\cite{TheLIGOScientific:2016wfe,TheLIGOScientific:2016pea}. 
Future observations of strong GW signals will allow to test general relativity through consistency tests between inspiral, merger and ringdown~\cite{Ghosh:2016qgn},
significantly improving upon~\cite{TheLIGOScientific:2016src}.

Apart from GW observations, the final state of a BBH merger is astrophysically interesting in itself,
e.g. for the computation of merger trees~\cite{Lacey:1993iv,Roukema:1997ed,Volonteri:2002vz,Bogdanovic:2007hp,Tichy:2008du,Barausse:2012fy,Sesana:2014bea,Klein:2015hvg}.
The mass and spin of BHs surrounded by matter,
e.g. accretion disks,
may also be inferred from electromagnetic observations
(see~\cite{McClintock:2011zq, Nielsen:2016kyw} for stellar-mass BHs
and~\cite{Brenneman:2013oba,Reynolds:2013rva,Peterson:2014smbh} for supermassive BHs).

For this paper, we concentrate on nonprecessing quasicircular BBHs,
where the black hole spins are parallel or antiparallel to the total orbital angular momentum of the binary.
These configurations are fully described in a three-dimensional parameter space:
given the masses $m_{1,2}$ and physical spins $S_{1,2}$, we use
the two component spins \mbox{$\chi_1=S_1/m_1^2$} and \mbox{$\chi_2=S_2/m_2^2$}
and the mass ratio, given either as \mbox{$q=m_1/m_2$} with the convention \mbox{$m_1>m_2$},
or as the symmetric mass ratio \mbox{$\eta = (m_1 m_2)/(m_1 + m_2)^2  = q/(1+q)^2$}.
The total mass is only a scaling factor, and here we work in units of \mbox{$m_1+m_2=1$}.

As suggested by the post-Newtonian (PN) results~\cite{Damour:2001tu,Racine:2008qv,Bohe:2013cla,Marsat:2013caa} for radiated energy and angular momentum,
and confirmed by many previous studies of NR-calibrated models~\cite{Ajith:2009bn,Santamaria:2010yb,Ajith:2011ec,Husa:2015iqa},
the two dominant parameter dependencies of both final spin and final mass are on the mass ratio and on some appropriately chosen {\em effective} spin parameter of the binary, 
whereas the effects of any difference between the two individual spins are much smaller.
However, only by accurately modeling these small unequal-spin effects can the full two-spin information be extracted from GW observations,
disentangling any true physical degeneracies from systematic effects due to limitations of the waveform models.

A similar, but more complex, situation is encountered in the calibration of phenomenological models to precessing binaries,
where so far a simple waveform model based on a single effective precession parameter~\cite{Hannam:2013oca,T1500602,PhenomPv2Paper}
has successfully been employed to analyze the first gravitational wave detections~\cite{TheLIGOScientific:2016wfe,TheLIGOScientific:2016pea}. 
(A complementary precessing analysis~\cite{Abbott:2016izl} based on the model of~\cite{Pan:2013rra,Babak:2016tgq} has also been published recently.)
The current work, while mainly a first step to develop models for the full three-dimensional parameter space of nonprecessing waveforms,
can also be considered as a toy model for the numerical calibration of subdominant effects in generic binaries, such as precession and higher modes.

For the current three-dimensional aligned-spin parameter space,
in addition to the hierarchy of mass ratio, effective spin and spin difference,
the sampling by available NR simulations still displays significant bias toward simple subsets:
namely the one-dimensional subspaces of nonspinning cases ($\eta$ dependence only)
and equal-mass, equal-spin cases (\mbox{$\eta=0.25$}, \mbox{$q=1$}, \mbox{$\chi_1=\chi_2$}, thus effective-spin dependence only)
are covered particularly well by existing NR catalogs,
while few simulations exist for unequal spins, high spins, and/or very unequal masses.
Independently, the \emrl (\mbox{$\eta \rightarrow 0$, $q \rightarrow \infty$}, arbitrary spins) is known analytically~\cite{Bardeen:1972fi}.

Here we exploit and investigate this structure
by parametrizing spin effects in terms of an effective spin and a spin-difference parameter.
As the effective spin we choose
\begin{align}
 \label{eq:Stot3}
 \Seff=\frac{\Stot}{m_1^2+m_2^2}, && \text{with} \; \Stot=m_1^2\,\chi_1+m_2^2\,\chi_2 \,,
\end{align}
which has already been found to work well for final-state quantities in~\cite{Husa:2015iqa}.
We discuss other possible choices for the effective spin,
for which our method also works robustly, in Appendix~\ref{sec:appendix-spinpar}.
For spin difference, we use simply \mbox{$\chidiff=\chi_1-\chi_2$},
which makes no assumptions on how spin-difference effects depend on mass ratio.

\begin{figure}[t]
 \includegraphics[width=\columnwidth]{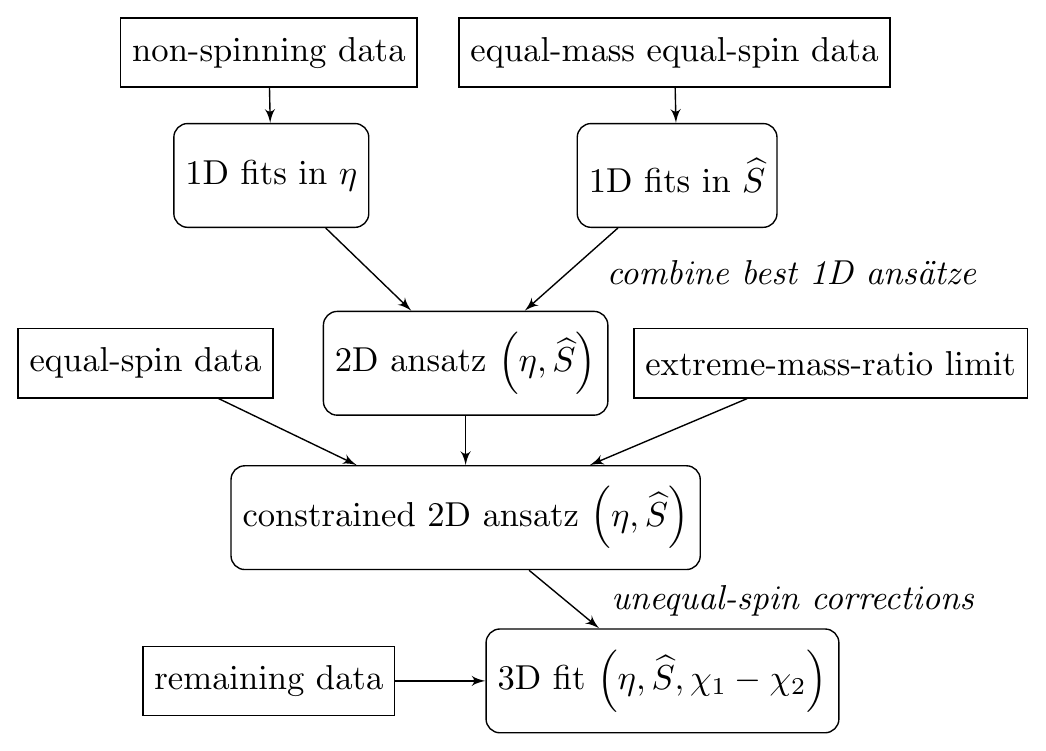}
 \caption{
  \label{fig:flowchart}
  Flowchart of the hierarchical step-by-step construction leading to a three-dimensional ansatz and fit
  for the quantity of interest over the \mbox{$\left(\eta,\chi_1,\chi_2\right)\equiv\left(\eta,\Seff,\chidiff\right)$} space.
  \vspace{-\baselineskip}
 }
\end{figure}

We develop our hierarchical approach,
with the aim to ensure an accurate modeling of the subdominant spin-difference effects,
along the lines illustrated as a flowchart in \autoref{fig:flowchart}:
First we consider the one-dimensional subspaces of nonspinning and of \eqmeqS black holes.
We then combine and generalize these subspace fits,
adding additional degrees of freedom to cover the entire two-dimensional space of equal-spin black holes,
but constrain the generalized ansatz with information from the \emrl.
In a third step, we investigate the leading subdominant terms,
which are linear in the difference between spins,
and also identify additional nonlinear spin-difference terms.
We finally produce a three-dimensional fit to the complete data set with the hierarchically constructed ansatz.
This way, we can construct a full ansatz with a relatively low number of free fit coefficients and
avoid overfitting of spurious effects only due to small sample sizes,
while still capturing the essential physical effects that are known from the well-constrained regions.

At each step, we evaluate the performance of different fit choices by several quantitative measures:
by the overall residuals,
by the Akaike and Bayesian information criteria (AICc, BIC,~\cite{Akaike:1974,Schwarz:1978}),
and by how well determined the individual fit coefficients are.
The information criteria are model selection tools to choose between fits with comparable goodness of fit
but different degrees of complexity, i.e. they penalize high numbers of free coefficients.
See Appendix~\ref{sec:appendix-stats} for details on these statistical methods.

Previous published fits for final spin and/or mass
include~\cite{Buonanno:2007sv,Rezzolla:2007rd,Rezzolla:2007rz,Tichy:2008du,Barausse:2009uz,Barausse:2012qz,Healy:2014yta,Husa:2015iqa,Zlochower:2015wga,Zlochower:2015wga-erratum,Hofmann:2016yih},
and we will compare our results to the most recent results in the literature,
including both fits across the full three-dimensional nonprecessing parameter space,
and the effective-spin-based fits from \cite{Husa:2015iqa},
which were used in the aligned-spin IMRPhenomD waveform model~\cite{Husa:2015iqa,Khan:2015jqa}
and (with in-plane-spin corrections) in the precessing IMRPhenomPv2 model~\cite{Hannam:2013oca,T1500602,PhenomPv2Paper}.
These will be referred to as the ``PhenomD fits'' in the following.
The plan of the paper is as follows:
We first describe our data sets, built from several NR catalogs, in \autoref{sec:data},
together with the available \emrl information.
In \autoref{sec:spinfits} we develop the general fitting recipe for the example of the final spin.
We then apply our method also for final mass -- or equivalently, radiated energy -- in \autoref{sec:energyfits},
which illustrates some of the specific choices and adaptations required to apply the general method to each quantity.
We summarize our method and results in \autoref{sec:conclusions},
and give additional details about NR data, fit construction, and fit uncertainty estimates
in Appendices \ref{sec:appendix-data}--\ref{sec:appendix-uncertainties}.

\begin{figure}[t!]
 \includegraphics[width=\columnwidth]{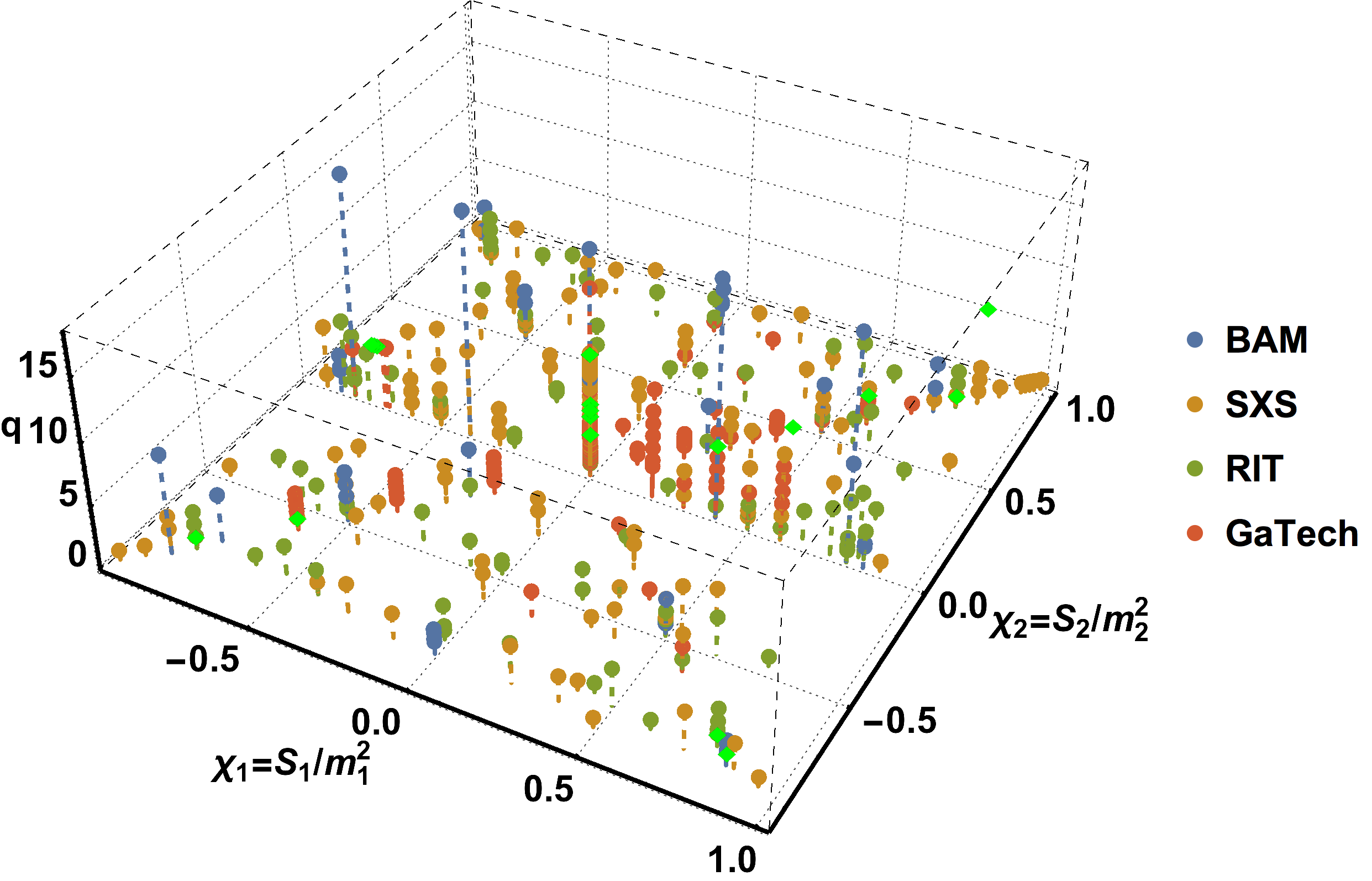}
 \caption{
  \label{fig:eta_chi1_chi2_NR}
  The NR data set used in this paper,
  over mass ratio \mbox{$q=m_1/m_2$} and the two dimensionless spin components $\chi_1$, $\chi_2$,
  with color indicating the source catalog.
  Bright green points are cases removed from the analysis for data-quality reasons,
  as discussed in Appendix~\ref{sec:appendix-data}.
  \vspace{-1.5\baselineskip}
 }
\end{figure}

Our fits for final spin and radiated energy are also provided in Mathematica and python formats
as supplementary material~\cite{Jimenez-Forteza:2016oae-suppl,*Jimenez-Forteza:2016oae-anc},
and implemented under the label ``UIB2016'' in the \texttt{nrutils.py} package of LALInference~\cite{Veitch:2014wba,lalsuite}.
This final version of the paper uses a larger NR calibration set than the initial arXiv submission (1611.00332v1),
with fit results fully consistent but better constrained.

\begin{figure*}[t!]
 \includegraphics[width=\textwidth]{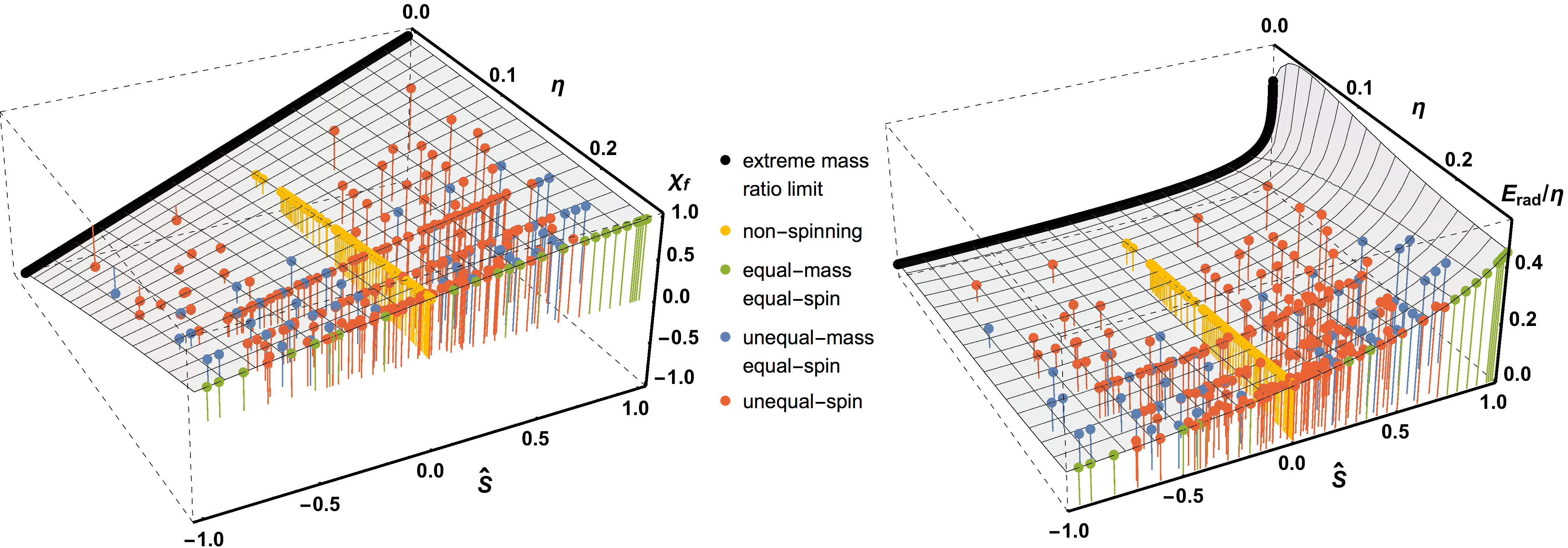}
 \caption{
  \label{fig:eta_S_af_NR_extreme}
  Input data plotted against symmetric mass ratio $\eta$ and effective spin $\Seff$.
  Data consist of the combined set of NR simulations (colored points)
  and the analytically known~\cite{Bardeen:1972fi} \emr behavior (black line).
  Left panel: final spin; right panel: radiated energy rescaled by $\eta$.
  Both $\af$ and $\Erad/\eta$ follow a smooth surface in this space,
  and the well-constrained 1D subspaces together already give a good indication of its curvature.
  \vspace{-\baselineskip}
 }
\end{figure*}

\section{Input data}
\label{sec:data}

\subsection{Numerical Relativity data sets}
 \label{sec:data-NR}
 
We combine four data sets of aligned-spin numerical relativity BBH simulations from independent codes and sources:
the public catalogs of
SXS~\cite{Mroue:2013xna,SXScatalog},
RIT~\cite{Healy:2014yta,Healy:2016lce}
and GaTech~\cite{Jani:2016wkt,GaTechcatalog};
as well as a set of our own simulations with the \BAM code~\cite{Bruegmann:2006at,Husa:2007hp,Husa:2015iqa},
including \NRcountBAMnew new cases for which initial configurations and results are listed in Table~\ref{tbl:newBAM} in Appendix~\ref{sec:appendix-data}.
After removing \NRcountOutliers cases from the combined data set due to data quality considerations as discussed in Appendix~\ref{sec:appendix-data},
we have \NRcountSXS cases from the SXS catalog, \NRcountRIT from RIT, \NRcountGaT from GaTech and \NRcountBAM from \BAM;
for a total of \NRcount cases.
The sampling of our three-dimensional parameter space by the four data sets is shown in \autoref{fig:eta_chi1_chi2_NR}.
The initial arXiv version of this paper (1611.00332v1) used a smaller data set of 256 NR cases,
with the increase coming from an update of the public SXS catalog and new RIT results from~\cite{Healy:2016lce}.

To obtain a qualitative understanding of the hierarchical structure in the two-dimensional parameter space of mass ratio and effective spin,
in \autoref{fig:eta_S_af_NR_extreme} we show the NR data set over the \mbox{$(\eta,\Seff)$} plane together
with the analytical \emr results, discussed below in Sec.~\ref{sec:data-extreme}.
For both final spin and radiated energy, we find a reasonably smooth surface spanned by the NR data points.
These plots already suggests that --
together with the known \emr results to compensate the sparsity of NR simulations at increasingly unequal masses --
good one-dimensional fits in the two best-sampled one-dimensional subsets
(\eqmeqS and nonspinning BHs)
will significantly constrain any two-dimensional fits.

\begin{figure}[t!]
 \includegraphics[width=\columnwidth]{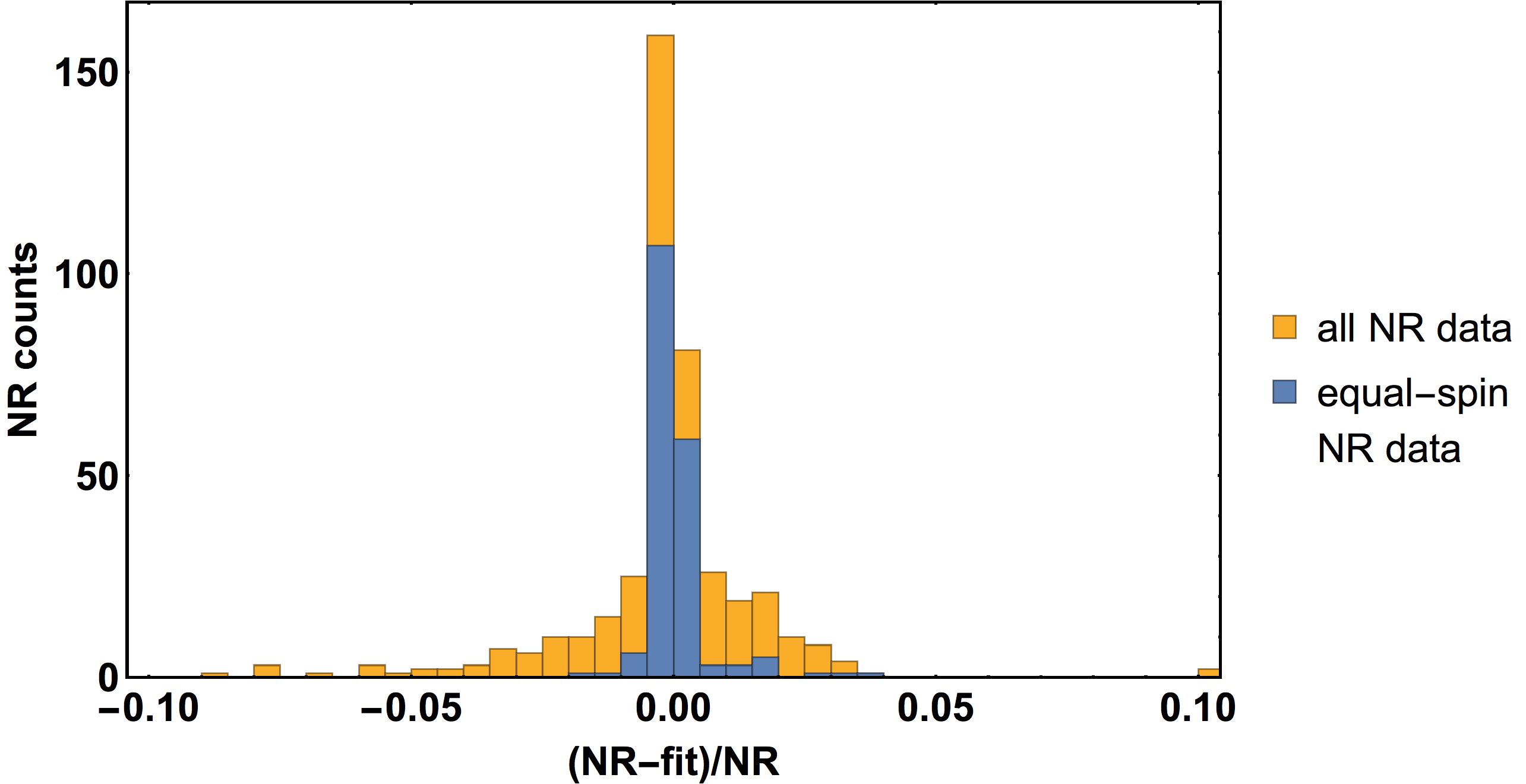}
 \caption{
  \label{fig:resid_PhenomD_af_Erad}
  Relative errors in final spin of the combined NR data set for this paper under the two-dimensional PhenomD fit~\cite{Husa:2015iqa}.
  \vspace{-\baselineskip}
 }
\end{figure}

Furthermore, as a first quantitative check that the assumption of a hierarchical structure in three-dimensional BBH parameter space holds,
with unequal-spin effects subdominant to the dependence on $\eta$ and $\Seff$,
we can study the residuals of this data set under the two-dimensional PhenomD fits~\cite{Husa:2015iqa}.
For final spin, we find that 90\% of relative errors are below 3\%.
(The only cases over 10\% are those with absolute values close to zero,
where relative error is not a good measure, and absolute errors (residuals) are limited to \mbox{$\Delta\af\leq0.025$}.)
Still, this comparison suggests that unequal-spin effects make a large contribution to these small errors,
as shown by four times smaller 90\% quantiles when restricting to equal-spin cases only.
See also \autoref{fig:resid_PhenomD_af_Erad} for histograms of these distributions.
For radiated energy, 90\% of relative errors are below 2\%,
with a reduction of that quantile by 1.4 for equal-spin cases only,
indicating that spin-difference effects are even smaller for this quantity,
which we will also see confirmed in our final results.

For details about
extraction of final-state quantities,
NR data quality
and weight assignment,
see Appendix~\ref{sec:appendix-data}.
As explained there,
we do not have a full set of NR error estimates available,
so we assign heuristic fit weights to each case based on
the expected accuracy of the respective NR code in that particular parameter space region.
For example, high-mass-ratio cases are down-weighted more for puncture codes.

\vspace*{-\baselineskip}

\subsection{Extreme-mass-ratio limit}
\label{sec:data-extreme}

The computational cost of numerical simulations of BH binaries in full general relativity diverges in the \emrl \mbox{$\eta \rightarrow0$}.
However, this limit is also equivalent to the much simpler case of a test particle orbiting a Kerr black hole.
The energy and orbital angular momentum for that configuration have long been known analytically~\cite{Bardeen:1972fi}:
inserting the radius of the innermost stable circular orbit (ISCO) from Eq.~(2.21) of~\cite{Bardeen:1972fi} into Eqs.~(2.12) and (2.13) of the same reference
yields the test-particle energy (equivalent to the radiated energy) and orbital angular momentum at ISCO:
\begin{subequations}
\label{eq:extreme_mass_ratio_Eisco_Lisco}
\begin{align}
 \Eisco(\eta,\chi) &=& \eta \left( 1- \sqrt{1-\frac{2}{3\,\rhoisco(\af)}} \right) \,, \label{eq:extreme_mass_ratio_Eisco} \\[0.5\baselineskip]
 \Lisco(\eta,\chi) &=& \frac{2 \eta \left( 3\sqrt{\rhoisco(\chi)} - 2\chi \right)}{\sqrt{3\,\rhoisco(\chi)}} \,, \label{eq:extreme_mass_ratio_Lisco}
\end{align}
\end{subequations}
with
\begin{subequations}
\begin{eqnarray}
\label{eq:extreme_mass_ratio_aux}
 \rhoisco(\chi) &=& 3+Z_2-\mathrm{sgn}(\chi)\sqrt{(3-Z_1)(3+Z_1+2 \, Z_2)}, \quad \\
 Z_1(\chi)      &=& 1+(1-\chi^2)^{1/3} \left[(1+\chi)^{1/3}+(1-\chi)^{1/3}\right], \\
 Z_2(\chi)      &=& \sqrt{3\chi^2+Z_1^2} \,.
\end{eqnarray}
\end{subequations}
Note that both $\Eisco$ and $\Lisco$ depend linearly on $\eta$. 

In the test-particle limit, the small BH plunges after reaching the ISCO,
and further mass loss scales with $\eta^2$ \cite{Davis:1971gg}.
Similar to previous work~\cite{Hughes:2002ei,Buonanno:2007sv,Healy:2014yta,Hofmann:2016yih},
we will exploit this fact to compute the final spin and radiated energy to linear order in $\eta$
from the analytical expressions, ~\autoref{eq:extreme_mass_ratio_Eisco_Lisco}, holding at the ISCO.
To linear order in $\eta$, we thus simply have \mbox{$\Erad =  \Eisco$} or \mbox{$\Mf = 1-\Eisco$} for the final mass,
and for the final spin $\af$ we obtain the implicit equation
\begin{equation}
 \label{eq:af_extreme_limit}
 \af \, \Mf(\eta,\af)^2 = \Lisco(\eta,\af) + S_1 + S_2 \, ,
\end{equation}
where the individual BH spins can be written in terms of our effective spin as
\begin{equation}
 \label{eq:Seff_extreme_limit}
 S_1 + S_2 = (1- 2 \eta) \, \Seff \, .
\end{equation}
Equation~\ref{eq:af_extreme_limit} can then be solved numerically for the final spin $\af$ as a function of $\eta$ and of the effective spin $\Seff$.
Since this result holds to linear order in $\eta$,
and assuming that the final spin and mass are regular functions of $\eta$,
we have thus essentially computed the derivatives $\partial \Erad/\partial\eta$ and $\partial \af/\partial\eta$ at \mbox{$\eta=0$},
in addition to the values at \mbox{$\eta=0$}, which are \mbox{$\Erad(0) = 0$} and \mbox{$\af(0) = S_1/M^2$}.

Additionally, assuming that the final state is indeed a Kerr BH, its final spin has to satisfy \mbox{$\af \leq 1$}.
One would also expect the final spin for maximal effective spin, \mbox{$\Seff=1$},
to decrease monotonically with increasing $\eta$.
To construct an accurate fit in a neighborhood of \mbox{$\Seff \rightarrow 1$}
that satisfies these expectations -- in particular the Kerr limit --
we will constrain our ansatz with the analytically computed value of \mbox{$\af'=\partial \af/\partial\eta$} at \mbox{$(\eta=0, \Seff=1)$}.
By perturbing \autoref{eq:af_extreme_limit} around \mbox{$\left\lbrace\eta \rightarrow 0,\af \rightarrow 1\right\rbrace$}
to linear order before taking the derivative in $\eta$ at the same point, we find 
\begin{equation}
\label{eq:afp_limit}
\af'\left(\eta \rightarrow 0,\Seff \rightarrow 1 \right) = 0 \,.
\end{equation} 
Several variations of this procedure have been used for previous final-spin fits,
and differences are due to previous works neglecting the radiated energy in \autoref{eq:af_extreme_limit}~\cite{Buonanno:2007sv,Healy:2014yta},
or not enforcing the derivative for satisfying the Kerr limit~\cite{Hofmann:2016yih}.

\section{Final spin}
\label{sec:spinfits}

We will now first develop the details of our hierarchical fitting procedure for the example of the final spin of BBH merger remnants,
giving more detail here than we will do for the radiated energy in \autoref{sec:energyfits}.

\subsection{Choice of fit quantity}
\label{sec:spinfits-quantity}

We first need to decide which quantity exactly we want to fit.
It appears natural to fit a quantity related to the ``final" orbital angular momentum $\trueLorb$ near merger,
i.e. separating out the known initial spins $S_i$.
This is particularly useful in connection with the \emrl,
since with \autoref{eq:extreme_mass_ratio_Lisco}, $\trueLorb$ is linear in  $\eta$ to leading order.
We can use the relation from~\autoref{eq:af_extreme_limit} between $\trueLorb$ and the dimensionless Kerr parameter $\af$ of the remnant BH,
\mbox{$\Mf^2 \, \af = \trueLorb + S_1 + S_2 = \trueLorb + S$},
also outside the \emrl.
Here $\Mf$ is the final mass of the remnant BH.

Instead of the actual angular momentum $\trueLorb$,
we take the liberty of fitting the quantity \mbox{$\ourLorb=M^2\,\af-S$},
where (as throughout the paper) $M$ is set to unity.
This way, all fit results are easily converted to the final Kerr parameter $\af$ by adding the total initial spin $S$,
and no correction for radiated energy has to be applied.

\subsection{One-dimensional subspace fits}
\label{sec:spinfits-1d}

Motivated by the the unequal sampling of the parameter space by NR simulations,
as visualized in \autoref{fig:eta_S_af_NR_extreme},
we start our  hierarchical fit development with the simplest and best-sampled subspaces of the NR data set,
constructing one-dimensional fits \mbox{$\LorboneDeta$} and \mbox{$\LorboneDS$}
over \NRcountNS nonspinning and \NRcountEqSqone \eqmeqS cases.
We do not restrict ourselves to polynomial fits, and also include \ansaetze in the form of rational functions.
We have also found good fits for more general functions, but omit these here since we have not explored that option systematically.

All fits are performed with Mathematica's \texttt{NonlinearModelFit} function,
but also partially cross-checked with the \texttt{nls} package of R.
Since rational functions can have singularities,
codes such as \texttt{NonlinearModelFit} may not converge to a valid solution without good starting values for the coefficients.
We solve this problem by first performing a sufficiently high-order polynomial fit,
from which we compute a \pade approximant at the desired order,
and use the coefficients of this approximant as starting values for the rational-function fit.
We denote rational functions with a numerator of polynomial order $m$ and denominator of polynomial order $k$ as an ansatz of order \mbox{$(m,k)$}.
Before fitting to the NR data, all \ansaetze are constrained by two facts:
For nonspinning BBHs, both $\af$ and $\ourLorb$ have to vanish for \mbox{$\eta \rightarrow 0$},
so that there can be no constant term in the ansatz.
Furthermore, from the \emr prediction \autoref{eq:extreme_mass_ratio_Eisco_Lisco},
it also follows that the spin-independent coefficient linear in $\eta$ is $2\sqrt{3}$ (see also ~\cite{Rezzolla:2008fs}).
We will include spin-dependent information linear in $\eta$ in Sec.~\ref{sec:spinfits-2d}.

Thus, we obtain \mbox{$\LorboneDeta$} fits for a large set of polynomial and rational functions.
Several of them produce competitive goodness of fit,
as measured by the root-mean-square-error (RMSE) or the full distribution of residuals.
However, we do not want to overfit the data,
which could induce spurious oscillations in the region of very unequal BH masses that is not covered by NR data.
Hence, we rank the fits by information criteria penalizing superfluous free coefficients.

\begin{figure}[t!]
 \includegraphics[width=\columnwidth]{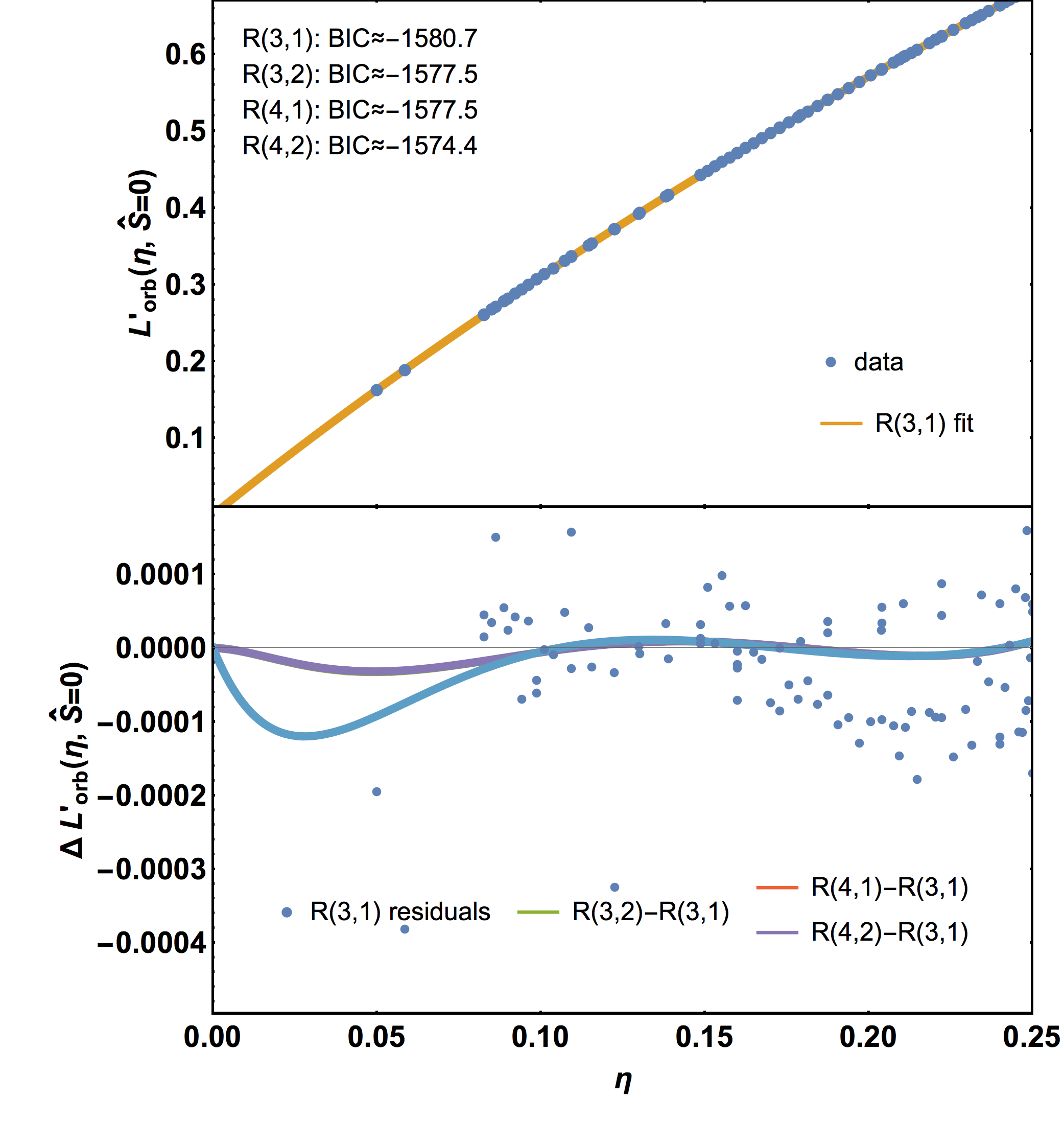}
 \caption{
  \label{fig:Lorb_eta_0}
  Nonspinning NR data and one-dimensional \mbox{$\LorboneDeta$} fit as a function of mass ratio $\eta$.
  Top panel: best fit in terms of the Bayesian information criterion (BIC), 
  a rational function R(3,1), see \autoref{eq:af_etaansatz}.
  Lower panel: residuals (\mbox{$\Delta\ourLorb=$\,data$-$fit}) of this fit (points)
  and differences from the three next-best-ranking fits in terms of BIC (lines).
  See also \autoref{fig:af_eta_fit_BIC} in Appendix~\ref{sec:appendix-stats} for an illustration of BIC ranking for this example.}
 \vspace{0.5\baselineskip}
 \begin{tabular}{lrrr}\hline\hline
  &$ \text{Estimate} $&$ \text{Standard error} $&$ \text{Relative error [$\%$]} $\\\hline$
 a_2 $&$  3.833            $&$ 0.085            $&$ 2.2 $\\$
 a_3 $&$ -9.49\hphantom{3} $&$ 0.24\hphantom{3} $&$ 2.5 $\\$
 a_5 $&$  2.513            $&$ 0.046            $&$ 1.8 $\\
\hline\hline\end{tabular}
 \captionof{table}{
  \label{tbl:af_eta_fit_coeffs}
  Fit coefficients for the one-dimensional nonspinning \mbox{$\LorboneDeta$} fit
  over \NRcountNS NR cases,
  along with their uncertainties (standard errors)
  and relative errors (Std.err./estimate).
  \vspace{-\baselineskip}
 }
\end{figure}

\begin{figure}[t!]
 \includegraphics[width=\columnwidth]{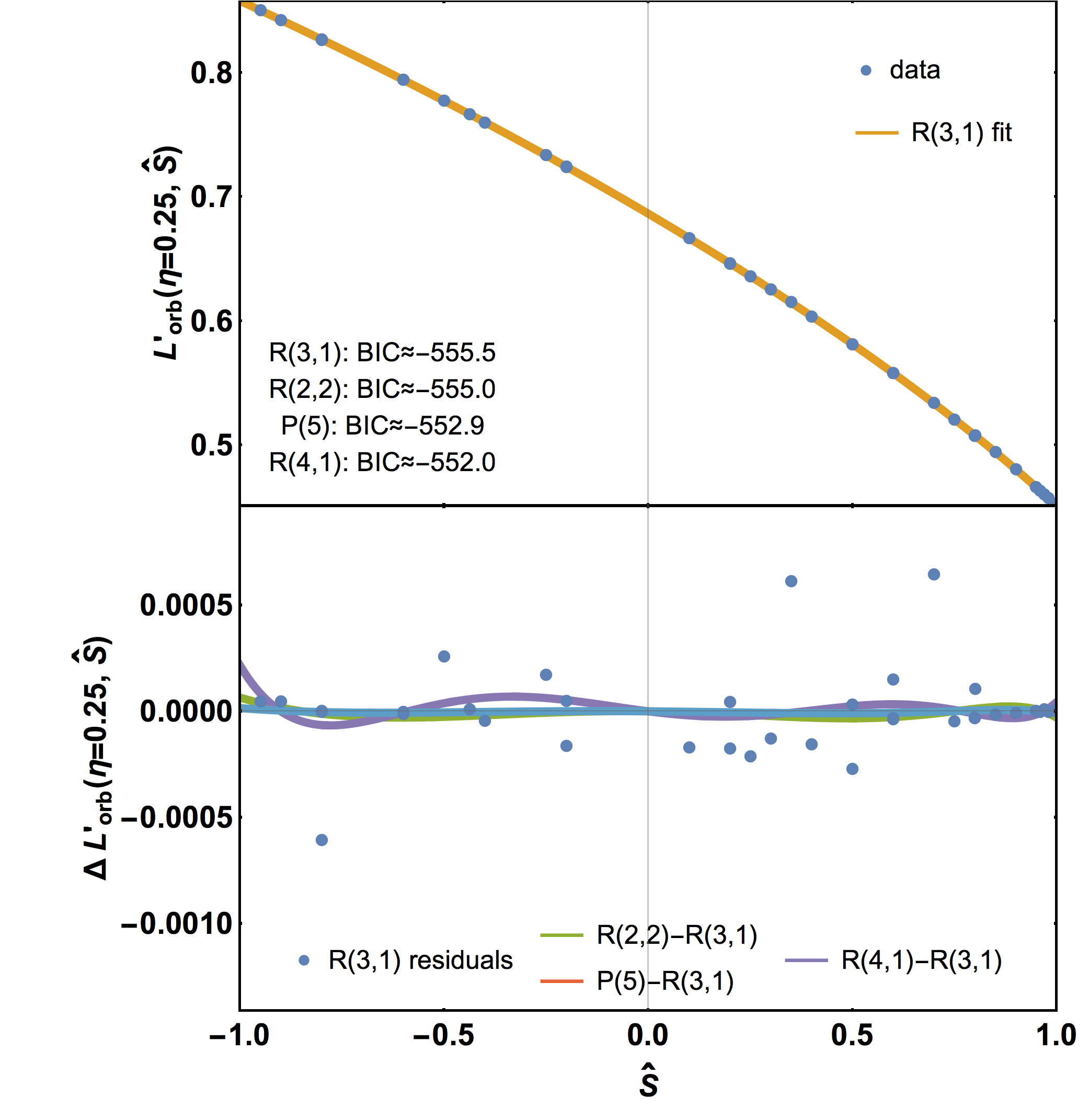}
 \caption{
  \label{fig:Lorb_s_025}
  \EqmeqS NR data and one-dimensional fits of \mbox{$\LorboneDS$} as a function of effective spin $\Seff$.
  Top panel: best fit in terms of BIC,
  a rational function R(3,1), see \autoref{eq:af_Sansatz}.
  Lower panel: residuals of this fit (points)
  and differences from the three next-best-ranking fits in terms of BIC (lines).
 }
 \vspace{0.5\baselineskip}
 \begin{tabular}{lrrr}\hline\hline
  &$ \text{Estimate} $&$ \text{Standard error} $&$ \text{Relative error [$\%$]} $\\\hline$
 b_1 $&$ 1.00096            $&$ 0.00068            $&$  0.1 $\\$
 b_2 $&$ 0.788\hphantom{45} $&$ 0.042\hphantom{45} $&$  5.3 $\\$
 b_3 $&$ 0.654\hphantom{45} $&$ 0.074\hphantom{45} $&$ 11.4 $\\$
 b_5 $&$ 0.840\hphantom{45} $&$ 0.030\hphantom{45} $&$  3.6 $\\
\hline\hline\end{tabular}
 \captionof{table}{
  \label{tbl:af_S_fit_coeffs}
  Fit coefficients for the one-dimensional \eqmeqS \mbox{$\LorboneDS$} fit
  over \NRcountEqSqone NR cases.
  \vspace{-\baselineskip}
 }
\end{figure}

Figure~\ref{fig:Lorb_eta_0} shows the top-ranked fit in terms of Schwarz's Bayesian information criterion (BIC),
which is a rational function of order (3,1):
\begin{equation}
 \label{eq:af_etaansatz}
 \LorboneDeta = \frac{1.3 a_3 \eta ^3+5.24 a_2 \eta ^2+2 \sqrt{3} \eta }{2.88 a_5 \eta +1} \,.
\end{equation}
The fit coefficients $a_i$ along with their uncertainties are given in Table~\ref{tbl:af_eta_fit_coeffs};
all are well determined.
The exact ranking of fits can depend on the choice of fit weights (see Appendix~\ref{sec:appendix-data}) and on the ranking criterion,
but we find that \autoref{eq:af_etaansatz} is top-ranked by both BIC and AICc.
While only ranked 6th by RMSE, none of the considered fits is better than \autoref{eq:af_etaansatz} by more than 6\% in that metric either.
Additionally, under variations of the weighting scheme,
this is robustly the fit among the top-ranked group -- by all three criteria --
with the lowest number of fitting coefficients,
indicating it is a robust choice.
For comparison, a simple third-order polynomial (two free coefficients) is disfavored clearly,
by more than a factor of 8 in RMSE and an offset of +452 in BIC,
and a fourth-order polynomial (three free coefficients, just as \autoref{eq:af_etaansatz}) by almost a factor of 2 and by +332 respectively.

The lower panel of \autoref{fig:Lorb_eta_0} also compares the preferred fit both to the NR data and to the three next-best ranking fits by BIC.
We find that the residuals are centered around zero with no major trends,
while the differences among high-ranked fits are much smaller than the scatter of residuals for the well-covered high-$\eta$ range,
and that the ``systematic uncertainty'', as indicated by the difference of high-ranked fits,
is still at the same level even in the extrapolatory low-$\eta$ region.
The BIC ranking for this example is also illustrated in \autoref{fig:af_eta_fit_BIC} in Appendix~\ref{sec:appendix-stats}.

Before the second 1D fit in the effective spin parameter $\Seff$,
we need to decide how we are later going to construct a 2D ansatz combining both 1D fits.
We can either subtract or divide the NR data by the nonspinning fit,
and find that subtraction exhibits a simpler functional form.
Thus we decide to construct a 2D ansatz as the sum of nonspinning and spinning contributions.
(We will choose a product ansatz for the radiated energy discussed in the next section.)
We thus constrain the constant term of the 1D ansatz in $\Seff$
to reproduce the \mbox{$\eta=0.25$} nonspinning result,
i.e. \mbox{$\ourLorb(\eta=0.25,\Seff=0)$} must be identical for both 1D fits.

The top-ranked \mbox{$\LorboneDS$} fits by BIC are shown in \autoref{fig:Lorb_s_025},
and again we find a unique top-ranked fit by both AICc and BIC,
a rational function of order (3,1):
\begin{equation}
 \label{eq:af_Sansatz}
 \small
 \LorboneDS =  \frac{0.00954 b_3 \widehat{S}^3+0.0851 b_2 \widehat{S}^2-0.194 b_1 \widehat{S}}{1-0.579 b_5 \widehat{S}}+0.68637 \,,
\end{equation}
with four fit coefficients $b_i$ as listed in Table~\ref{tbl:af_S_fit_coeffs}.
This ansatz is ranked 8th by RMSE,
but with only 3\% difference from the lowest RMSE,
which is attained by a P(5) fit with one more coefficients,
marginally disfavored by about +1.7 AICc and +2.6 in BIC.
The best three-coefficient fit R(2,1) is significantly worse,
with differences of over +40 in BIC/AICc and 40\% in RMSE.
Again, the distribution of residuals is well behaved,
and differences between the four top-ranked fits by BIC are smaller than the scatter of residuals.

\begin{figure}[t!]
 \includegraphics[width=\columnwidth]{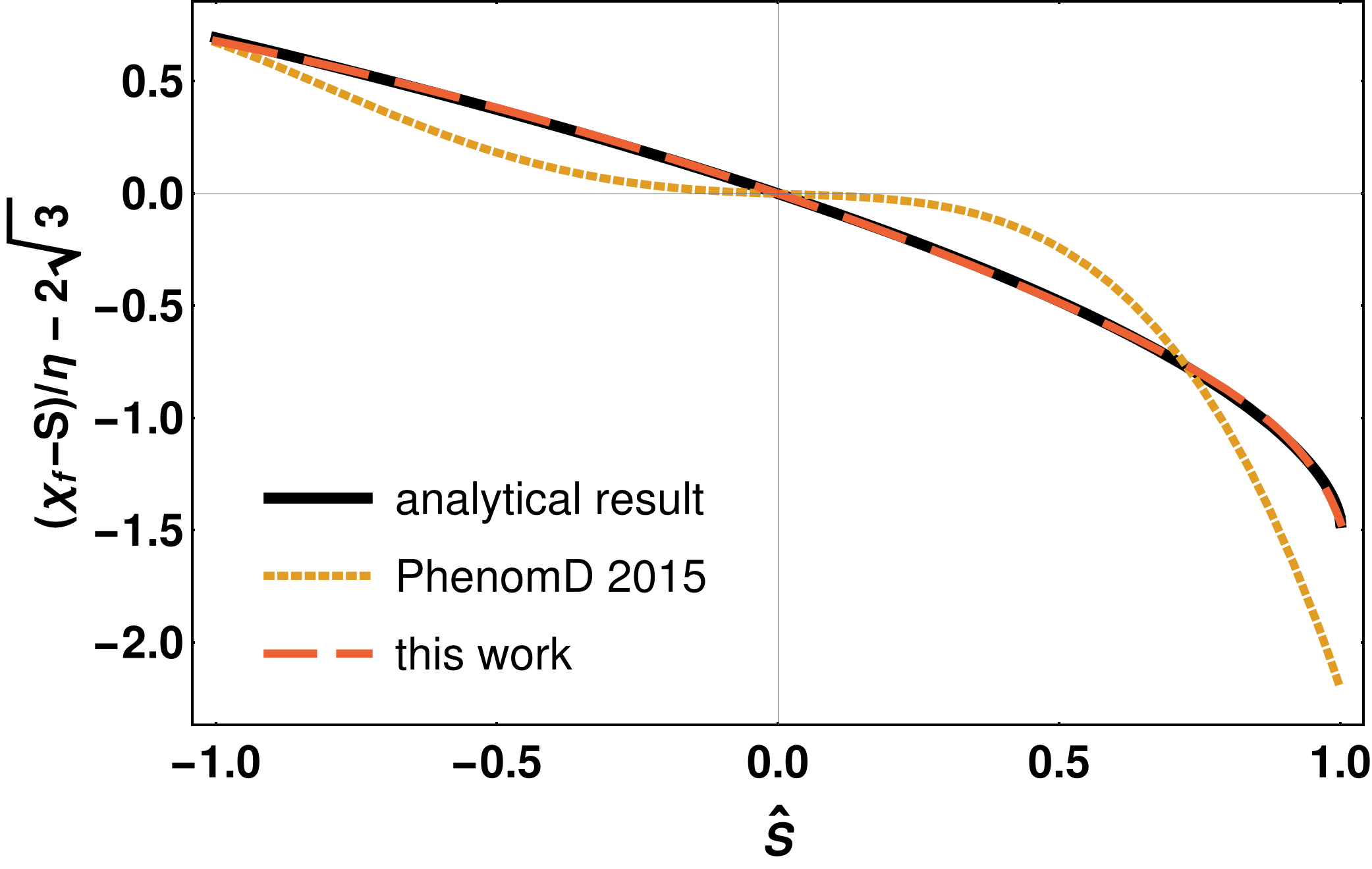}
 \caption{
  \label{fig:extreme_mass_ratio_af}
  \Emr comparison of the rescaled final spin: analytical results from solving \autoref{eq:af_extreme_limit},
  the previous PhenomD final-spin fit of~\cite{Husa:2015iqa}, and this work.
 }
 \vspace{0.5\baselineskip}
 \begin{tabular}{lrrr}\hline\hline
  &$ \text{Estimate} $&$ \text{Standard error} $&$ \text{Relative error [$\%$]} $\\\hline$
 f_{21} $&$  8.774\hphantom{4} $&$ 0.019\hphantom{4} $&$ 0.2 $\\$
 f_{31} $&$ 22.83\hphantom{34} $&$ 0.27\hphantom{34} $&$ 1.2 $\\$
 f_{50} $&$  1.8805            $&$ 0.0025            $&$ 0.1 $\\$
 f_{11} $&$  4.4092            $&$ 0.0047            $&$ 0.1 $\\
\hline\hline\end{tabular}
 \captionof{table}{
  \label{tbl:af_eta0_coeffs}
  Fit coefficients for the \emrl of the final spin,
  fitted to discretized analytical results.
  The fourth coefficient, $f_{11}$, is fixed by the derivative constraint in~\autoref{eq:af_derivconstr_fitted}
  and its estimate and error computed from the others.
  \vspace{-\baselineskip}
 }
\end{figure}

\subsection{Two-dimensional fits}
\label{sec:spinfits-2d}

Next, we want to construct a two-dimensional fit covering the \mbox{$(\eta,\Seff)$} space,
as it was illustrated in \autoref{fig:eta_S_af_NR_extreme},
by combining both the 1D subspace fits and the \emrl.
As discussed above, we take the sum of \autoref{eq:af_etaansatz} and the spin-dependent terms of \autoref{eq:af_Sansatz}.
We introduce the necessary flexibility to describe 2D curvature and the \emrl by generalizing the $\Seff$-dependent terms,
inserting a polynomial of order $J$ in $\eta$ for each $b_i$ through the substitution
\begin{equation}
 \label{eq:2D_substitution}
 b_i \rightarrow b_i \sum_{j=0}^{j=J} f_{ij} \, \eta^j \,.
\end{equation}

\begin{figure}[thbp]
 \includegraphics[width=\columnwidth]{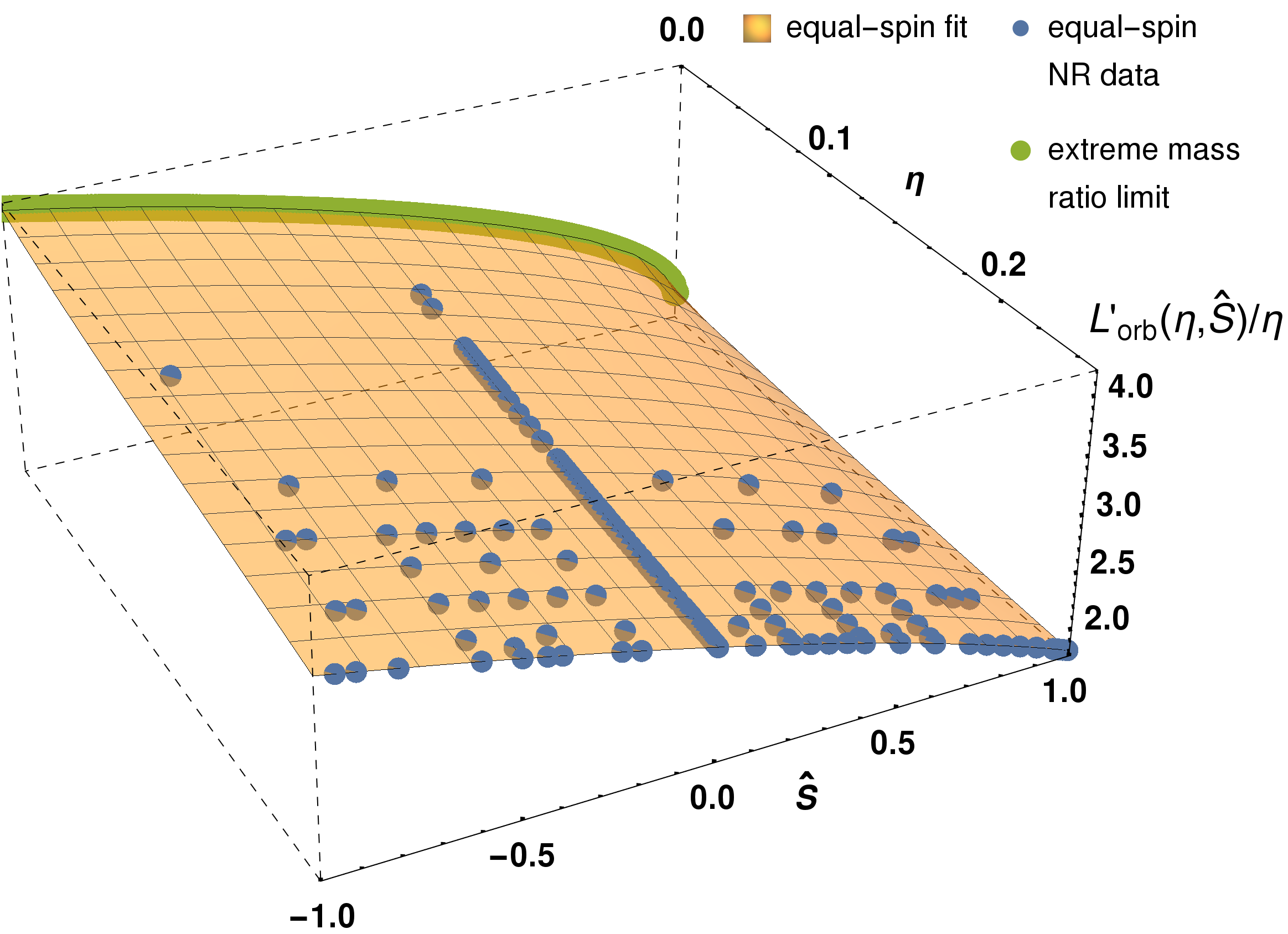}
 \caption{
  \label{fig:Lorb_2dfit}
  Two-dimensional \mbox{$\LorbtwoD$} fit,
  visualized as \mbox{$\LorbtwoD/\eta$}.
  Application of the \emrl helps in avoiding extrapolation artifacts
  which would otherwise appear in low-$\eta$, high-$|\Seff|$ regions that are uncovered by NR simulations.
  }
 \end{figure}

The general 2D ansatz is thus
\begin{equation}
 \label{eq:af_2Dansatz}
 \LorbtwoD = \ourLorb\left(\eta,0\right) \, + \, \ourLorb\left(0.25,\Seff,f_{ij}\right) \, - \, \ourLorb\left(0.25,0\right) \,.
\end{equation}

\begin{figure*}[t!]
 \includegraphics[width=0.9\columnwidth]{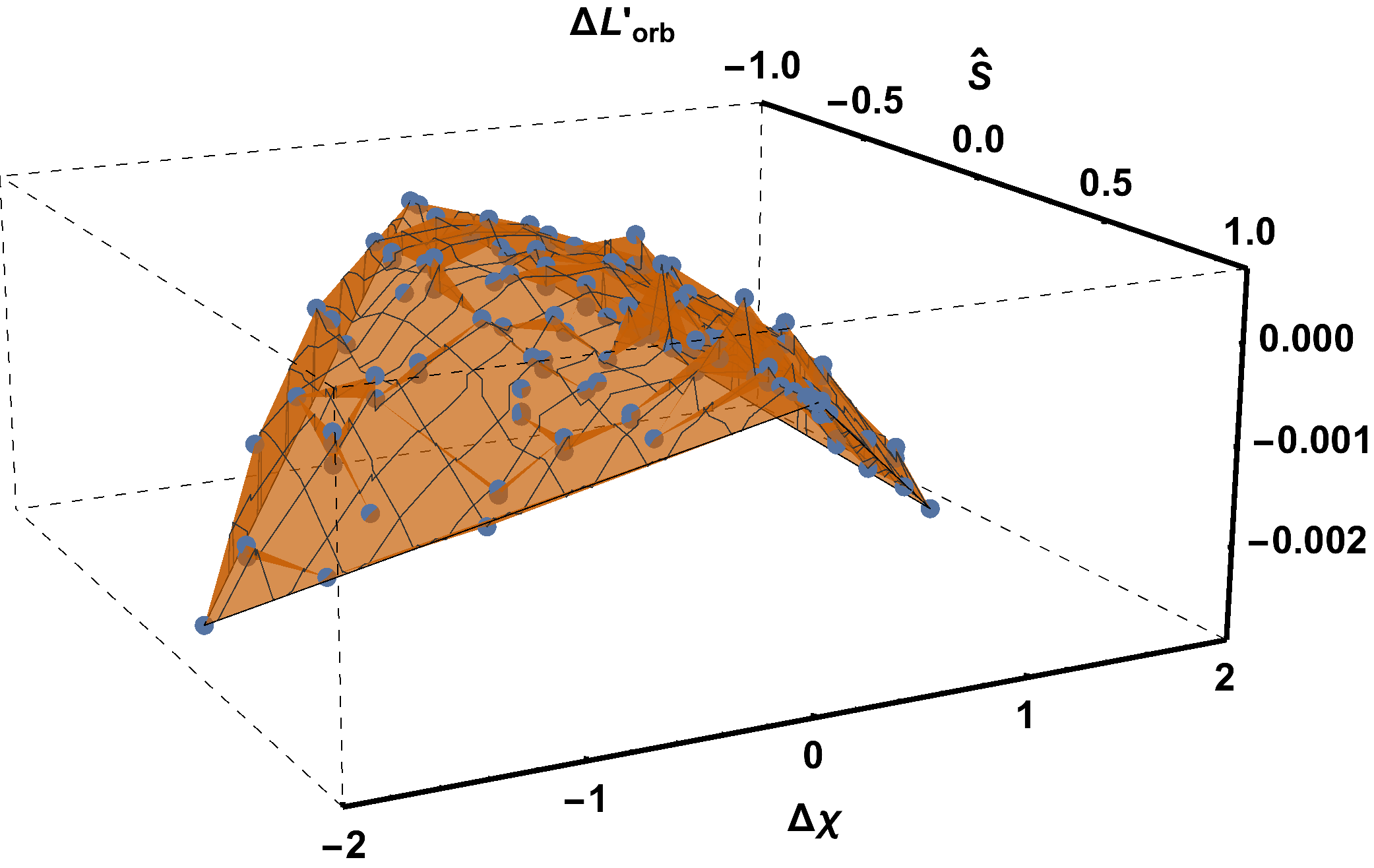} \hspace{1cm}
 \includegraphics[width=0.9\columnwidth]{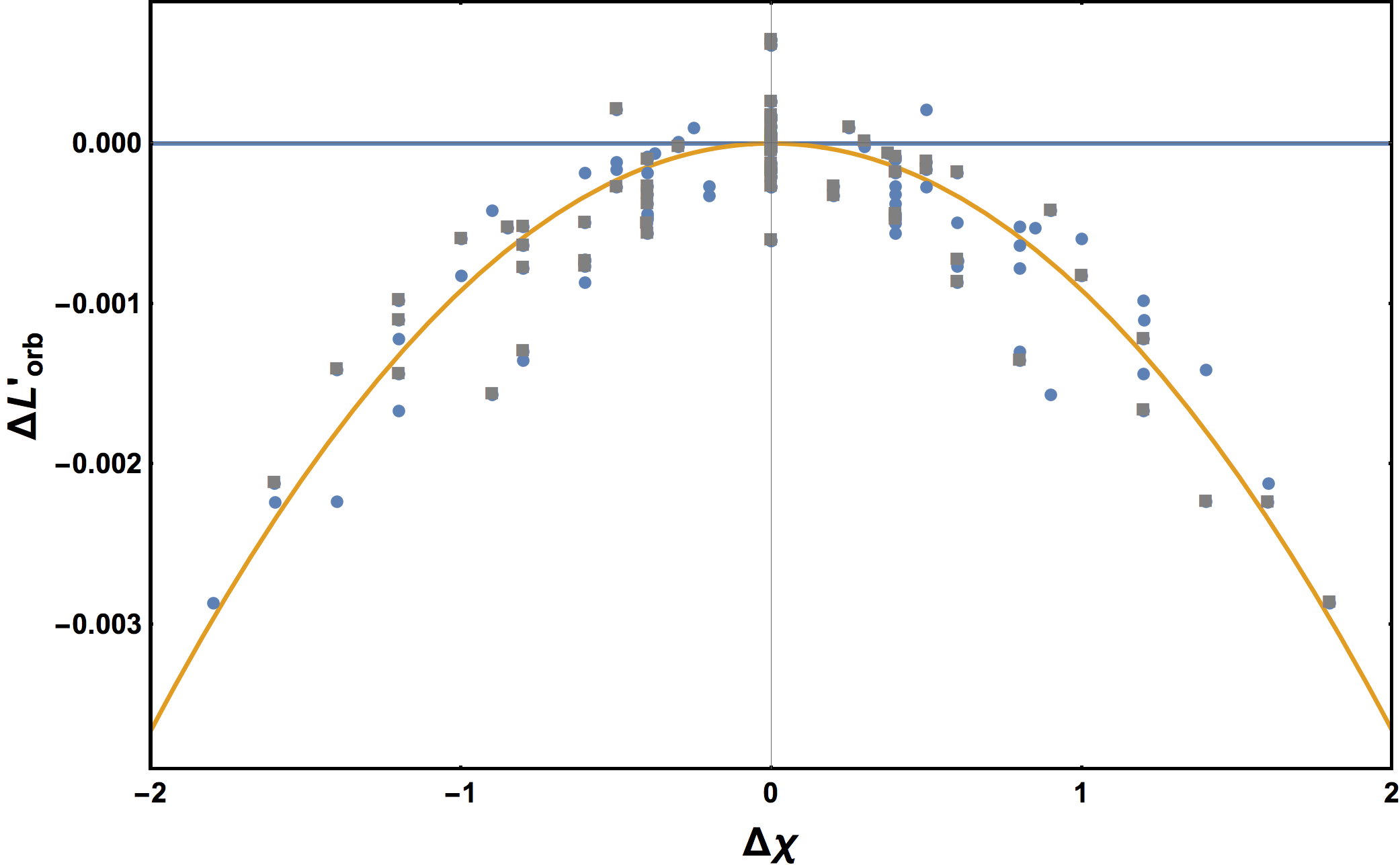} \\[-\baselineskip]
 \includegraphics[width=0.9\columnwidth]{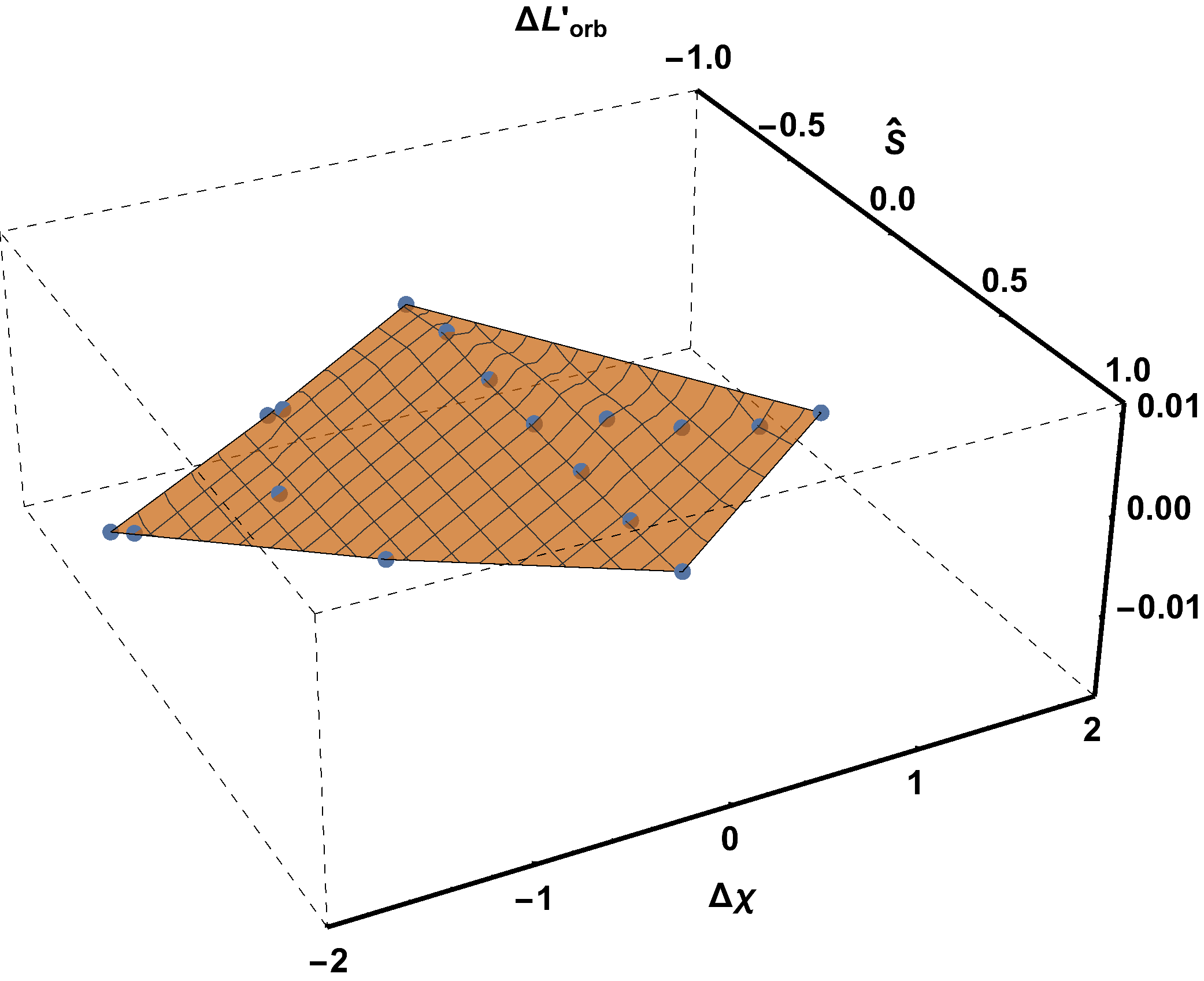} \hspace{1cm}
 \includegraphics[width=0.9\columnwidth]{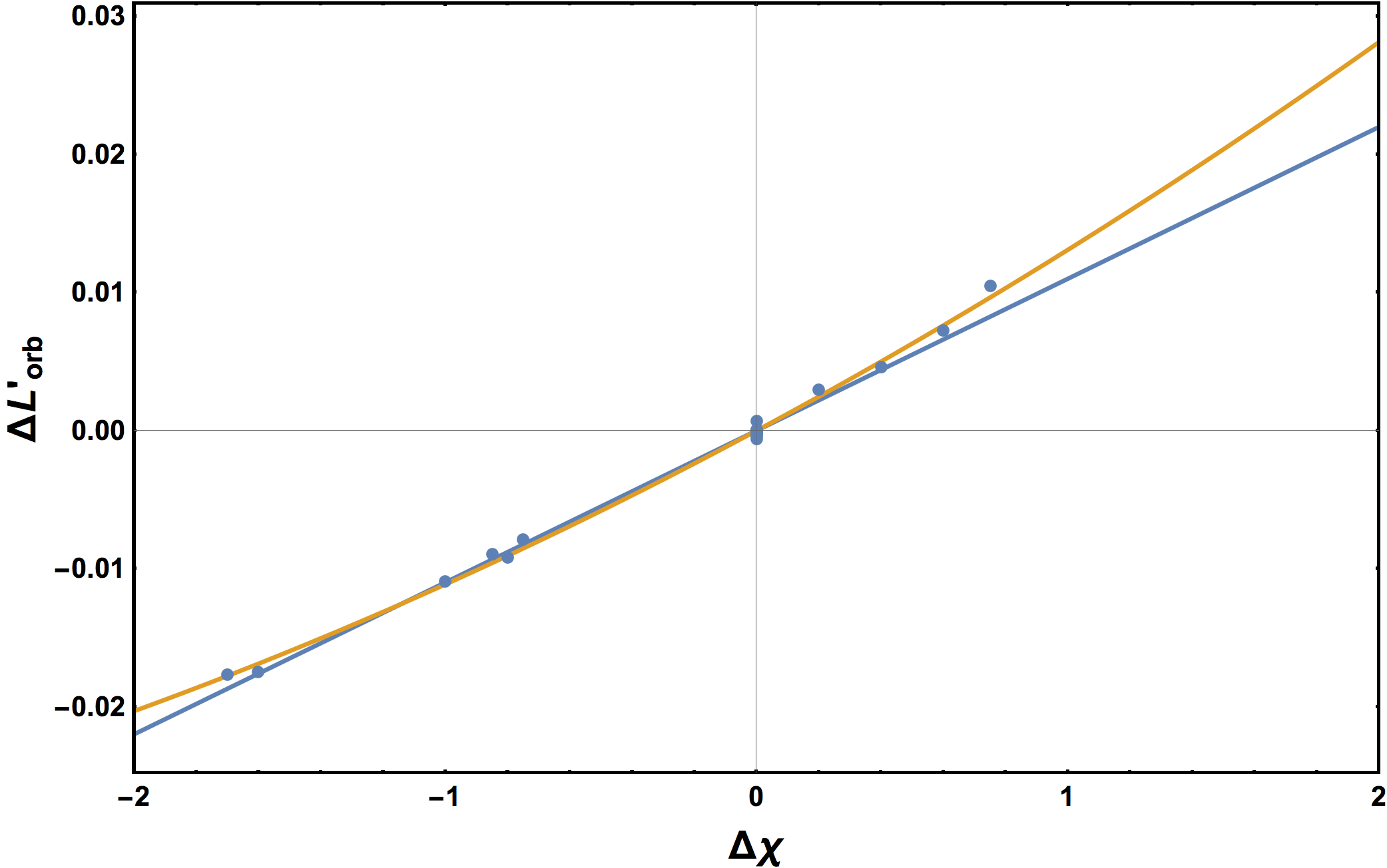}
 \caption{
  \label{fig:spin_diff_per_q}
  Examples of spin-difference behavior at fixed mass ratios,
  for residuals $\DLorb$ after subtracting the two-dimensional \mbox{$\LorbtwoD$} fit, as defined in \autoref{eq:af_deltalorb}.
  Top row: \mbox{$q=1$};
  lower row: \mbox{$q=4$};
  left column: surfaces in \mbox{$\left(\Seff,\chidiff,\DLorb\right)$} space;
  right column: projections onto the $\chidiff$ axis with linear and quadratic fits.
  At equal mass, the surface is parabolic,
  with the linear term (blue line) and mixture term (not shown) vanishing,
  but a clear quadratic dependence (orange line).
  At \mbox{$q=4$} and other intermediate mass ratios, the surface is very close to flat and the linear term dominates.
  \vspace*{-\baselineskip}
 }
\end{figure*}

Here we choose to expand to third order in $\eta$ (\mbox{$J=3$}),
which is the lowest order leaving enough freedom to incorporate all available constraints from the 1D fits and the \emrl,
and, as evidenced by the residuals we find below, also high enough to adequately model this data set.
Of the resulting 16 coefficients, the three $f_{i0}$ in the numerator must vanish to preserve the \mbox{$\ourLorb\left(\eta=0,\Seff\right)=0$} limit,
while consistency with the equal-mass fit from \autoref{eq:af_Sansatz} provides four constraints which we use to fix the $f_{i3}$ terms:
\begin{equation}
 \label{eq:af_2dconstraints}
 f_{i3} = 64 - 64 f_{i0} - 16 f_{i1} - 4 f_{i2} \,.
\end{equation}
Four more coefficients are fixed by the \emr information discussed in~\autoref{sec:data-extreme}:
we re-express~\autoref{eq:af_extreme_limit} in terms of \mbox{$\ourLorb / \eta$} and fit the discretized quantity
\begin{equation}
 \label{eq:af_extreme}
 \lim\limits_{\eta\rightarrow0} \frac{\LorbtwoD}{\eta} - 2 \sqrt{3}
  =
 \lim\limits_{\eta\rightarrow0} \frac{\af\,\left(\eta,S\right) - S}{\eta} - 2 \sqrt{3}
\end{equation}
where $2 \sqrt{3}$ is the linear contribution from the nonspinning part (cf.~\autoref{eq:af_etaansatz})
and the \mbox{$\af\,(\eta\rightarrow0,S)$} values are obtained by solving \autoref{eq:af_extreme_limit} numerically for small $\eta$.
Before fitting, we apply the derivative constraint from \autoref{eq:afp_limit},
which for the sum ansatz \autoref{eq:af_2Dansatz} implies a coefficient constraint
\begin{align}
 \label{eq:af_derivconstr_fitted}
 f_{11}\to 0.345225 f_{21}+0.0321306 f_{31}-3.66556 f_{50}+7.5397.
\end{align}

We find this extra physical constraint to be essential in avoiding superextremal $\af$ results due to fitting artifacts.
The \emrl fit coefficients are listed in Table~\ref{tbl:af_eta0_coeffs},
and the improved agreement between analytical results and this new fit,
as compared with the previous fit of~\cite{Husa:2015iqa},
is illustrated in \autoref{fig:extreme_mass_ratio_af}.

In summary, after constraining to the well-covered one-dimensional NR data subsets and the analytically known \emrl,
the 2D ansatz from \autoref{eq:af_2Dansatz} has reduced from 16 to 5 free coefficients.
We fit it to the \NRcountEqSrest remaining equal-spin NR cases that were not yet included in the 1D subsets.
To remove possible singularities in the \mbox{$\left(\eta,\Seff\right)$} plane for this rather general rational ansatz,
we set the least-constrained denominator coefficient $f_{52}$ to zero.
Thus we obtain a smooth four-coefficient fit as shown in \autoref{fig:Lorb_2dfit}.
It has RMSE four times larger than the 1D $\eta$ fit and twice as high as the $\Seff$ fit,
but these residuals are still smaller than expected unequal-spin effects
and rather noisily distributed without any clear parameter-dependent trends,
thus indicating that the 2D fit sufficiently captures the dominant $\left(\eta,\Seff\right)$ dependence to form the basis of
studying subdominant spin-difference effects.
The least-constrained coefficient at this point has a relative error of about 25\%,
which is good enough to keep it in the ansatz for the next, 3D step where we will refit to a larger data set.

\subsection{Unequal-spin contributions and 3D fit}
\label{sec:spinfits-3d}

\begin{figure*}[thbp]
 \includegraphics[width=0.9\columnwidth]{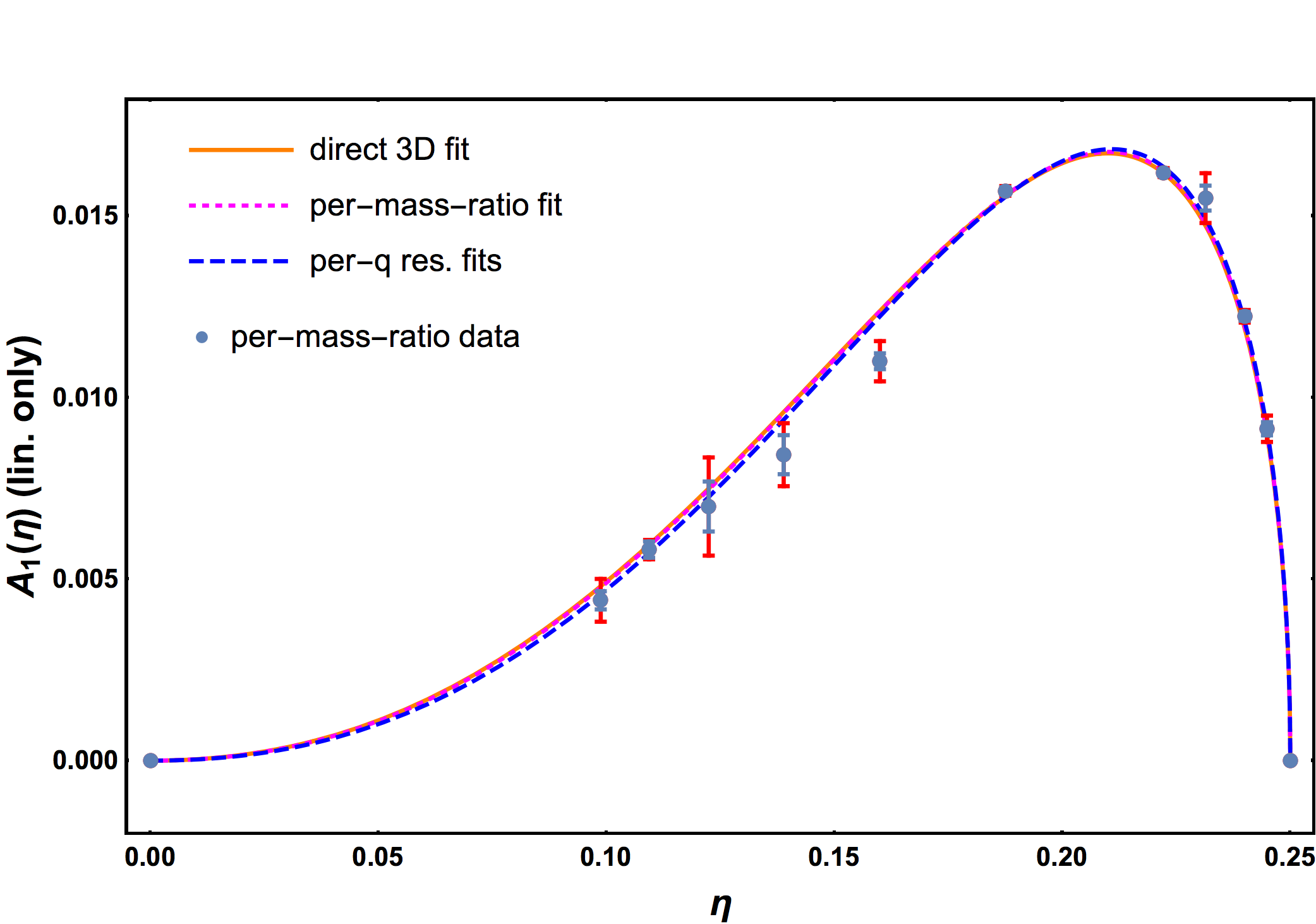} \hspace{1cm}
 \includegraphics[width=0.9\columnwidth]{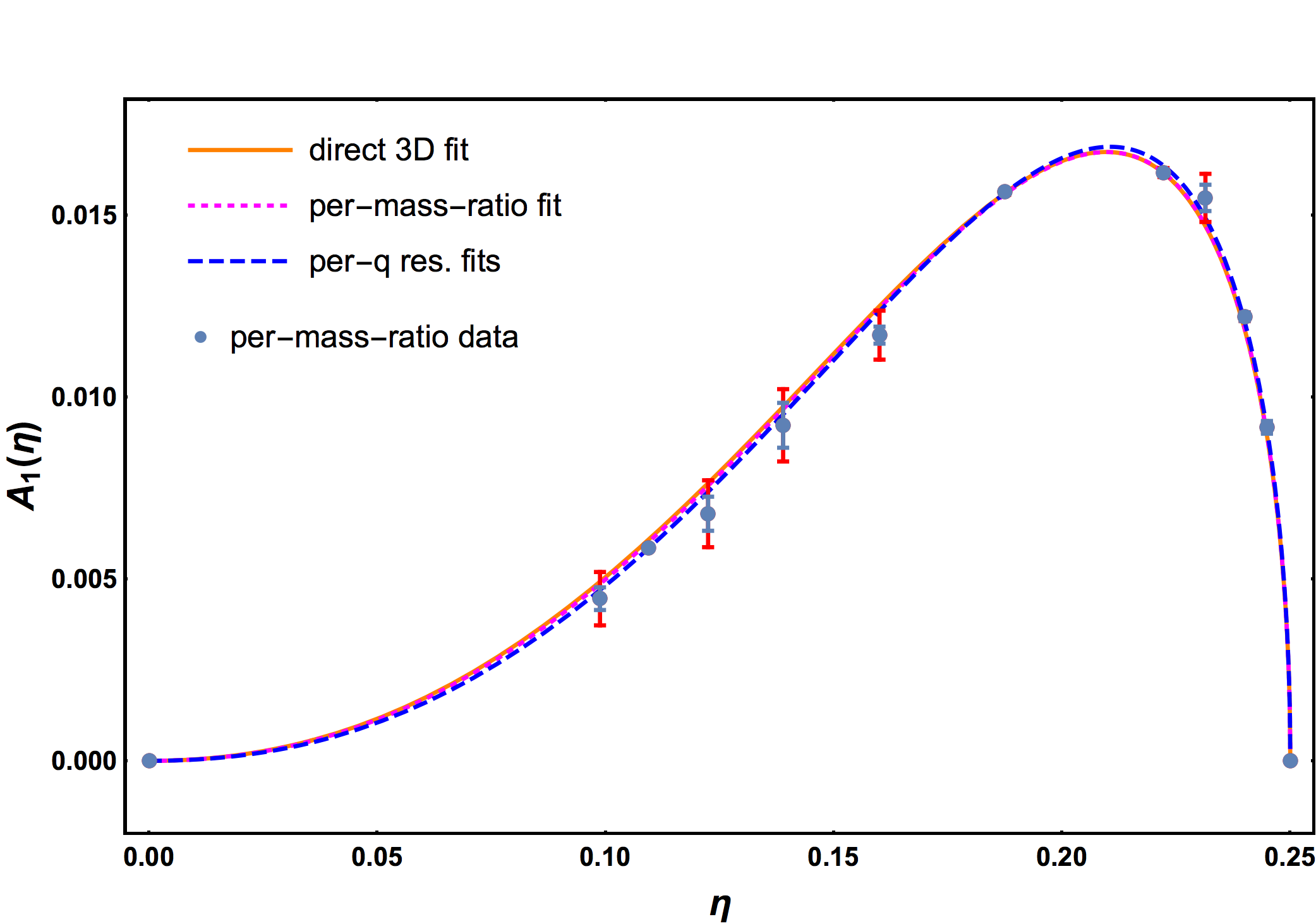} \\[-\baselineskip]
 \includegraphics[width=0.9\columnwidth]{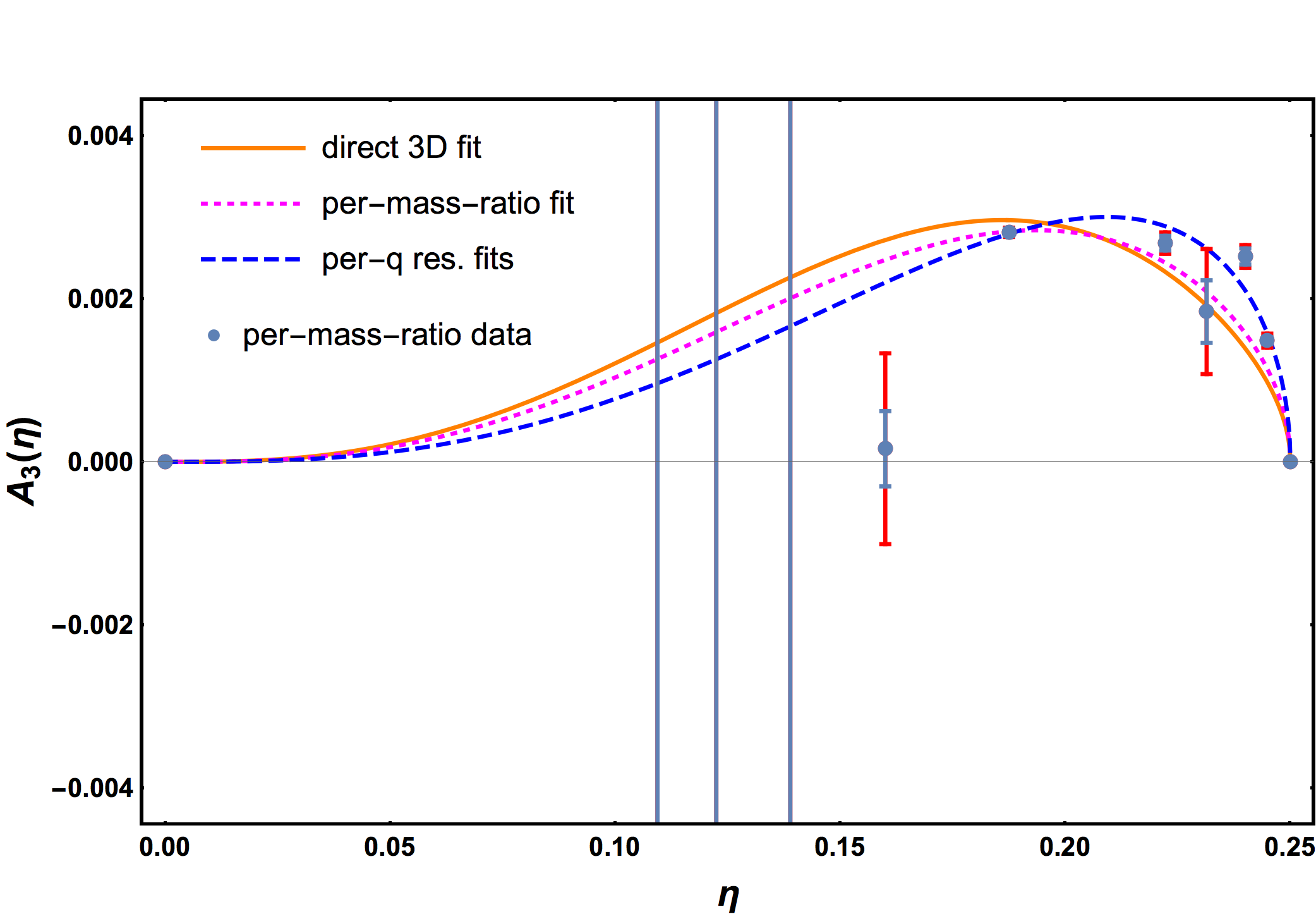} \hspace{1cm}
 \includegraphics[width=0.9\columnwidth]{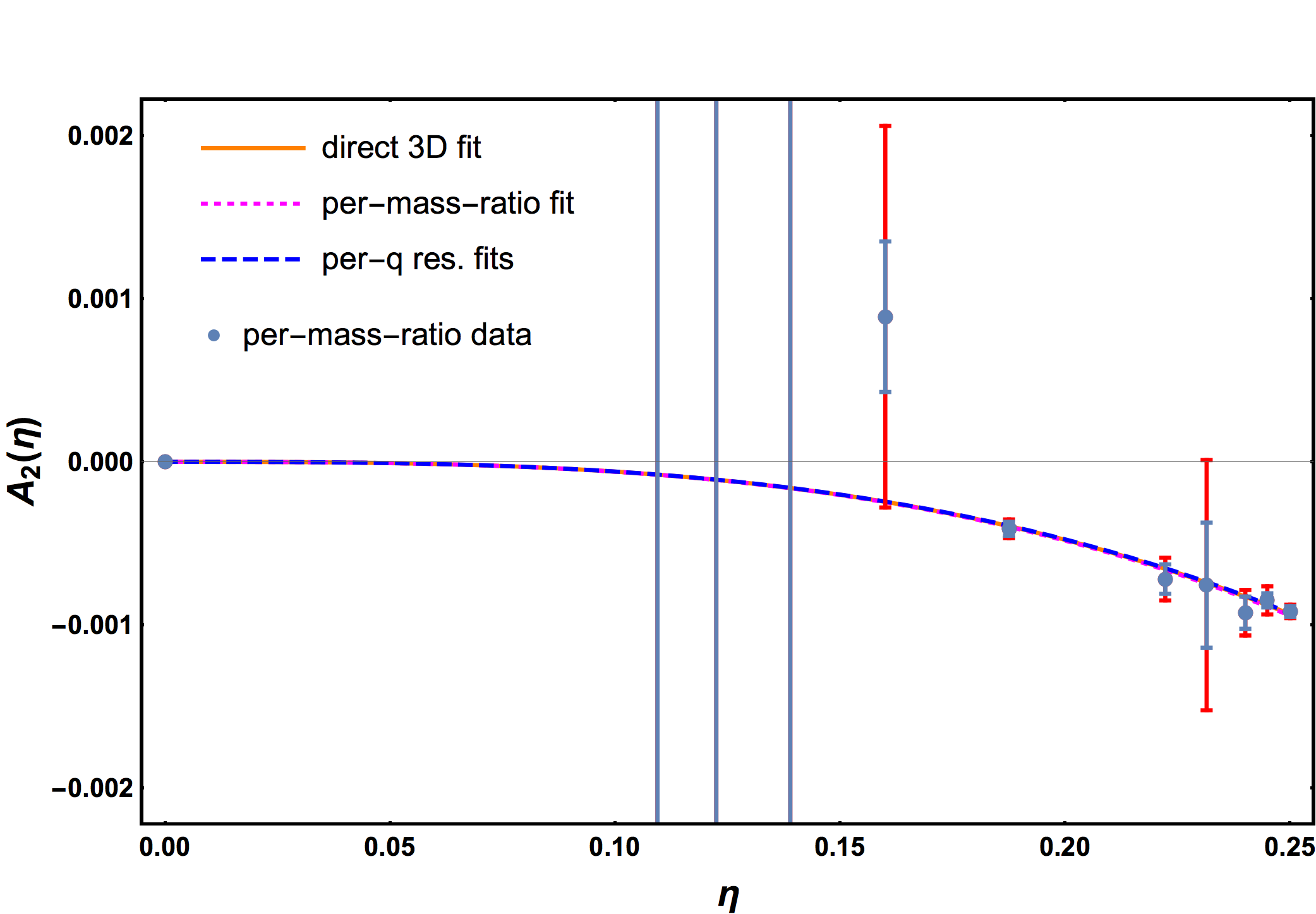}
 \caption{
  \label{fig:af_spin_diff}
  Spin-difference behavior of final-spin data after subtraction of the two-dimensional \mbox{$\LorbtwoD$} fit,
  showing the results of fits as in \autoref{fig:spin_diff_per_q} at $\eta$ steps
  corresponding to \mbox{$q=\{1,\,1.33,\,1.5,\,1.75,\,2,\,3,\,4,\,5,\,6,\,7,\,8\}$}
  and three estimates for the three ansatz functions $A_i(\eta)$ from Eqs.~\ref{eq:af_chidiff_ansatz} and~\ref{eq:af_spindiff_terms}:
  (i) unequal-spin part of the final 3D fit from \autoref{eq:af_finalansatz} (``direct 3D fit''),
  (ii) fit of the unequal-spin terms from \autoref{eq:af_spindiff_terms} (``fit to residuals'')
       to the residuals of the 2D fit from \autoref{eq:af_2Dansatz} over all mass ratios,
  (iii) fits of \autoref{eq:af_spindiff_terms} to the per-mass-ratio results.
  Top-left panel: linear term $A_1$ only.
  The remaining panels are for the combined linear+quadratic+mixture fit, in clockwise order:
  linear term $A_1$,
  quadratic term $A_2$ and
  mixture term $A_3$.
  The $A_1$ results from the combined fit are very similar to those from the linear-only fit,
  demonstrating the robustness of extracting leading-order spin-difference effects.
  For the two lower panels, data points for low $\eta$ are outside the displayed range,
  but the error bars are huge and hence this region does not contribute significantly to the weighted per-mass-ratio fits.
  In the direct 3D fit to the full data set, however, low-$\eta$ information can be better incorporated,
  leading to the somewhat different shape of the mixture-term fit.
  See \autoref{sec:spinfits-assess} for more discussion of how well constrained these shapes actually are with the current data set.
  \vspace{-\baselineskip}
 }
\end{figure*}

Now the final step in the hierarchical procedure is to explore the subdominant effects of unequal spins,
parametrized by the spin difference \mbox{$\chidiff=\chi_1-\chi_2$}.
We first study the residuals of the \NRcountUneqS unequal-spin NR cases under the equal-spin 2D fit:
\begin{equation}
 \label{eq:af_deltalorb}
 \DLorbthreeD := \ourLorbNR\threeDparams \, - \, \ourLorb|_{\mathrm{eqSpinFit}}\left(\eta,\Seff\right) \,.
\end{equation}
We do this at fixed steps in mass ratio,
having sufficient numbers of NR cases for this analysis
at mass ratios \mbox{$q=\{1,\,1.33,\,1.5,\,1.75,\,2,\,3,\,4,\,5,\,6,\,7,\,8\}$}.
This per-mass-ratio analysis is only used to guide the construction of the full 3D ansatz and as a consistency check,
while the final full 3D fit will consist of fitting the constrained 2D ansatz
plus spin-difference terms directly to the full data set.

At each mass ratio, we visually inspect the residuals, which span 2D surfaces in \mbox{$\left(\chi_1,\chi_2,\ourLorb\right)$}
or, equivalently, \mbox{$\left(\Seff,\chidiff,\ourLorb\right)$} space.
As illustrated in \autoref{fig:spin_diff_per_q},
we find surfaces close to a plane,
indicating a dominant linear dependence on $\chidiff$ and possibly a mixture term $\Seff\,\chidiff$.
The exception is at equal masses, 
where quadratic curvature in the $\chidiff$ dimension dominates.
In this case, exchange of $\chi_1$ and $\chi_2$ yields an identical binary configuration,
so that terms linear in $\chidiff$ indeed have to vanish.
We have also exploited this fact in the \mbox{$q=1$} analysis by adding mirror duplicates of each NR data point.
Motivated by these empirical findings and symmetry argument,
we introduce up to three spin-difference terms, 
\begin{equation}
 \label{eq:af_chidiff_ansatz}
 \DLorbthreeD =   A_1(\eta) \, \chidiff
                + A_2(\eta) \, \chidiff^2
                + A_3(\eta) \, \Seff \chidiff \,.
\end{equation}
The full 3D ansatz is then simply the sum of Eqs.~\ref{eq:af_2Dansatz} and~\ref{eq:af_chidiff_ansatz}:
\begin{equation}
 \label{eq:af_finalansatz}
 \LorbthreeD = \LorbtwoD + \DLorbthreeD \,.
\end{equation}
Adding higher orders in the effective spin or spin difference is not supported by visual inspection.
At each mass ratio, we now perform four fits in $\chidiff$ for the values of the $A_i$:
linear, linear+quadratic, linear+mixed, or the sum of all three terms.
Examples are also shown in \autoref{fig:spin_diff_per_q}.

We then collect the coefficients of each of these fits and use them as data $A_i(\eta)$ to be fitted as functions of mass ratio
(see the 'per-mass-ratio data' in~\autoref{fig:af_spin_diff}),
using as weights the fit uncertainty from each mass ratio rescaled by the average data weight for that mass ratio.
We also apply what we know about the \emr and equal-mass limits:
all three $A_i(\eta)$ have to vanish in the limit \mbox{$\eta=0$},
and the $A_1$, $A_3$ linear in $\chidiff$ have to vanish for \mbox{$\eta=0.25$}. 
We thus choose \ansaetze of the form
\begin{equation}
 \label{eq:af_chidiff_ansatz_lin_mix}
 A_i = d_{i0} \, \eta^{p_i} \left(\sqrt{1 - 4 \eta}\,\right)^{q_i}  \left( 1 + d_{i1} \eta \right)
\end{equation}
for $A_{i=1,3}$ linear in $\chidiff$,
where the factor
\mbox{$\left(\sqrt{1 - 4 \eta}\,\right)^{q_i}$}
is motivated from post-Newtonian (PN) results~\cite{Bohe:2013cla,Marsat:2013caa},
and 
\begin{equation}
 \label{eq:af_chidiff_ansatz_quad}
 A_{2}   = d_{20} \, \eta^{p_2} \left( 1 + d_{21} \left(\sqrt{1 - 4 \eta}\,\right)^{q_2} \right)
\end{equation}
for the term quadratic in $\chidiff$.
We find that the data can be well fit without any higher-order terms and
by reducing some of the freedom of these three terms
exploratory fits keeping all coefficients free give results close to integer numbers for the $p_i$, \mbox{$q_i=1$} and \mbox{$d_{21}=0$}.
Hence we choose the three parsimonious \ansaetze
\begin{subequations}
 \label{eq:af_spindiff_terms}
 \begin{align}
  A_1(\eta)&=d_{10} (1-4 \eta )^{0.5} \eta ^2 \left(d_{11} \eta +1\right) \\
A_2(\eta)&=d_{20} \eta ^3 \\
A_3(\eta)&=d_{30} (1-4 \eta )^{0.5} \eta ^3 \left(d_{31} \eta +1\right) \,.
 \end{align}
\end{subequations}

The blue points and lines in \autoref{fig:af_spin_diff} show these per-mass-ratio results.
The shape and numerical results of the dominant linear term $A_1$ are quite stable under adding one or two of the other terms.
Fitting two terms, either linear+quadratic or linear+mixture,
yields quadratic/mixture effects of very similar magnitude,
with the quadratic term following the same basic shape (an intermediate-mass-ratio bulge) as the other two.
However, combining all three terms, the results match better with the expectations from symmetry detailed before,
with the bulge shape limited to the linear and mixture terms
while the quadratic term provides a correction mostly at similar masses.

Using again the \mbox{$q=1$}, \mbox{$\Seff=0$} and \mbox{$\eta\rightarrow0$} constraints on the general ansatz from \autoref{eq:af_finalansatz},
we end up with a total of nine free coefficients in this final step.
We now fit to \NRcountNZSUM cases with arbitrary spins not yet used in the 1D fits,
with results given in Table~\ref{tbl:af_final_fit_coeffs}.
Together with the coefficients from Tables \ref{tbl:af_eta_fit_coeffs}--\ref{tbl:af_eta0_coeffs},
these fully determine the fit.
To convert back from our fit quantity $\ourLorb$ to the actual dimensionless final spin $\af$,
just add the total initial spin
\mbox{$\Stot=m_1^2\,\chi_1+m_2^2\;\chi_2$}.

\begin{table}[t!]
 \begin{tabular}{lrrr}\hline\hline
  &$ \text{Estimate} $&$ \text{Standard error} $&$ \text{Relative error [$\%$]} $\\\hline$
 d_{10} $&$    0.322\hphantom{3} $&$ 0.020\hphantom{4}  $&$  6.2 $\\$
 d_{11} $&$    9.33\hphantom{34} $&$ 0.87\hphantom{34}  $&$  9.3 $\\$
 d_{20} $&$   -0.0598            $&$ 0.0021             $&$  3.5 $\\$
 d_{30} $&$    2.32\hphantom{34} $&$ 0.28\hphantom{34}  $&$ 12.1 $\\$
 d_{31} $&$   -3.26\hphantom{34} $&$ 0.20\hphantom{34}  $&$  6.1 $\\$
 f_{12} $&$    0.512\hphantom{3} $&$ 0.085\hphantom{4}  $&$ 16.7 $\\$
 f_{22} $&$  -32.1\hphantom{234} $&$ 3.6\hphantom{234}  $&$ 11.3 $\\$
 f_{32} $&$ -154\hphantom{.1234} $&$ 10\hphantom{.1234} $&$  6.5 $\\$
 f_{51} $&$   -4.77\hphantom{34} $&$ 0.34\hphantom{34}  $&$  7.1 $\\
\hline\hline\end{tabular}
 \caption{
  \label{tbl:af_final_fit_coeffs}
  Fit coefficients for the final 3D step of the $\ourLorb$ fit to \NRcountNZSUM cases not yet used in the 1D fits of \autoref{sec:spinfits-1d}.
  \vspace{-\baselineskip}
 }
\end{table}

We find that the data set is sufficiently large and clean,
and the equal-spin part modeled well enough from the 2D step,
to confidently extract the linear spin-difference term and its $\eta$-dependence,
which is stable when adding the other terms;
and to find some evidence for the combined mixture and quadratic terms,
whose shape however is not fully constrained yet. 

\subsection{Fit assessment}
\label{sec:spinfits-assess}

In \autoref{fig:af_spin_diff} we also compare the spin-difference terms from this final ``direct 3D'' fit
to those obtained from the per-mass-ratio residuals analysis.
The linear term is fully consistent,
confirming that it is well determined by the data,
while for the quadratic and mixture terms both approaches agree on the qualitative shape,
but do not match as closely.
Under the chosen \ansaetze, the 3D fit coefficients even for those terms are tightly determined (see Table~\ref{tbl:af_final_fit_coeffs}).
However, we have explicitly chosen the spin-difference terms in \autoref{eq:af_spindiff_terms} to achieve this goal,
while several other ansatz choices
(changing the fixed exponents of the multiplicative $\eta$ or \mbox{$\sqrt{1-4\eta}$} terms,
or adding more terms with free coefficients in the $\eta$ polynomials)
can produce fits that are indistinguishable by summary statistics (AICc, BIC, RMSE).
Still, most of these have some strongly degenerate and underconstrained coefficients,
while the reported fit has the desirable property of sufficient complexity to be within the plateau region of summary statistics
while not having any degenerate coefficients.

Yet, the shape of the functions $A_2(\eta)$ and $A_3(\eta)$ for the mixture and quadratic terms is not actually as closely
constrained from the current data set as the coefficient uncertainties alone seem to imply,
due to this ambiguity in ansatz selection.
This becomes clear from the comparison of direct 3D fit and per-mass-ratio analysis in \autoref{fig:af_spin_diff}.
The per-mass-ratio analysis also demonstrates that the data at mass ratios \mbox{$\eta<0.16$}
are not yet constraining enough to help characterize these terms.
(The error bars are so large, and hence the weights so low, that they effectively do not contribute to the fit.)
It also becomes clear that additional unequal-spin data at intermediate mass ratios
would be very useful in constraining the $A_{2,3}(\eta)$ functions.
Meanwhile, it is important to note again that the leading linear spin-difference term
is already determined much more narrowly and robustly with the current data set.

\begin{figure}[t!]
 \includegraphics[width=\columnwidth]{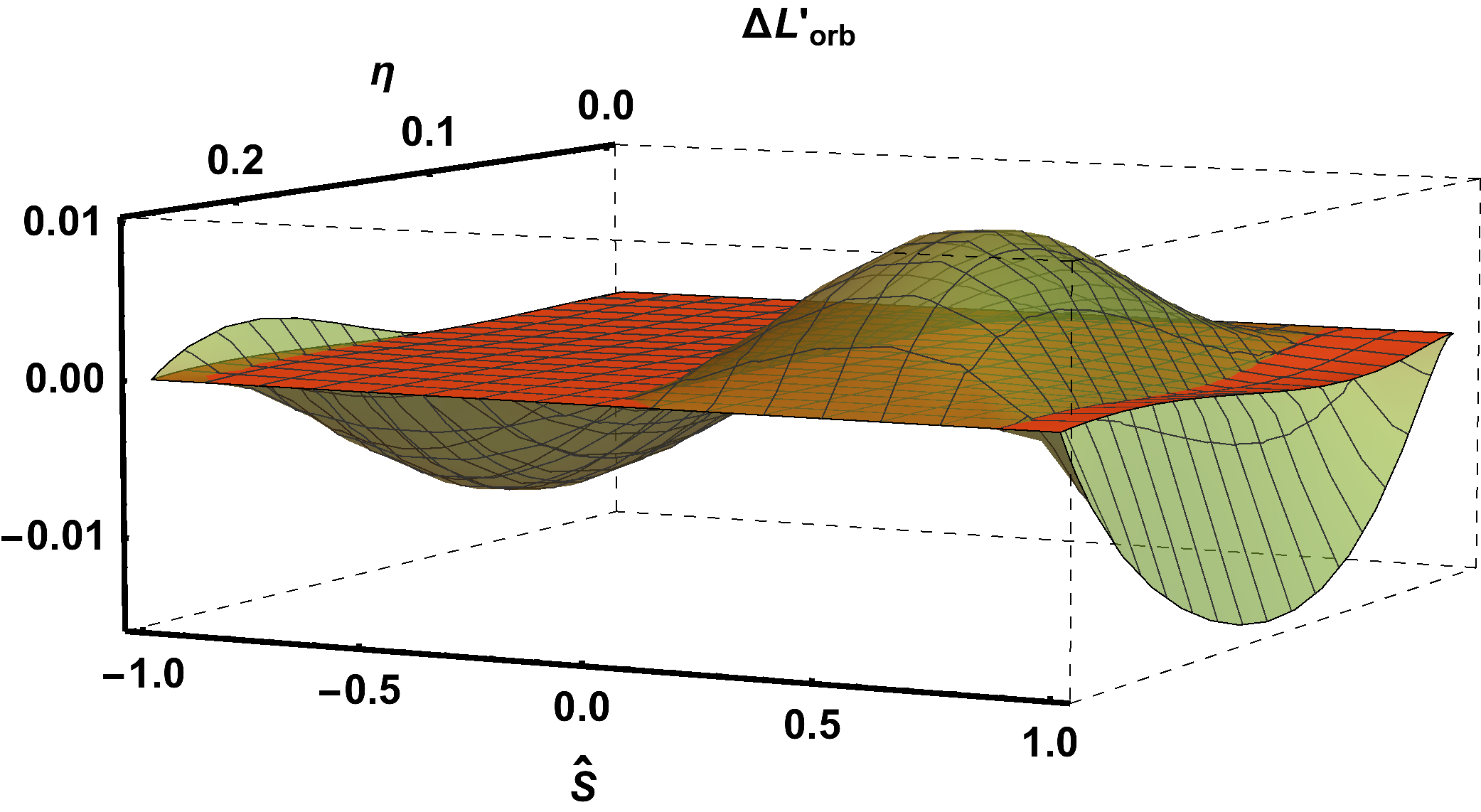}
 \caption{
  \label{fig:af_2deq_2dall_3d}
  Green:
  Difference $\Delta\ourLorb$ of
  a 2D fit (equal-spin physics only) to the full data set
  minus the 2D fit to equal-spin cases only,
  both including \emr constraints.
  The strong curvature at intermediate mass ratios and nonzero spins
  is due to the equal-spin-physics-only fit trying to compensate for the addition of unequal-spin NR cases.\\
  Orange:
  Difference $\Delta\ourLorb$ of
  the 2D part of the 3D fit to the full data set
  minus the 2D-only fit to equal-spin data.
  The bulk of the parameter space is no longer distorted,
  and only at high effective-spin magnitudes a small opposite effect
  to the $\eta$-dependent behavior of the spin-difference terms (cf. \autoref{fig:af_spin_diff})
  can be seen.
 }
 \vspace{0.5\baselineskip}
 \begin{tabular}{lrcccc}\hline\hline
  &$ N_{\text{data}} $&$ N_{\text{coeff}} $&$ \text{RMSE} $&$ \text{AICc} $&$ \text{BIC} $\\\hline$
 \text{1D $\eta $} $&$ 92 $&$ 3 $&$ 9.41\times 10^{-5} $&$ -1590.8 $&$ -1580.7 $\\$
 \text{1D }\hat{S} $&$ 37 $&$ 4 $&$ 2.05\times 10^{-4} $&$ -\hphantom{1}563.6 $&$ -\hphantom{1}555.5 $\\$
 \left.\text{2D (}\chi _1=\chi _2\right) $&$ 60 $&$ 4 $&$ 3.90\times 10^{-4} $&$ -\hphantom{1}880.5 $&$ -\hphantom{1}870.8 $\\$
 \text{2D all} $&$ 298 $&$ 4 $&$ 8.05\times 10^{-3} $&$ -2247.4 $&$ -2229.0 $\\$
 \text{3D lin} $&$ 298 $&$ 6 $&$ 9.20\times 10^{-4} $&$ -3628.4 $&$ -3602.9 $\\$
 \text{3D lin+quad} $&$ 298 $&$ 7 $&$ 8.28\times 10^{-4} $&$ -3765.0 $&$ -3735.8 $\\$
 \text{3D lin+mix} $&$ 298 $&$ 8 $&$ 8.11\times 10^{-4} $&$ -3693.4 $&$ -3660.6 $\\$
 \text{3D lin+quad+mix} $&$ 298 $&$ 9 $&$ 6.10\times 10^{-4} $&$ -4087.3 $&$ -4050.9 $\\
\hline\hline\end{tabular}
 \captionof{table}{
  \label{tbl:af_summary}
  Summary statistics for the various steps of the hierarchical final-spin fit.
  Note that it is not meaningful to compare AICc and BIC between data subsets of different sizes.
  There is statistical preference for the 3D fit including all three linear+mixture+quadratic terms,
  although many different choices of the $A_i(\eta)$ ansatz functions yield similar results
  with just $\pm$ a few percent in RMSE and $\pm$ a few in AICc/BIC,
  so that the shape of the mixture and quadratic terms is not yet fully constrained.
  \vspace{-\baselineskip}
 }
\end{figure}

We can further assess the success of the hierarchical 3D fitting procedure by comparing\vspace{-0.25\baselineskip}
\begin{itemize}
 \item a 2D fit (equal-spin physics only) to equal-spin NR cases only (same as in \autoref{fig:Lorb_2dfit}),\vspace{-0.5\baselineskip}
 \item a 2D fit (equal-spin physics only) to all NR data,\vspace{-0.5\baselineskip}
 \item and the 2D part of the full 3D fit.
\end{itemize}
As shown in \autoref{fig:af_2deq_2dall_3d},
fitting the 2D equal-spin ansatz to the full data set induces strong curvature in the \mbox{$(\eta,\Seff)$} plane,
which the full 3D fit is able to correct by the additional degrees of freedom in the spin-difference dimension.
This is how it was possible to pull out the subdominant spin-difference effects with this enlarged data set.
The same conclusion is supported by the comparison of summary statistics between the various steps and 2D/3D fit variants in Table~\ref{tbl:af_summary},
showing that the RMSE only increases by 50\% from the 2D equal-spin case to the full 3D fit using all data.

\begin{figure}[thbp]
 \includegraphics[width=\columnwidth]{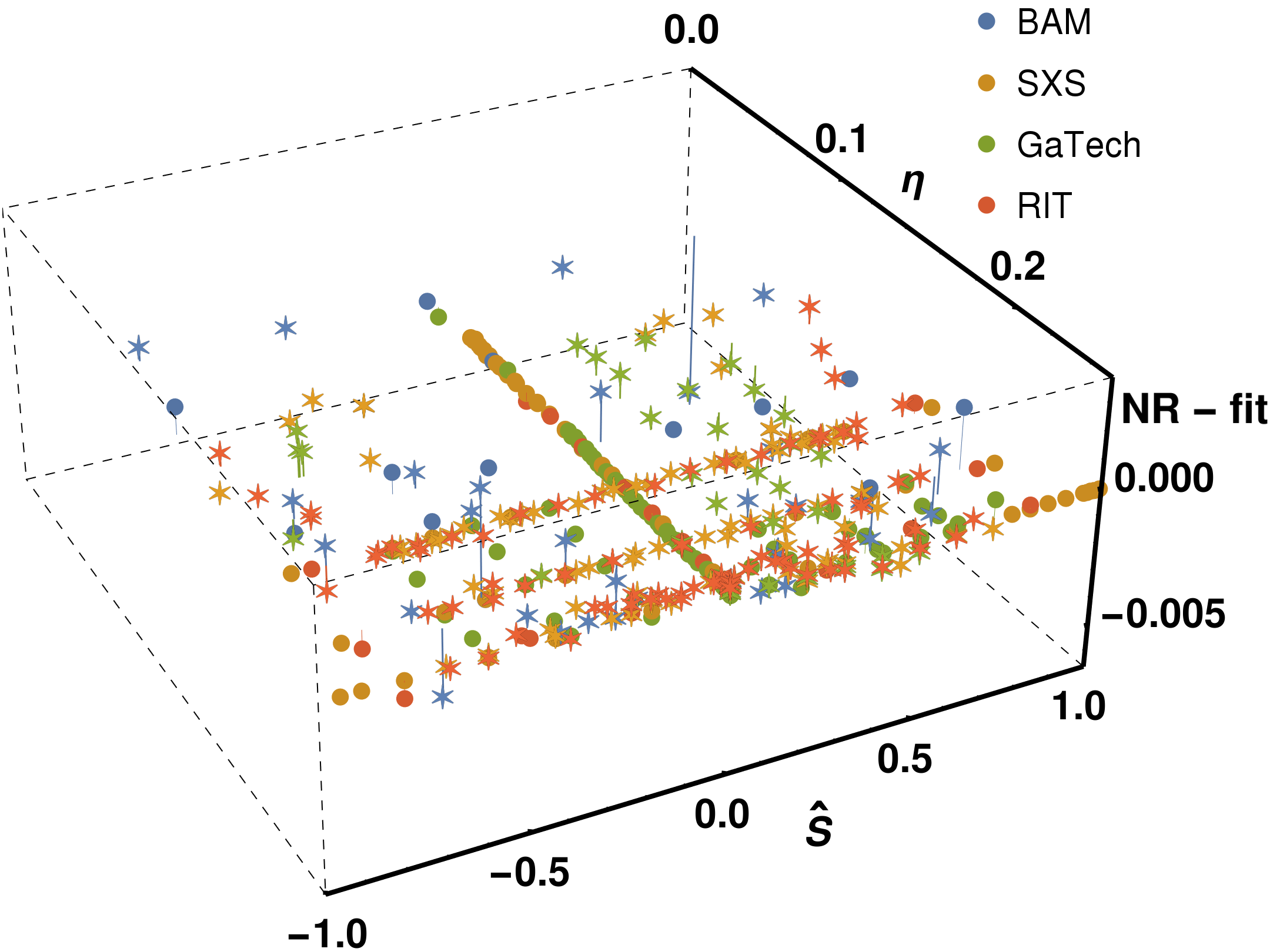}
 \caption{
  \label{fig:af_residuals_paramspace}
  Residuals of the new 3D final-spin fit, projected to the 2D parameter space of $\eta$ and $\Seff$.
  The four NR data sets are distinguished by colors,
  and unequal-spin points highlighted with stars.
  }
\vspace{\baselineskip}
 \includegraphics[width=\columnwidth]{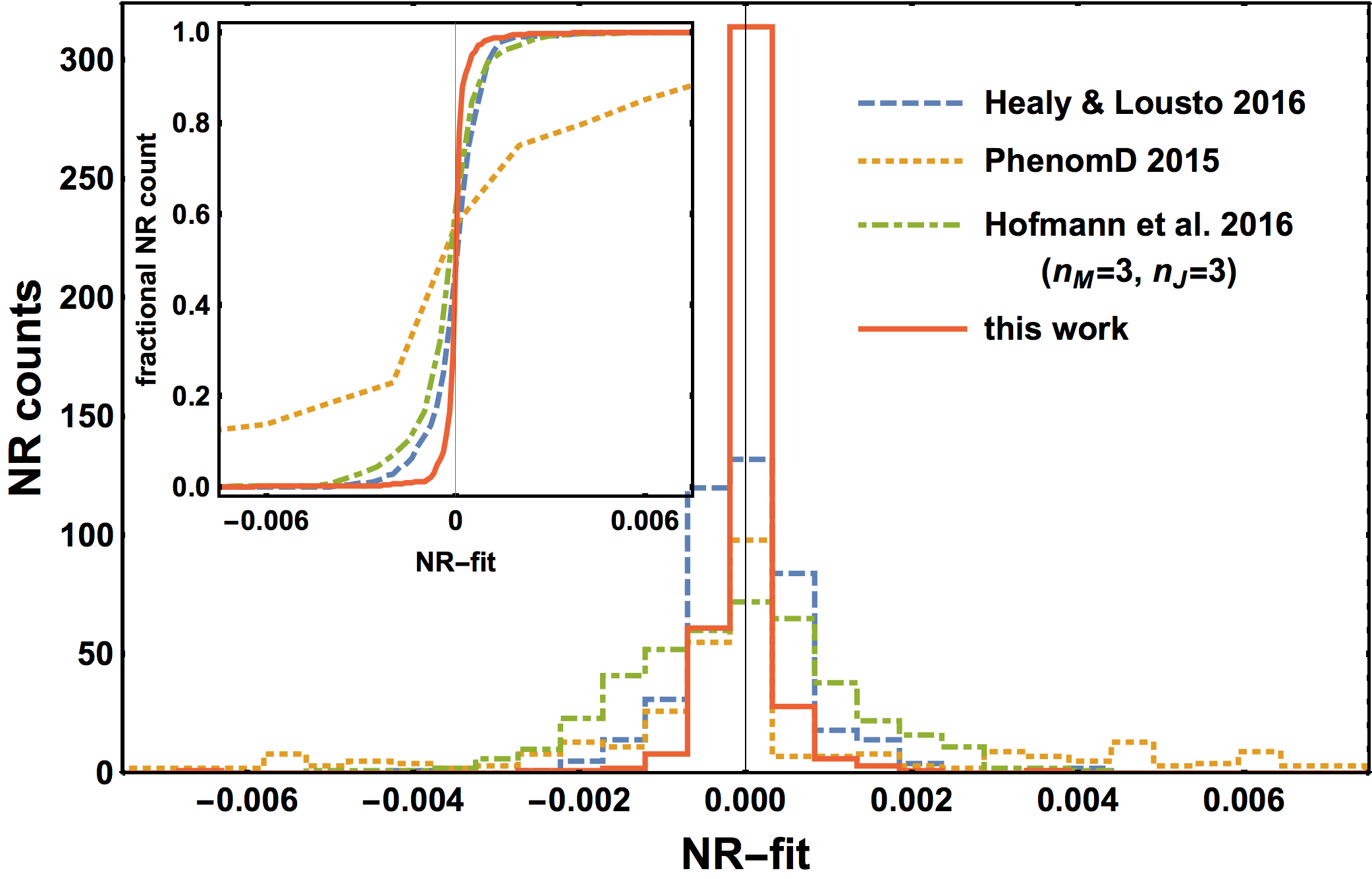}
 \vspace{-1.25\baselineskip}
 \captionof{figure}{
  \label{fig:af_residuals_hist}
  Fit residuals of the final spin $\af$,
  for this work and for previously published fits~\cite{Healy:2016lce,Husa:2015iqa,Hofmann:2016yih},
  evaluated over the set of \NRcount NR simulations shown in \autoref{fig:eta_chi1_chi2_NR}.
  Main panel: histograms, with 102 outliers for PhenomD with \mbox{$|\text{NR}-\text{fit}|>0.0075$} outside of the plot range.
  Inset: cumulative distributions over the same range.
 }
 \vspace{0.5\baselineskip}
 \begin{tabular}{lrrccc}\hline\hline
  &$ N_{\text{coef}} $&$ \text{mean} $&$ \text{stdev} $&$ \text{AICc} $&$ \text{BIC} $\\\hline$
 \text{HLZ2014~\cite{Healy:2014yta}} $&$ 19 $&$ -4.8\times 10^{-5} $&$ 8.9\times 10^{-4} $&$ -5141.0 $&$ -5061.7 $\\$
 \text{HL2016~\cite{Healy:2016lce}} $&$ 19 $&$ 8.1\times 10^{-7} $&$ 7.9\times 10^{-4} $&$ -5358.1 $&$ -5278.9 $\\$
 \text{PhenomD~\cite{Husa:2015iqa}} $&$ 11 $&$ -4.7\times 10^{-5} $&$ 7.2\times 10^{-3} $&$ -3309.0 $&$ -3260.9 $\\$
 \text{ (refit)} $&$ 11 $&$ -1.7\times 10^{-4} $&$ 7.0\times 10^{-3} $&$ -3334.5 $&$ -3286.5 $\\$
 \text{HBR2016~\cite{Hofmann:2016yih}} $&$ 6 $&$ -1.2\times 10^{-4} $&$ 1.4\times 10^{-3} $&$ -4717.2 $&$ -4689.0 $\\$
 \text{ (refit)} $&$ 6 $&$ -1.4\times 10^{-4} $&$ 1.3\times 10^{-3} $&$ -4791.4 $&$ -4763.2 $\\$
 \text{HBR2016~\cite{Hofmann:2016yih}} $&$ 16 $&$ -2.8\times 10^{-4} $&$ 1.2\times 10^{-3} $&$ -4877.3 $&$ -4809.7 $\\$
 \text{ (refit)} $&$ 16 $&$ -1.4\times 10^{-5} $&$ 1.0\times 10^{-3} $&$ -4975.8 $&$ -4908.2 $\\$
 \text{This work} $&$ 16 $&$ -2.3\times 10^{-5} $&$ 5.2\times 10^{-4} $&$ -5991.5 $&$ -5923.9 $\\$
 \text{ (refit)} $&$ 16 $&$ -2.1\times 10^{-5} $&$ 5.1\times 10^{-4} $&$ -6011.3 $&$ -5943.6 $\\$
 \text{ (uniform)} $&$ 16 $&$ -1.2\times 10^{-5} $&$ 5.0\times 10^{-4} $&$ -5240.1 $&$ -5172.5 $\\$
 \text{ (uniform refit)} $&$ 16 $&$ -6.9\times 10^{-6} $&$ 4.9\times 10^{-4} $&$ -5256.8 $&$ -5189.2 $\\
\hline\hline\end{tabular}
 \vspace{-0.75\baselineskip}
 \captionof{table}{
  \label{tbl:af_residuals}
  Summary statistics for the new final-spin fit
  compared with previous fits~\cite{Healy:2014yta,Husa:2015iqa,Hofmann:2016yih,Healy:2016lce},
  evaluated over the \NRcount NR simulations shown in \autoref{fig:eta_chi1_chi2_NR}.
  For Hofmann \textit{et al.}~\cite{Hofmann:2016yih},
  both the \mbox{$(n_M=1, n_J=2)$} fit (6 coefficients) and the \mbox{$(n_M=3, n_J=3)$} version (16 coefficients) are listed.
  The new fit has a total of 16 coefficients calibrated to NR,
  corresponding to Tables~\ref{tbl:af_eta_fit_coeffs}, \ref{tbl:af_S_fit_coeffs} and \ref{tbl:af_final_fit_coeffs},
  not counting those constrained from the \emrl.
  We also show results for refitting previous \ansaetze to the present NR data set,
  for a refit of our hierarchically obtained ansatz directly using the full data set,
  and for the same fitting procedure, but using uniform weights.
  \vspace{-\baselineskip}
 }
\end{figure}

\begin{figure}[t!hbp]
 \includegraphics[width=\columnwidth]{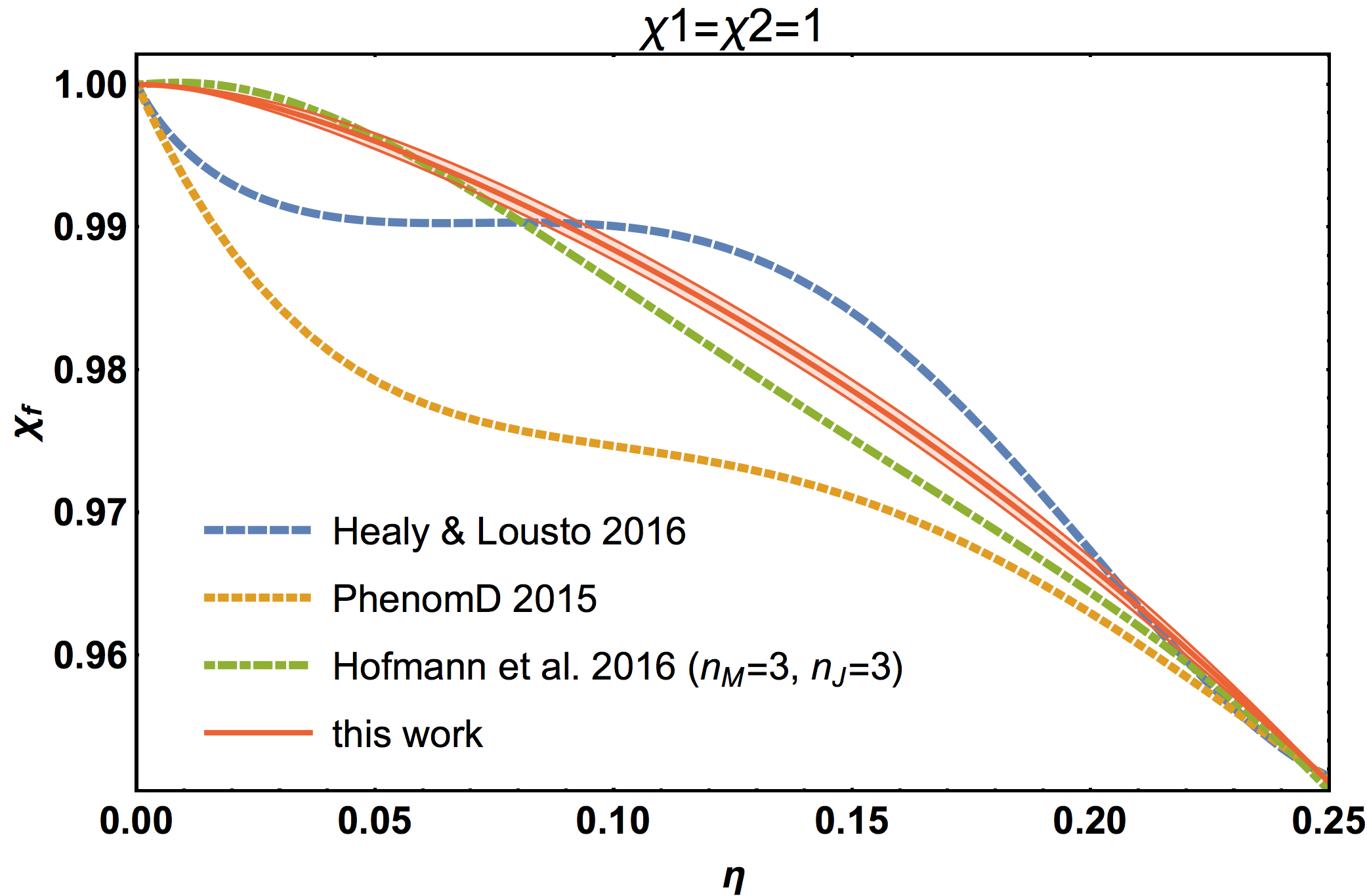}
 \caption{
  \label{fig:af_extreme_spin}
  Comparison of this work with previously published fits~\cite{Healy:2016lce,Husa:2015iqa,Hofmann:2016yih}
  in the limit of extremal aligned spins,
  \mbox{$\chi_1=\chi_2=1$}.
  The shaded region shows our fit's 90\% confidence interval,
  which is narrow enough to indicate that discrepancies with the referenced fits
  are significant and due to the different ansatz constructions, especially in the \emrl (cf.~\autoref{sec:data-extreme}),
  and not just a consequence of insufficient data.
  \vspace{-\baselineskip}
 }
\end{figure}

The distribution of fit residuals for the full data set,
projected onto the \mbox{$(\eta,\Seff)$} plane,
is shown in \autoref{fig:af_residuals_paramspace},
and a comparison of fit residuals with other previously published fits,
over the calibration data set of the current work,
is shown as histograms in \autoref{fig:af_residuals_hist}
and summarized in Table~\ref{tbl:af_residuals} along with AICc and BIC metrics.
The shape of the distributions is consistent,
and for all fits the means are much smaller than the standard deviations,
showing no evidence for any systematic bias.
Our new fit improves significantly over the previous fit~\cite{Husa:2015iqa}
used in the calibration of the IMRPhenomD waveform model~\cite{Khan:2015jqa},
and also yields some improvement over recent fits from other groups~\cite{Healy:2016lce,Hofmann:2016yih},
even when those \ansaetze are refit to our present NR data set.

Refitting our final hierarchically obtained ansatz directly to the full data set produces slightly better summary statistics,
but also allows uncertainties from the less well-controlled unequal-spin set to influence the other parts of the fit,
while the stepwise fit gives better control over the \emr behavior and better-determined coefficients for the well-constrained subspaces.

As a further test of robustness, we have repeated the hierarchical fitting procedure with uniform weights
instead of the weights used so far and discussed in Appendix~\ref{sec:appendix-data}.
This yields a fit consistent with our main result, though slightly less well constrained, but still improving over previous fits,
thus demonstrating the robustness of the hierarchical fit construction under weighting choice.

We have also verified that our new fit
does not violate the \mbox{$\af\leq1$} Kerr bound,
particularly in the extreme-spin limit (\mbox{$\Seff=1$}) and at low $\eta$,
see \autoref{fig:af_extreme_spin}.

\subsection{Precessing binaries}
\label{sec:spinfits-prec}

While some existing final-spin fits~\cite{Zlochower:2015wga,Zlochower:2015wga-erratum,Hofmann:2016yih}
also include a calibration to precessing cases,
it is also possible to use a simple ``augmentation'' procedure~\cite{Rezzolla:2007rz} (see also~\cite{Hughes:2002ei})
for aligned-spin-only calibrations
by adding the contribution of in-plane spins in quadrature to the aligned-spin fit result:
\begin{equation}
 \label{eq:af_augmented}
 \af^\text{aug} = \sqrt{\left(\af^\text{aligned}\right)^2 + \left(S^\text{in-plane}/M^2\right)^2} \;.
\end{equation}
This procedure is known to significantly improve accuracy and reduce bias for precessing binaries.
For example, it has been applied to the aligned-spin PhenomD fit~\cite{Husa:2015iqa} for the precessing PhenomPv2 model~\cite{T1500602,PhenomPv2Paper},
and to the RIT fit~\cite{Healy:2014yta} in recent parameter estimation work of the LIGO-Virgo collaboration~\cite{TheLIGOScientific:2016pea,T1600168,Abbott:2016izl}
(including spin evolution according to~\cite{Ajith:2011ec}).

Applying \autoref{eq:af_augmented} to our aligned-spin fit,
we find a small overshooting of the \mbox{$\left|\af\right|\leq1$} Kerr bound for mass ratios \mbox{$q\gtrsim24$},
when the spin magnitude of the heavier BH is very close to extremal,
and for certain orientation angles $\theta_i$ of the black holes' spins to the angular momentum.
The worst cases give an excess in $\af$ of about 0.12\% at \mbox{$q\sim60$} and intermediate opening angles,
comparable to the aligned-spin fit residuals.
No overshooting occurs if only the linear-in-$\eta$ term in the final spin is used.
Such a small inaccuracy when extending the aligned-spin fit to precessing cases is in principle not surprising,
as this parameter-space region is not covered with NR simulations
and hence the fit slope in this region is purely determined by extrapolation between the NR data and the \emrl,
which we have ensured to be smooth with a flat approach to \mbox{$\af=1$} at \mbox{$\left(\eta=0,\,\chi_1=1\right)$}
(see \autoref{sec:data-extreme} and \autoref{fig:af_extreme_spin}).
Very small inaccuracies in the intermediate-$\eta$ extrapolation region can thus lead to a minimal Kerr violation
when adding the in-plane spins according to \autoref{eq:af_augmented}.
A clean solution to this issue would require more calibration NR simulations in the critical region
and a study of precessing spin contributions in the \emrl.

However, as the overshooting is very small, we have investigated two easy \textit{ad hoc} solutions:
We could take the worst-case point and enforce our 3D fit
to be at or below the corresponding $\ourLorb$ value for the aligned-spin projection \mbox{$\chi_1=\cos(\theta_1)$}
by putting a constraint on one of the $f_{i2}$ coefficients.
This can remove the overshooting at the worst-case point and nearby,
but not over the whole problematic region,
as the fit still has enough freedom in other parameters.
But the accuracy of the aligned-spin fit already suffers from this one extra constraint,
and using constraints on more than one coefficient to pull down the augmented $\af$ over a wider parameter region is fully prohibitive
because insufficient freedom will remain in the fit to properly calibrate to the actual NR data.

\begin{figure}[t!hbp]
 \includegraphics[width=\columnwidth]{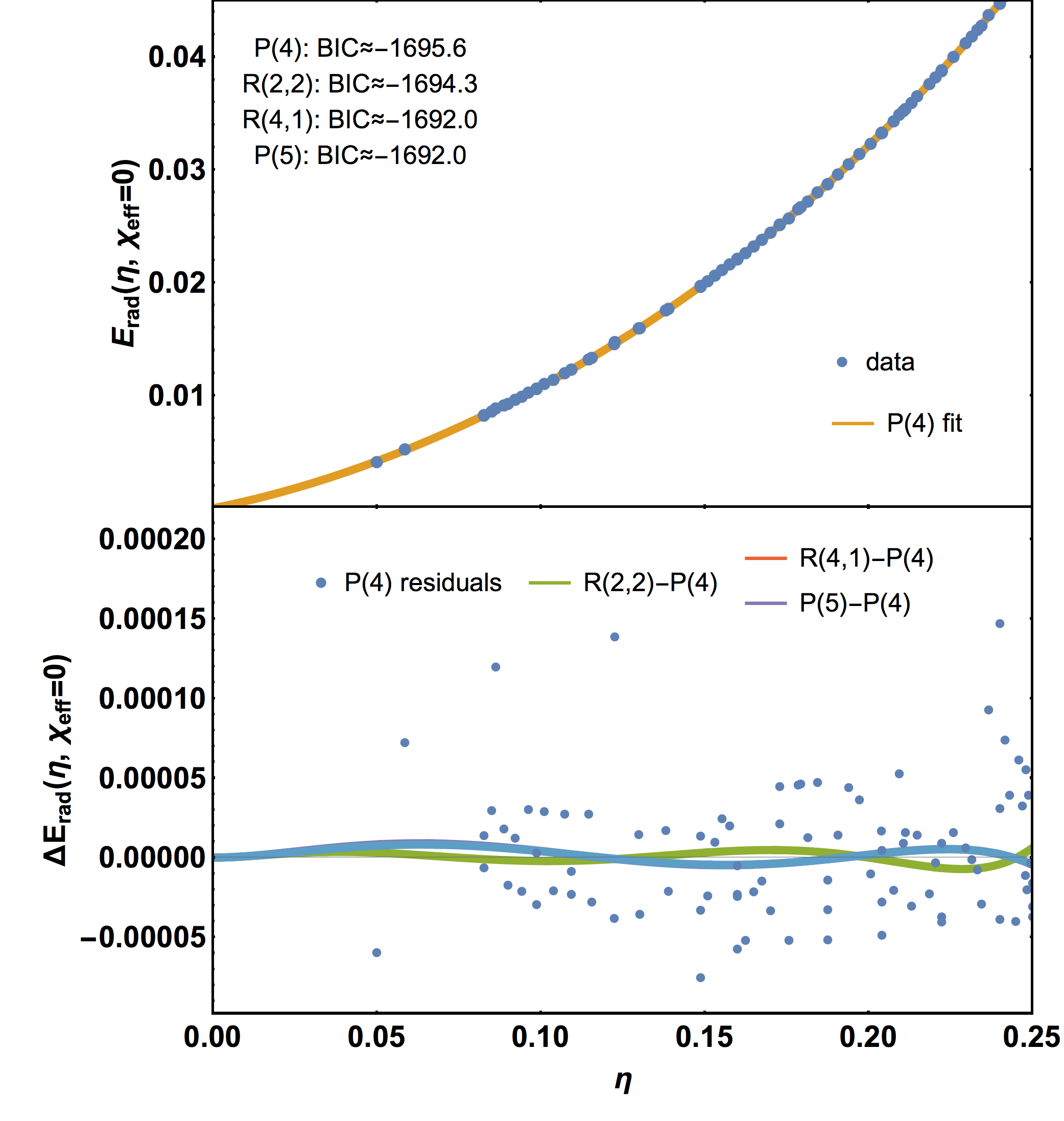}
 \caption{
  \label{fig:Erad_etafits}
  Nonspinning NR data and one-dimensional \mbox{$\EradoneDeta$} fit as a function of mass ratio $\eta$,
  Top panel: selected fit, a fourth-order polynomial P(4), see \autoref{eq:Erad_etaansatz}.
  Lower panel: residuals of this fit (points)
  and differences from the three next-best-ranking in terms of BIC (lines).}
 \vspace{0.5\baselineskip}
 \begin{tabular}{lrrr}\hline\hline
  &$ \text{Estimate} $&$ \text{Standard error} $&$ \text{Relative error [$\%$]} $\\\hline$
 a_2 $&$  0.5610            $&$ 0.0026            $&$ 0.5 $\\$
 a_3 $&$ -0.847\hphantom{4} $&$ 0.027\hphantom{4} $&$ 3.2 $\\$
 a_4 $&$  3.145\hphantom{4} $&$ 0.069\hphantom{4} $&$ 2.2 $\\
\hline\hline\end{tabular}
 \captionof{table}{
  \label{tbl:Erad_eta_fit_coeffs}
  Fit coefficients for the one-dimensional nonspinning \mbox{$\EradoneDeta$} fit
  over \NRcountNS NR cases.
  \vspace{-\baselineskip}
 }
\end{figure}

\begin{figure}[t!hbp]
 \includegraphics[width=\columnwidth]{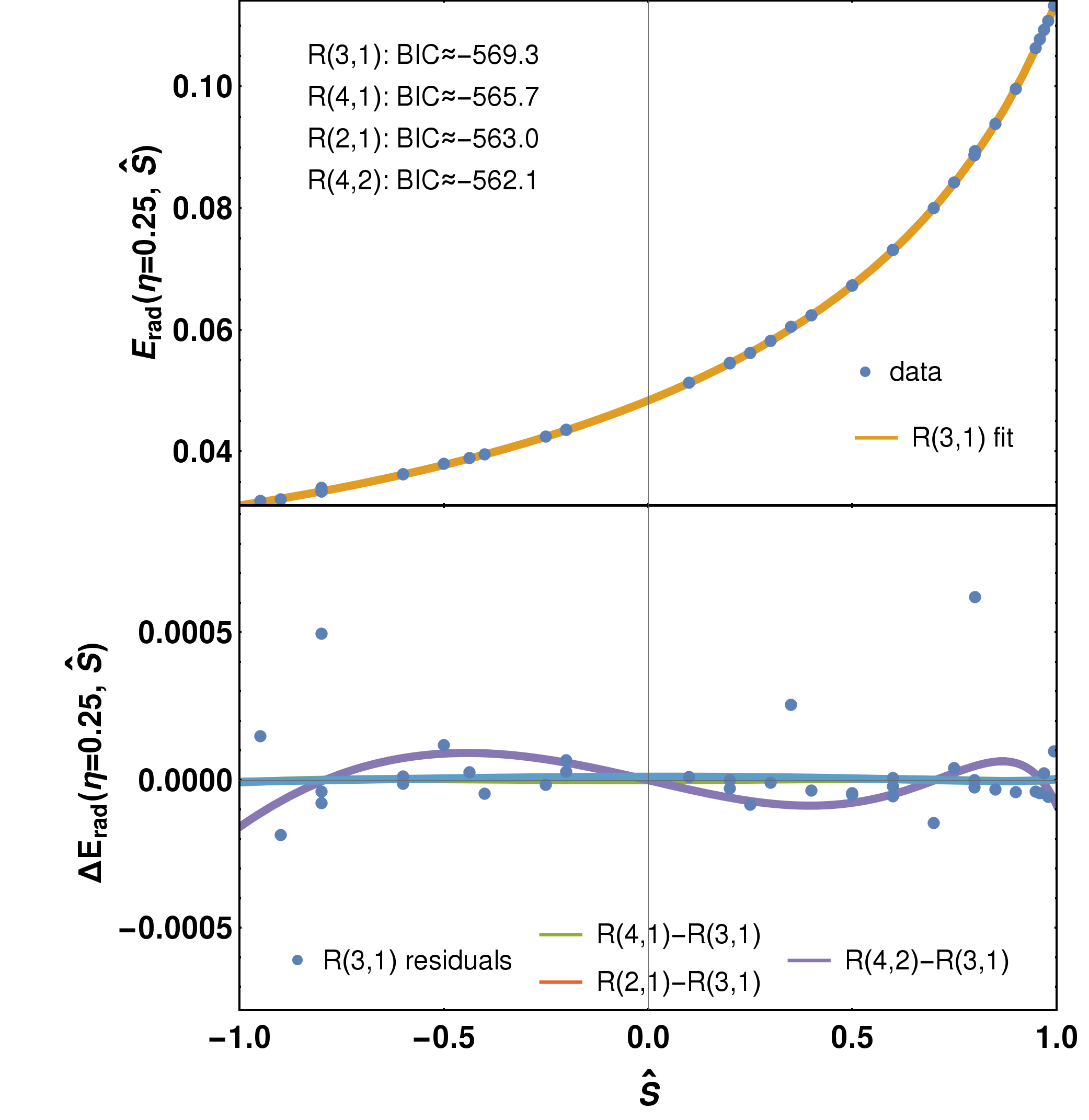}
 \caption{
  \label{fig:Erad_Sfits}  
  \EqmeqS NR data and one-dimensional fits of \mbox{$\EradoneDS$} as a function of effective spin $\Seff$.
  Top panel: selected fit, a rational function R(3,1), see \autoref{eq:Erad_Sansatz}.
  Lower panel: residuals of this fit (points)
  and differences from the three other top-ranking fits in terms of BIC (lines).
 }
 \vspace{0.5\baselineskip}
 \begin{tabular}{lrrr}\hline\hline
  &$ \text{Estimate} $&$ \text{Standard error} $&$ \text{Relative error [$\%$]} $\\\hline$
 b_1 $&$ -0.209 $&$ 0.016 $&$  7.6 $\\$
 b_2 $&$ -0.197 $&$ 0.026 $&$ 13.2 $\\$
 b_3 $&$ -0.159 $&$ 0.049 $&$ 31.1 $\\$
 b_5 $&$  2.985 $&$ 0.034 $&$  1.1 $\\
\hline\hline\end{tabular}
 \captionof{table}{
  \label{tbl:Erad_S_fit_coeffs}
  Fit coefficients for the one-dimensional \eqmeqS \mbox{$\EradoneDS$} fit
  over \NRcountEqSqone NR cases.
  \vspace{-\baselineskip}
 }
\end{figure}

Hence, we opt for an even simpler solution, truncating the augmentation from \autoref{eq:af_augmented} at unity:
$\af=\min\left(\af^\text{aug},1.0\right)$.
This is justified as the overshooting is very small, on the order of the fit residuals,
and limited to an extremal parameter-space region.
The need for this \textit{ad hoc} truncation will reduce or become obsolete
when low-$\eta$-high-spin NR simulations and/or precessing \emr information become available.
A detailed comparison of fit accuracies over a representative set of precessing NR runs is left to future work.

\section{Final mass and radiated energy}
\label{sec:energyfits}

To fit the final mass of remnant BHs from BBH mergers,
we use the same hierarchical approach as for the final spin, summarized in \autoref{fig:flowchart},
with only minor modifications.
The quantity we are going to fit here is the dimensionless radiated energy,
\mbox{$\Erad=M-\Mf=1-\Mf$}.
For \mbox{$\eta=0$}, it has to vanish even in spinning cases,
while the analytical expectation for the leading order in $\eta$,
as \mbox{$\eta\rightarrow0$},
is \mbox{$\Erad(\eta)/\eta \sim 1-\left(2 \sqrt{2}\right)/3$}.
We construct the two-dimensional $\EradtwoD$ ansatz as a product of the 1D \ansaetze,
instead of a sum as in the final-spin case.

In principle, the fitting procedure is robust enough to use either a sum or product ansatz for either final-state quantity.
Actually, by carrying out the full procedure for $\Erad$ as described in the following subsections,
but using a sum of the 1D contributions, 
we found a fit statistically at least competitive with that obtained from a product.
However, in this case the sum ansatz tends to produce suspicious curvature in the \mbox{$\Seff=1$}, low-$\eta$ region,
which cannot be suppressed by the \emr information.
With additional NR data in this region, that problem might be alleviated, but with the current data set
we find the product ansatz to be more robust and able to yield a final fit that is both
accurate and well determined over the calibration region
and without obvious bad behavior in extrapolation.

\subsection{One-dimensional fits}
\label{sec:energyfits-1d}

For the nonspinning 1D fit in symmetric mass ratio $\eta$,
a simple fourth-order polynomial
\begin{equation}
 \label{eq:Erad_etaansatz}
 \EradoneDeta = a_4 \eta ^4+a_3 \eta ^3+a_2 \eta ^2+\left(1-\frac{2 \sqrt{2}}{3}\right) \eta
\end{equation}
with three free coefficients, listed in Table~\ref{tbl:Erad_eta_fit_coeffs},
is marginally preferred by both AICc and BIC.
More complicated rational functions are not able to yield any significant change in residuals (only up to 1\% in RMSE),
while the differences between \autoref{eq:Erad_etaansatz} and the next-ranked fits
are again much smaller than the remaining residuals,
as shown in \autoref{fig:Erad_etafits}.

For the effective-spin dependence,
again the value at \mbox{$\left(\eta=0.25,\,\Seff=0\right)$} is fixed from the $\eta$ fit.
A rational function of order (3,1) is top-ranked by AICc, BIC and RMSE
and thus unambiguously selected as the preferred ansatz:
\begin{equation}
 \label{eq:Erad_Sansatz}
 \small
 \EradoneDS =  \frac{0.0484161 \left(0.128 b_3 \widehat{S}^3+0.211 b_2 \widehat{S}^2+0.346 b_1 \widehat{S}+1\right)}{1-0.212 b_5 \widehat{S}}
\end{equation}
with four free coefficients listed in Table~\ref{tbl:Erad_S_fit_coeffs},
and well-behaved residuals as seen in \autoref{fig:Erad_Sfits}.

\subsection{Two-dimensional fits}
\label{sec:energyfits-2d}

\begin{figure}[t!hbp]
 \includegraphics[width=\columnwidth]{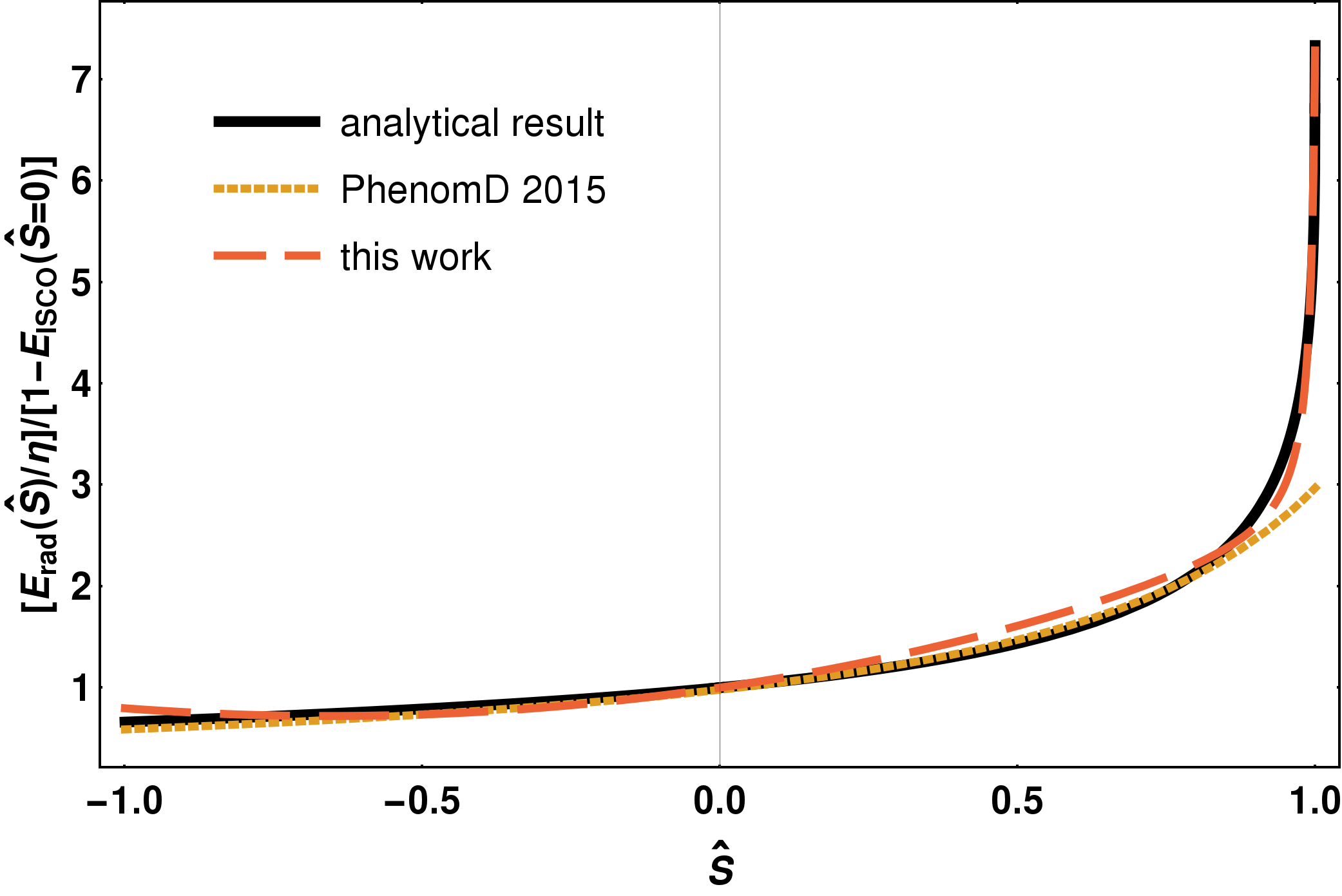}
 \caption{
  \label{fig:extreme_mass_ratio_Erad}
  Radiated energy: 
  \emr comparison of analytical results,
  the previous PhenomD radiated-energy fit of~\cite{Husa:2015iqa}, and this work.
  Note that, constrained to capture the steep rise at \mbox{$\Seff\rightarrow+1$},
  the new fit actually deviates slightly more from the analytical result at low positive $\Seff$.
  This could be avoided with a more complex 1D $\Seff$ ansatz, which is however disfavored by the current NR data set.
 }
 \vspace{0.5\baselineskip}
 \begin{tabular}{lrrr}\hline\hline
  &$ \text{Estimate} $&$ \text{Standard error} $&$ \text{Relative error [$\%$]} $\\\hline$
 f_{20} $&$  4.27\hphantom{345} $&$ 0.38\hphantom{345} $&$ 8.9\hphantom{2} $\\$
 f_{30} $&$ 31.09\hphantom{345} $&$ 0.71\hphantom{345} $&$ 2.3\hphantom{2} $\\$
 f_{50} $&$  1.56735            $&$ 0.00032            $&$ 0.02 $\\$
 f_{10} $&$  1.81\hphantom{345} $&$ 0.15\hphantom{345} $&$ 8.2\hphantom{2} $\\
\hline\hline\end{tabular}
 \captionof{table}{
  \label{tbl:Erad_eta0_coeffs}
  Fit coefficients for the \emrl of the radiated energy,
  fitted to discretized analytical results.
  The fourth coefficient, $f_{10}$, is fixed by the constraint at \mbox{$\Seff=1$},
  cf.~\autoref{eq:Erad_derivconstr_fitted},
  and its estimate and error are computed from the others.
 }
\end{figure}

\begin{figure}[t!hbp]
 \includegraphics[width=\columnwidth]{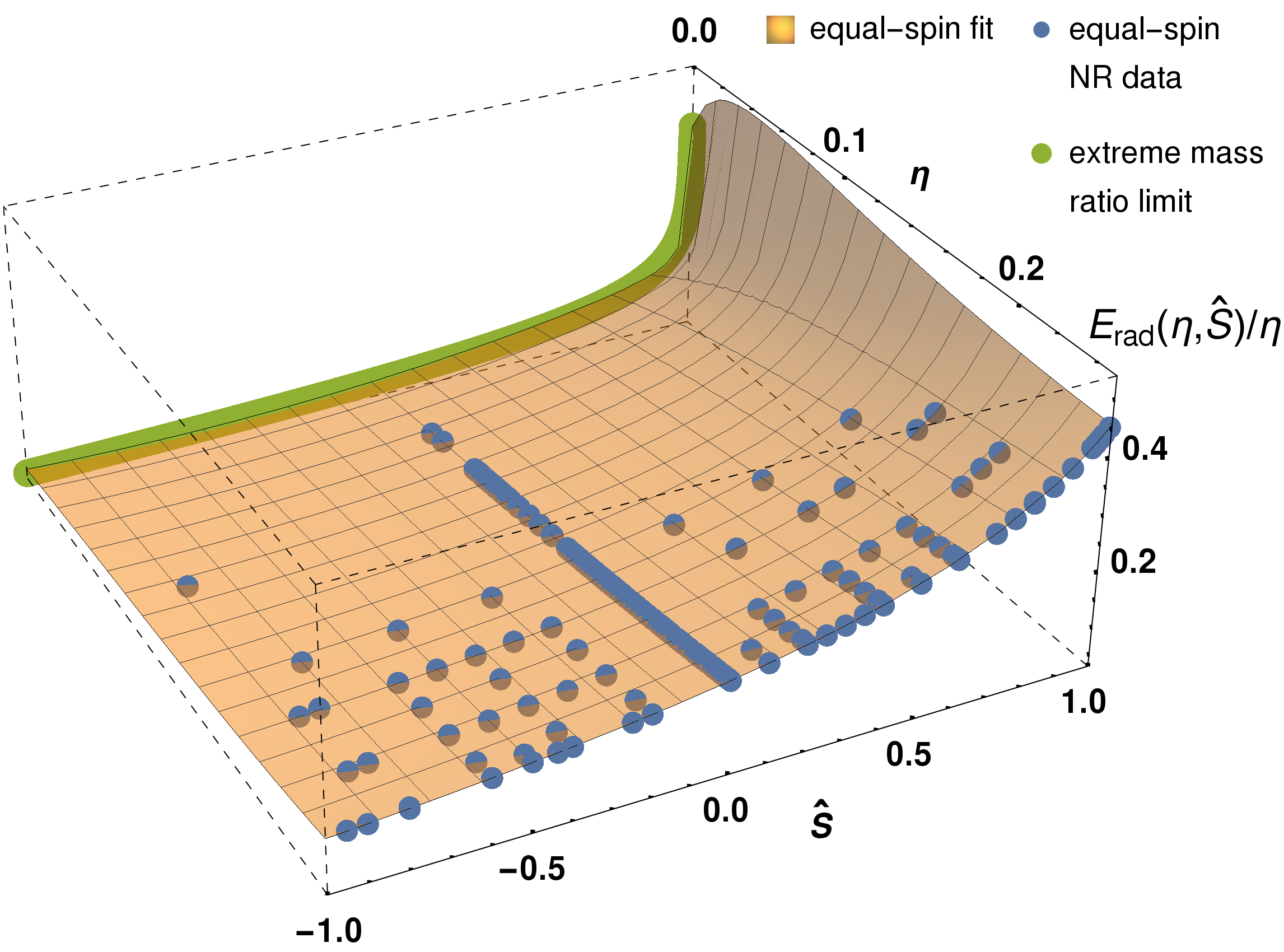}
 \caption{
  \label{fig:Erad_2dfit}
  Two-dimensional \mbox{$\EradtwoD$} fit,
  visualized as \mbox{$\EradtwoD/\eta$}.
  Application of the \emrl helps in avoiding extrapolation artifacts
  which would otherwise appear at low-$\eta$, high-$|\Seff|$ regions that are uncovered by NR simulations.
  }
\end{figure}

For the 2D ansatz, we combine the two 1D fits from Eqs.~\ref{eq:Erad_etaansatz} and~\ref{eq:Erad_Sansatz},
expanding each $\Seff$-dependent term with a polynomial in $\eta$,
according to \autoref{eq:2D_substitution},
and removing the \mbox{$\left(\eta=0.25,\,\Seff=0\right)$} value from the spin ansatz before multiplying with the $\eta$ terms:
\begin{equation}
 \label{eq:Erad_2Dansatz}
  \EradtwoD = \Erad\left(\eta,0\right) \frac{\Erad\left(0.25,\Seff,f_{ij}\right)}{\Erad\left(0.25,0\right)} \,.
\end{equation}

Contrary to the sum ansatz for $\af$ in \autoref{eq:af_2Dansatz},
we do not need to set the $\eta$-independent coefficients $f_{i0}$ of the $\Seff$ terms to zero,
as the \mbox{$\EradtwoD = \Erad\left(\eta,0\right)\left(1+\dots\right)$} form of \autoref{eq:Erad_2Dansatz} already guarantees the correct \mbox{$\eta=0$} limit.
Hence an expansion up to third order in $\eta$ of each $\Seff$ term, as we chose for the $\af$ fit, would yield too many free coefficients,
and instead we only expand up to second order.
The four $f_{i2}$ coefficients are again fixed by the equal-mass boundary conditions:
\begin{equation}
 \label{eq:Erad_2dconstraints}
 f_{i2} = 16 - 16 f_{i0} -  4 f_{i1} \,.
\end{equation}

Similar to the procedure for $\af$, we can use the \emrl to fix the four coefficients $f_{i0}$ of the linear-in-$\eta$ terms.
Using the analytic result from \autoref{eq:extreme_mass_ratio_Eisco},
we force the fit to satisfy the equality
\begin{equation}
 \label{eq:Erad_emrl}
 \Erad(\eta\rightarrow0,\Seff)
  = 1 - \Eisco(\Seff)
\end{equation}

\begin{figure*}[t]
 \includegraphics[width=0.9\columnwidth]{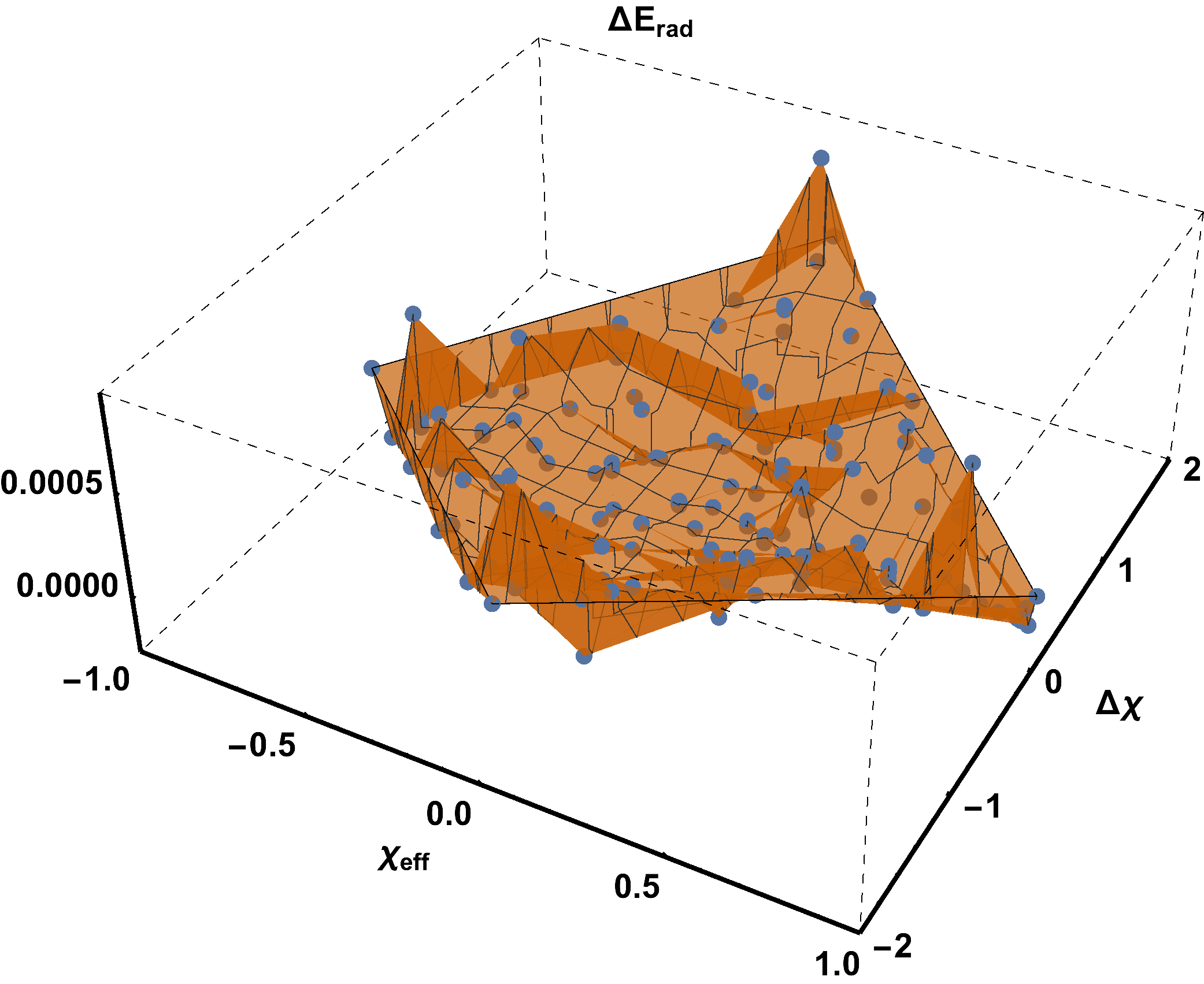} \hspace{1cm}
 \includegraphics[width=0.9\columnwidth]{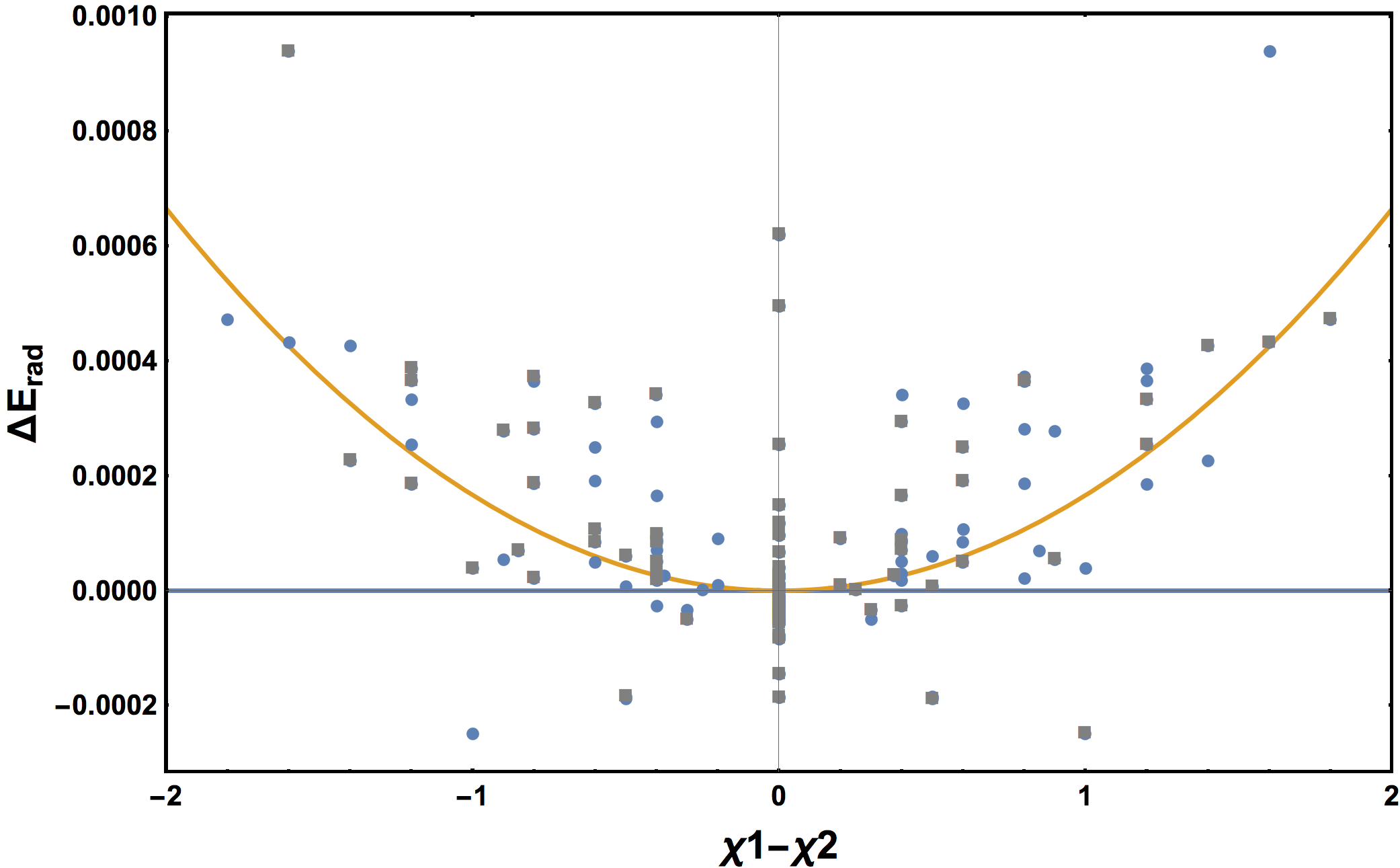} \\[-\baselineskip]
 \includegraphics[width=0.9\columnwidth]{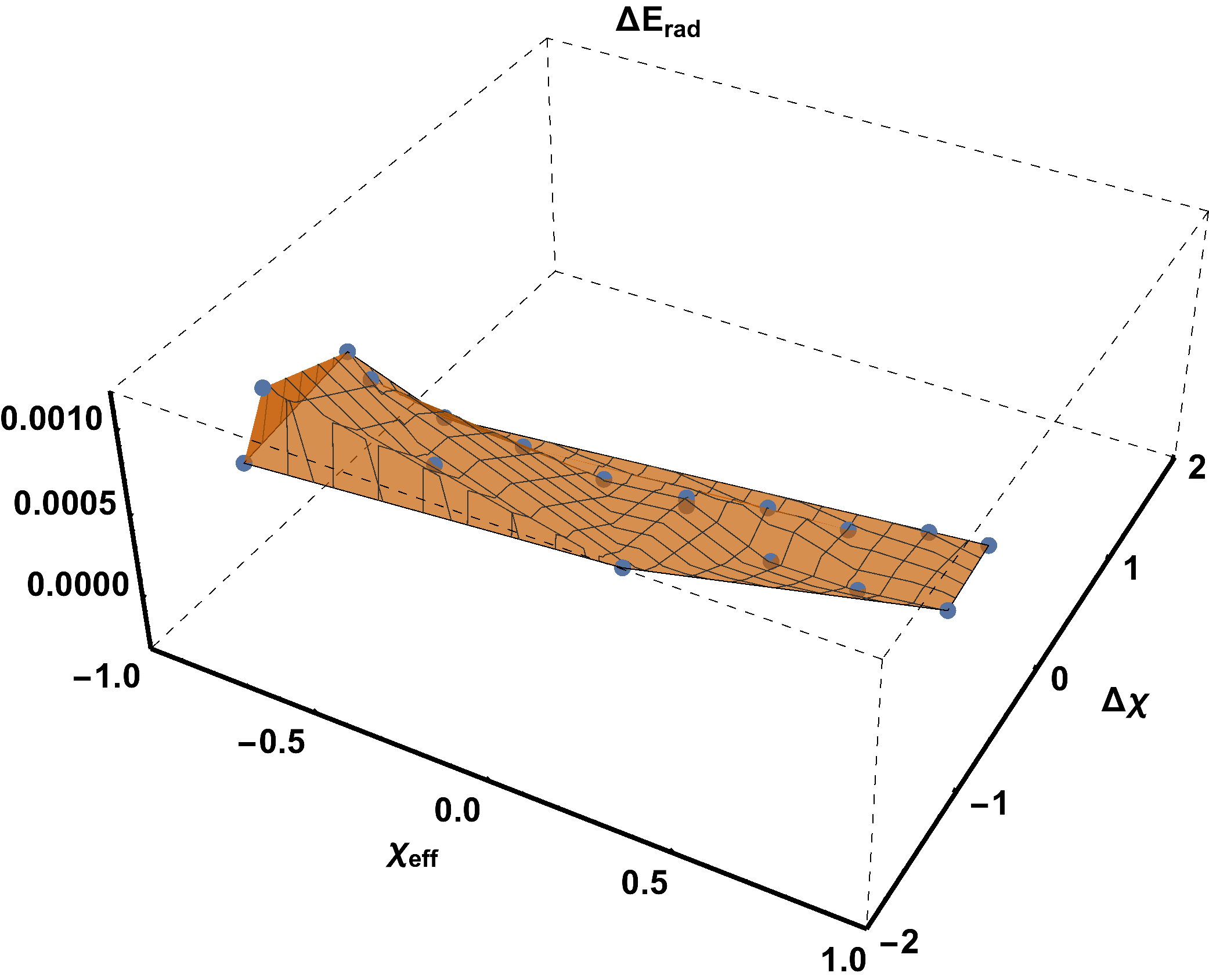} \hspace{1cm}
 \includegraphics[width=0.9\columnwidth]{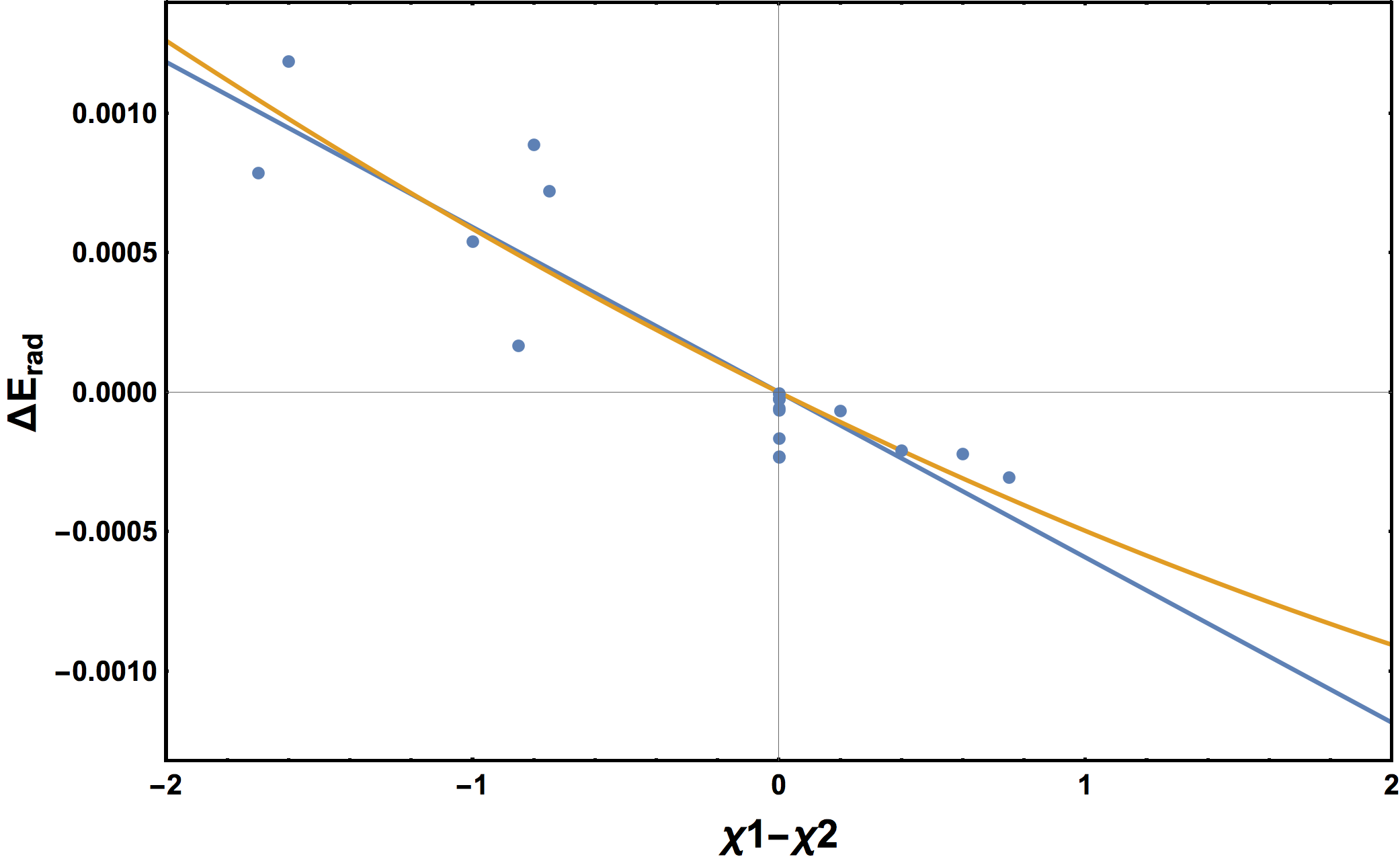}
 \caption{
  \label{fig:Erad_spin_diff_per_q}
  Examples of spin-difference behavior of the radiated energy at fixed mass ratios,
  for residuals $\DErad$ after subtracting the two-dimensional \mbox{$\EradtwoD$} fit.
  Top row: \mbox{$q=1$} (mirror-duplicated data points shown in gray);
  lower row: \mbox{$q=4$};
  left column: surfaces in \mbox{$\left(\Seff,\chidiff,\DErad\right)$} space;
  right column: projections unto the $\chidiff$ axis with linear and quadratic fits.
  At equal mass, the linear term and mixture term vanish,
  but the expected quadratic dependence (parabolic surface) is less clearly pulled out
  from rather noisy residuals than for the final spin (cf.~\autoref{fig:spin_diff_per_q}).
  At \mbox{$q=4$} and other intermediate mass ratios,
  the surface is not as close to flat as in the final-spin case,
  and the noisy data still shows some quadratic dependence.
  \vspace{-\baselineskip}
 }
\end{figure*}

\noindent and fit the corresponding leading-order $\eta$ dependence of our 2D ansatz to discretized values of this quantity.
Again we fix one of the four free coefficients of \mbox{$\Erad\left(\eta\rightarrow0,\Seff\right)$} by a constraint fixing the value at \mbox{$\Seff=1$},
which is necessary to capture the very steep rise of \autoref{eq:Erad_emrl} as \mbox{$\Seff\rightarrow+1$}:
\begin{align}
 \label{eq:Erad_derivconstr_fitted}
 f_{10}\to -0.574752 f_{20}-0.280958 f_{30}+64.6408 f_{50}-88.3165\,.
\end{align}

The agreement between discretized analytical result and fit is shown in \autoref{fig:extreme_mass_ratio_Erad},
and fit coefficients are listed in Table~\ref{tbl:Erad_eta0_coeffs}.

We thus have \mbox{$12-4-4=4$} free coefficients $f_{i1}$,
of which $f_{21}$ turns out to be extremely poorly constrained,
so that we set it to zero before refitting.
Results of the 2D fit, calibrated to equal-spin simulations only, are shown in \autoref{fig:Erad_2dfit},
which shows that the steep shape of the \emrl at high $\Seff$ is smoothly attained by the extrapolated fit. 
For the curvature at low $\eta$ and extremal \mbox{$\Seff=1$}, where there is no NR data,
there might be also a contribution from the small remaining fit issues in the \emrl (cf. \autoref{fig:extreme_mass_ratio_Erad}).
The residuals again have larger RMSE than the 1D fits in $\eta$ and $\Seff$,
by factors of 6.5 and 1.8 respectively,
but show no clear apparent trends,
allowing us to use this 2D fit as the basis for an unequal-spin residuals study in the next step.

\subsection{Unequal-spin contributions and 3D fit}
\label{sec:energyfits-3d}

The spin-difference dependence of unequal-spin residuals is less clear here than for the final spin:
As seen in the examples of \autoref{fig:Erad_spin_diff_per_q},
the general trend is the same with a quadratic dependence on $\chidiff$ at equal masses
and more dominant linear effects as $\eta$ decreases,
but the distributions are generally noisier and the second-order terms
(quadratic and mixture $\propto\Seff\chidiff$)
cannot be as cleanly separated.

For both the per-mass-ratio-step analysis and the direct 3D fit,
we use the same general functional forms for possible linear, quadratic and mixture terms
as in Eqs.~\ref{eq:af_chidiff_ansatz}, \ref{eq:af_chidiff_ansatz_lin_mix} and~\ref{eq:af_chidiff_ansatz_quad}.
After fixing ill-constrained coefficients to integer values, these reduce to
\begin{subequations}
 \label{eq:Erad_chidiff_ansatz_terms}
 \begin{align}
  A_1(\eta)&=d_{10} (1-4 \eta )^{0.5} \eta ^2 \left(d_{11} \eta +1\right) \\
A_2(\eta)&=d_{20} \eta ^3 \\
A_3(\eta)&=d_{30} (1-4 \eta )^{0.5} \eta  \left(d_{31} \eta +1\right) \,.
 \end{align}
\end{subequations}

\begin{figure*}[thbp]
 \includegraphics[width=0.9\columnwidth]{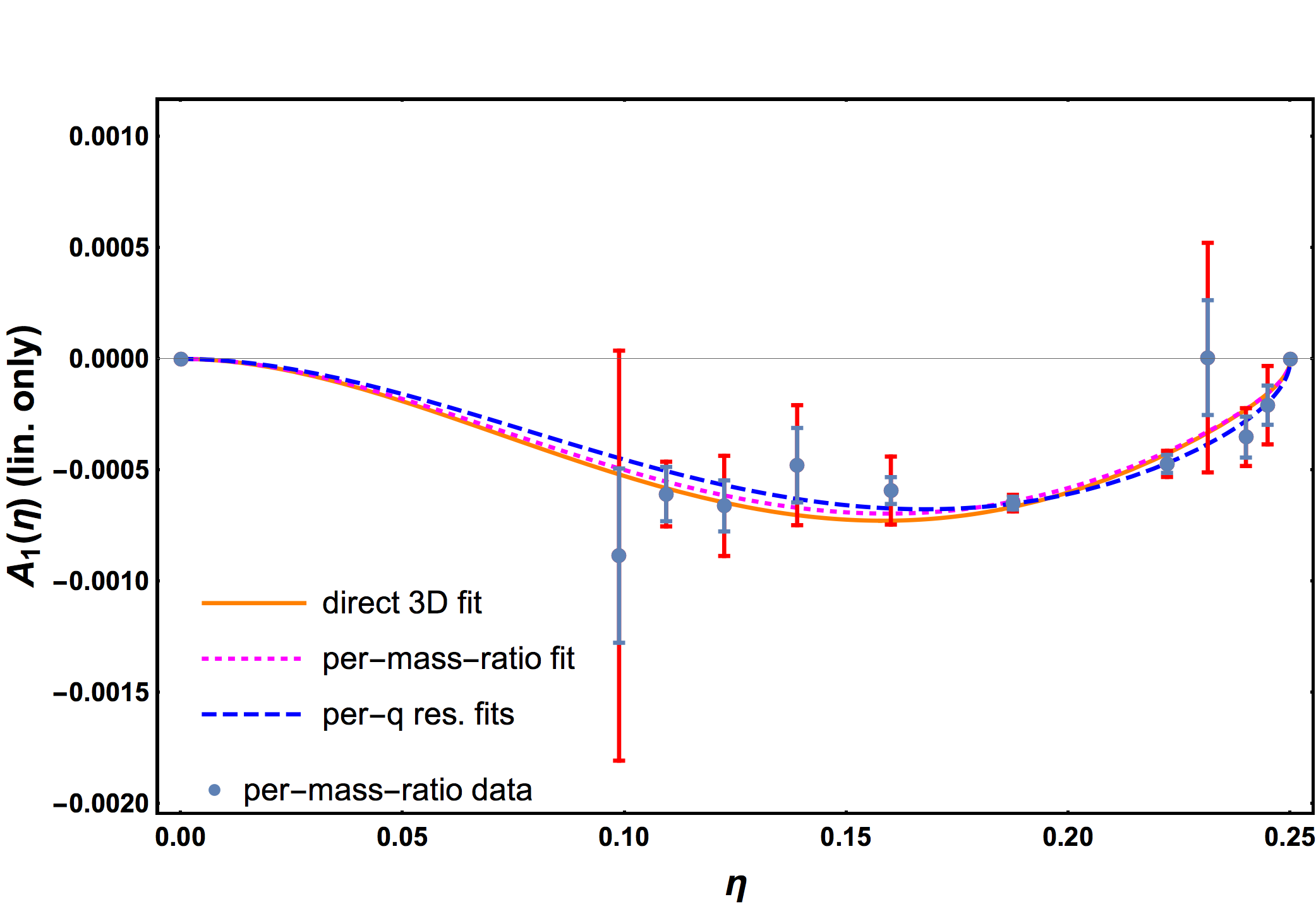} \hspace{1cm}
 \includegraphics[width=0.9\columnwidth]{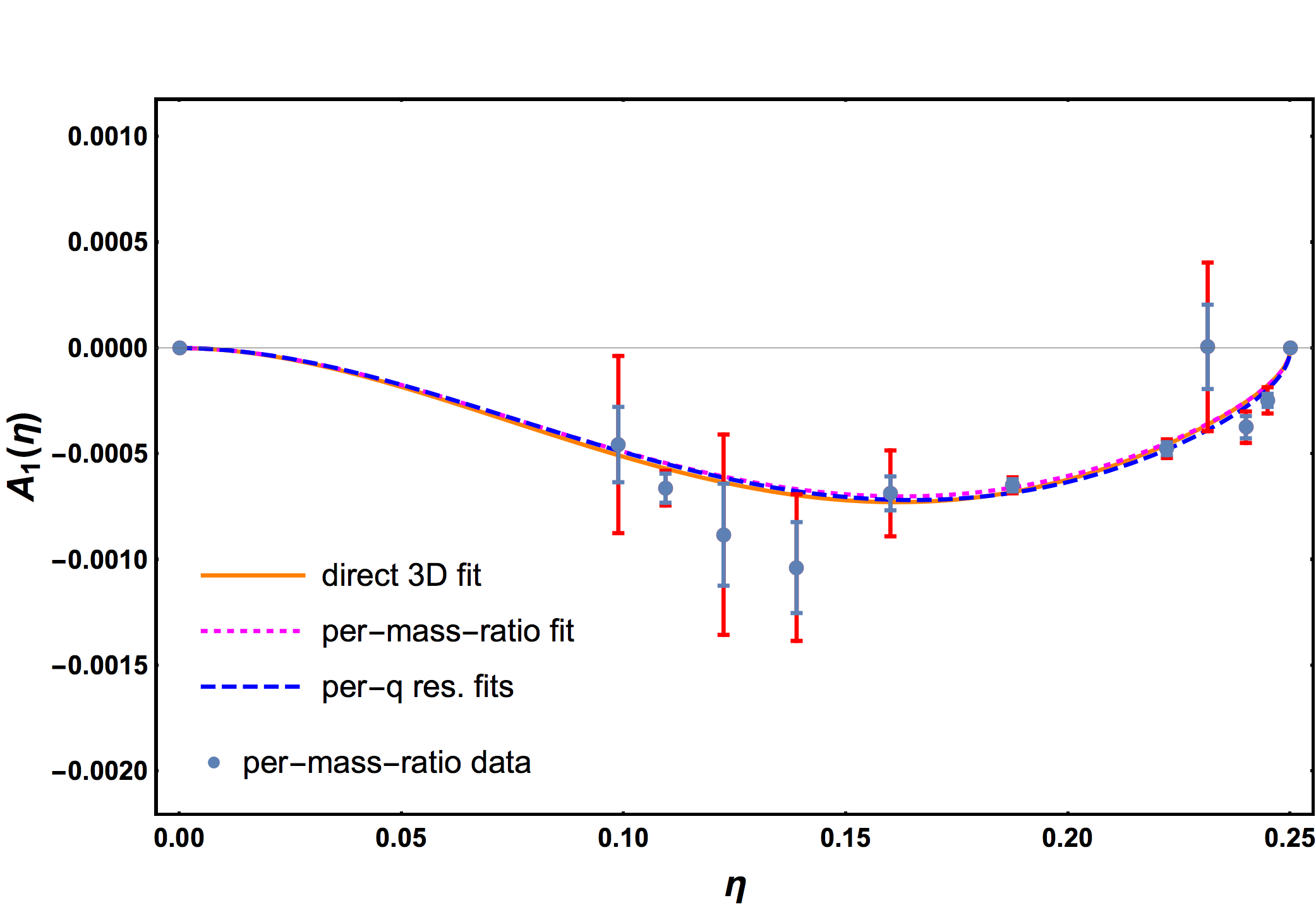} \\[-\baselineskip]
 \includegraphics[width=0.9\columnwidth]{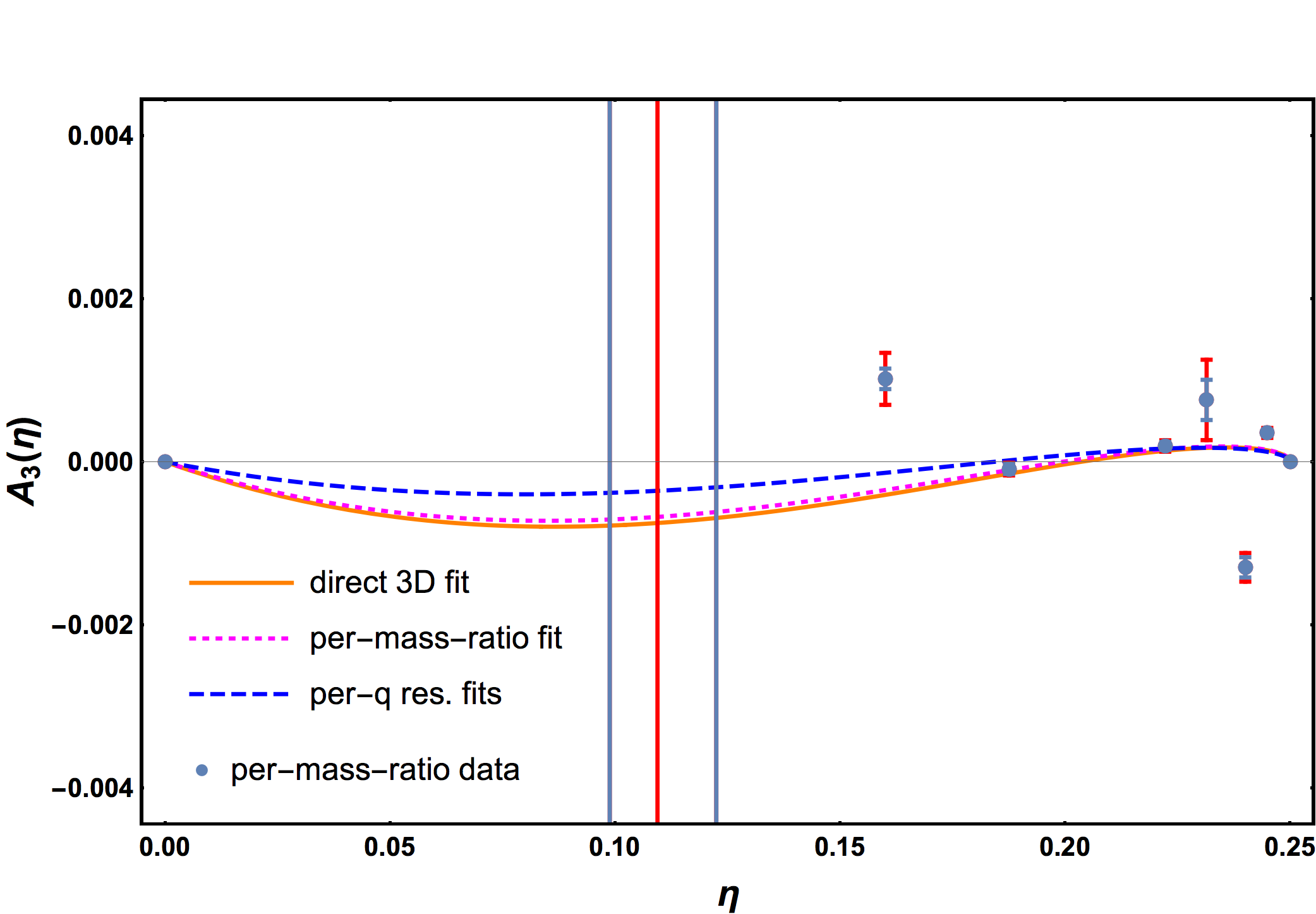} \hspace{1cm}
 \includegraphics[width=0.9\columnwidth]{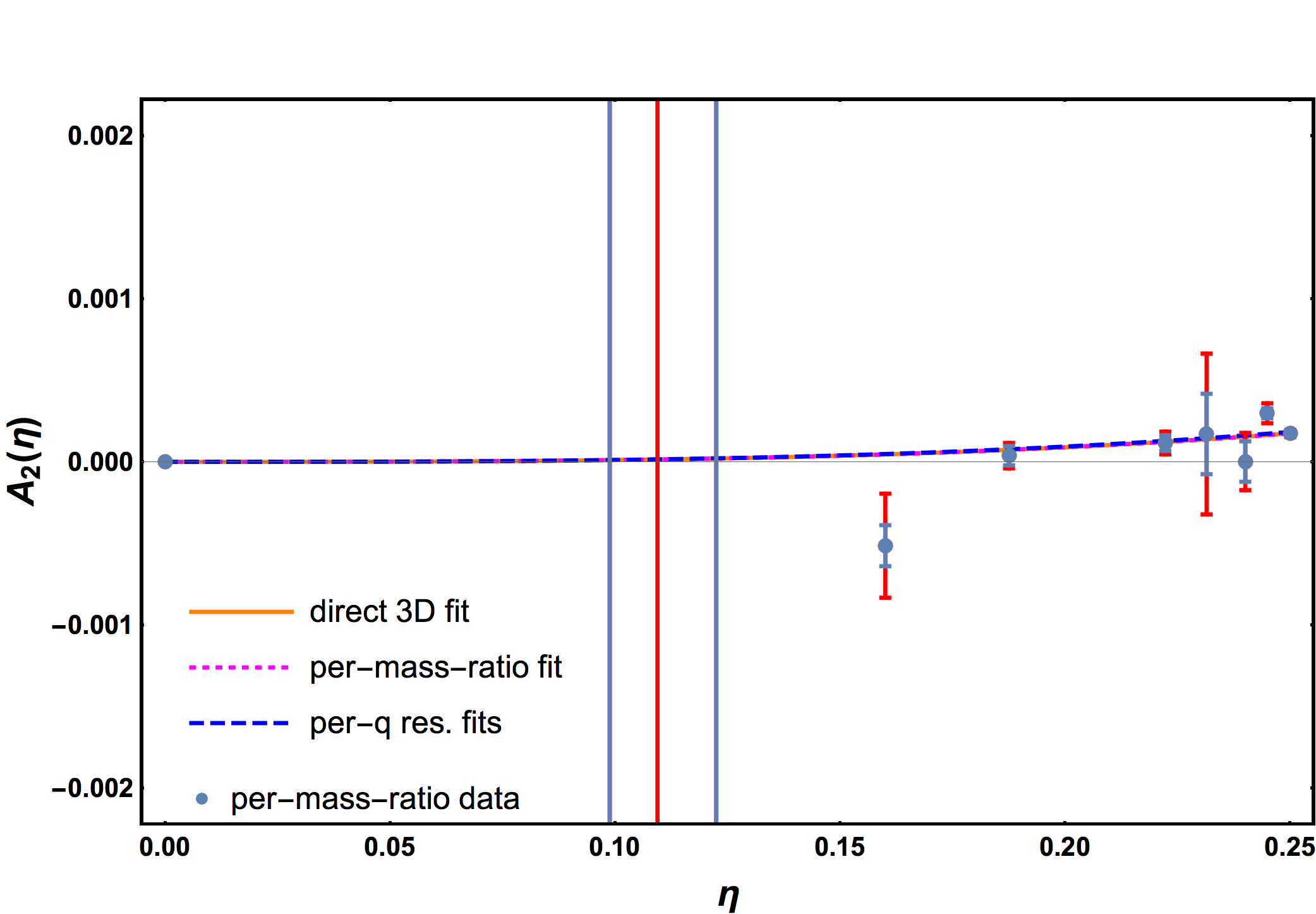}
 \caption{
  \label{fig:Erad_spin_diff}
  Spin-difference behavior of radiated-energy data after subtraction of the two-dimensional \mbox{$\EradtwoD$} fit,
  for the three ansatz functions $A_i(\eta)$ from \autoref{eq:Erad_chidiff_ansatz_terms},
  with the same mass-ratio steps and fits as in \autoref{fig:af_spin_diff}.
  Top-left panel: linear term $A_1$ only.
  The remaining panels are for the combined linear+quadratic+mixture fit, in clockwise order:
  linear term $A_1$,
  quadratic term $A_2$ and
  mixture term $A_3$.
  The $A_1$ results from the combined fit are very similar to those from the linear-only fit,
  demonstrating the robustness of extracting leading-order spin-difference effects.
  For the two lower panels, results are much more uncertain,
  and the error bars for low $\eta$ go far outside the displayed range,
  so that this region does not contribute significantly to the weighted per-mass-ratio fits.
  \vspace{-\baselineskip}
 }
\end{figure*}

Figure~\ref{fig:Erad_spin_diff} shows that the linear term is again robustly determined
and does not change shape much when adding the two additional terms,
but already the per-mass-ratio and direct-3D fits for this term
do not agree quite as closely as in the $\af$ fit.
The quadratic term is more noisy,
and for the mixture term the results are rather uncertain,
with an apparent sign change in the effect over $\eta$,
but the stepwise cross-checks at least agreeing on the overall shape.
Still, we will see below that inclusion of \mbox{both} these effects is statistically justified.

The full 3D ansatz for $\EradthreeD$ is then built up as
\begin{equation}
 \label{eq:Erad_finalansatz}
 \EradthreeD = \EradtwoD + \DEradthreeD \,,
\end{equation}
and this time has eight free coefficients (three from the 2D ansatz and five from the spin-difference terms).
Results for the fit to \NRcountNZSUM NR cases not previously used in the 1D fits are listed in Table~\ref{tbl:Erad_final_fit_coeffs}.

\begin{table}[thbp]
 \begin{tabular}{lrrr}\hline\hline
  &$ \text{Estimate} $&$ \text{Standard error} $&$ \text{Relative error [$\%$]} $\\\hline$
 d_{10} $&$   -0.098\hphantom{4} $&$ 0.011\hphantom{4} $&$ 11.3 $\\$
 d_{11} $&$   -3.23\hphantom{34} $&$ 0.18\hphantom{34} $&$  5.6 $\\$
 d_{20} $&$    0.0112            $&$ 0.0012            $&$ 10.5 $\\$
 d_{30} $&$   -0.0198            $&$ 0.0036            $&$ 18.4 $\\$
 d_{31} $&$   -4.92\hphantom{34} $&$ 0.19\hphantom{34} $&$  3.9 $\\$
 f_{11} $&$   15.7\hphantom{234} $&$ 1.2\hphantom{234} $&$  7.9 $\\$
 f_{31} $&$ -243.6\hphantom{234} $&$ 8.0\hphantom{234}  $&$  3.3 $\\$
 f_{51} $&$   -0.58\hphantom{34} $&$ 0.13\hphantom{34} $&$ 21.6 $\\
\hline\hline\end{tabular}
 \caption{
  \label{tbl:Erad_final_fit_coeffs}
  $\Erad$ fit coefficients for the final 3D step, using \NRcountNZSUM cases.
  \vspace{-\baselineskip}
 }
\end{table}

\subsection{Fit assessment}
\label{sec:energyfits-assess}

\autoref{tbl:Erad_summary} gives a statistical summary of the various 1D, 2D and 3D fits for $\Erad$ as discussed in this section.
We find that the full linear+quadratic+mixture fit to all data has better RMSE than the simpler versions,
and actually on the same level as the 2D equal-spin fit to equal-spin data only.
The additional coefficients are also justified by yielding the best AICc and BIC,
though not as clearly as for the final-spin case in Table~\ref{tbl:af_summary}.
Thus, we choose this three-term ansatz, matching the final-spin choice.
Also, from Table~\ref{tbl:Erad_final_fit_coeffs} we see that all coefficients of this ansatz are sufficiently well constrained,
with a worst standard error of 21.6\%,
to be useful for applications,
at least under the assumptions made for the ansatz terms in \autoref{eq:Erad_chidiff_ansatz_terms}.
Just as was the case for the final-spin fit,
and even more so because of the noisier data,
we caution that ambiguity over the exact shape of the spin-difference terms remains,
because different choices of the exponents and expansion orders in \autoref{eq:Erad_chidiff_ansatz_terms}
can yield statistically indistinguishable fits.

\begin{figure}[t!hbp]
 \includegraphics[width=\columnwidth]{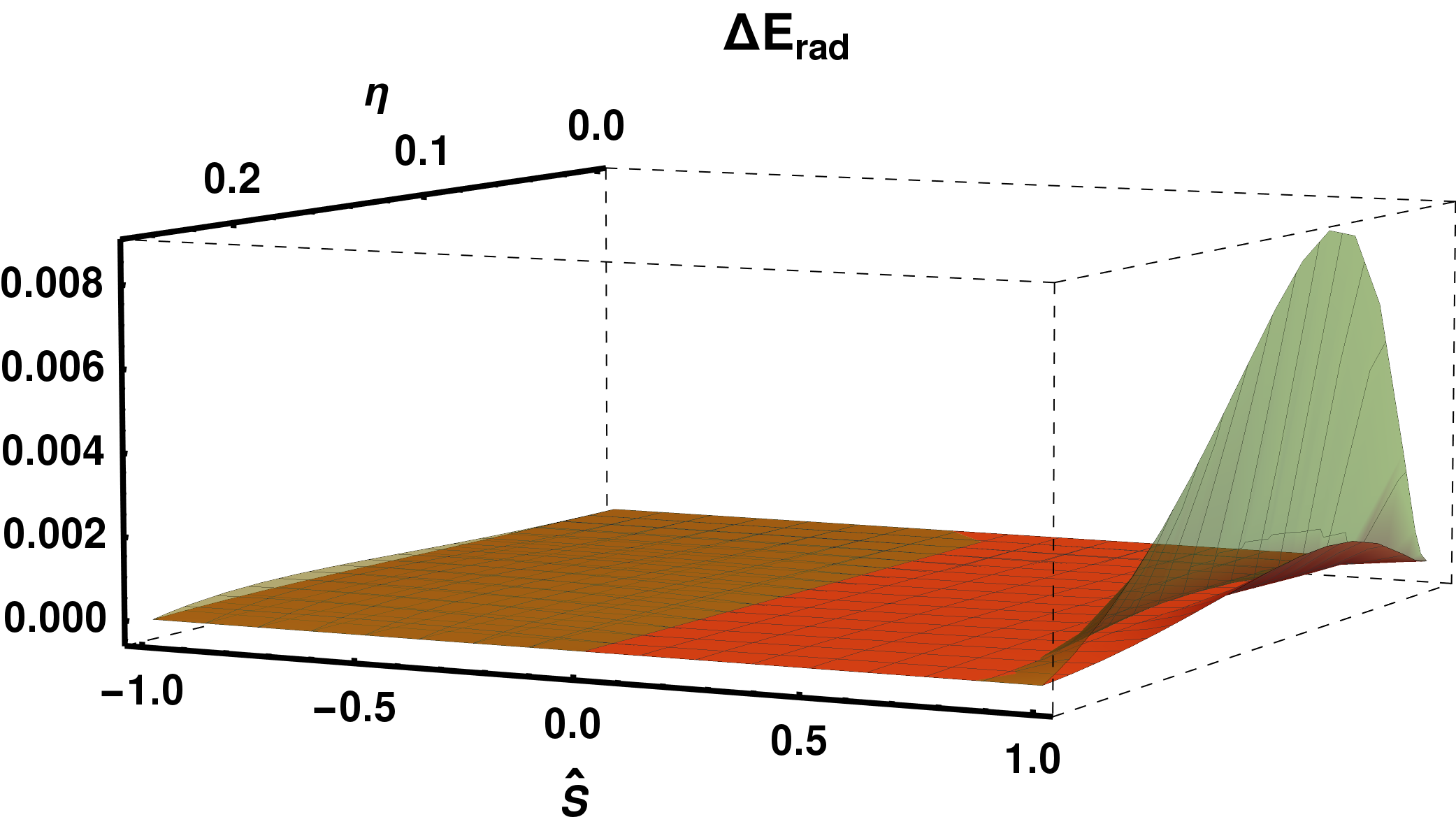}
 \caption{
  \label{fig:Erad_2deq_2dall_3d}
  Green:
  Difference $\Delta\Erad$ of a 2D fit to the full radiated energy data set minus the 2D fit to equal-spin cases only,
  both including \emr constraints.
  Orange:
  Difference $\Delta\Erad$ of the 2D part of the 3D fit to the full data set minus the 2D-only fit to equal-spin data.
  \vspace{-\baselineskip}
 }
\end{figure}

As an additional check on the overall functional behavior of the fit,
in \autoref{fig:Erad_2deq_2dall_3d} we show again comparisons between
the intermediate 2D fit calibrated to equal-spin data only,
a 2D fit to all data (not working well, as expected)
and the 2D part of the final 3D fit.
The equal-spin 2D fit and 3D fit agree well,
with the strong excess curvature of the all-data 2D fit in the high-$\Seff$ region
reduced significantly thanks to the unequal-spin terms.

\begin{table}[htbp]
 \begin{tabular}{lrcccc}\hline\hline
  &$ N_{\text{data}} $&$ N_{\text{coeff}} $&$ \text{RMSE} $&$ \text{AICc} $&$ \text{BIC} $\\\hline$
 \text{1D $\eta $} $&$ 92 $&$ 3 $&$ 4.14\times 10^{-5} $&$ -1705.7 $&$ -1695.6 $\\$
 \text{1D }\hat{S} $&$ 37 $&$ 4 $&$ 1.51\times 10^{-4} $&$ -\hphantom{1}577.3 $&$ -\hphantom{1}569.3 $\\$
 \left.\text{2D (}\chi _1=\chi _2\right) $&$ 60 $&$ 3 $&$ 2.67\times 10^{-4} $&$ -\hphantom{1}875.3 $&$ -\hphantom{1}867.4 $\\$
 \text{2D all} $&$ 298 $&$ 3 $&$ 4.26\times 10^{-4} $&$ -4070.9 $&$ -4056.2 $\\$
 \text{3D lin} $&$ 298 $&$ 5 $&$ 3.24\times 10^{-4} $&$ -4282.9 $&$ -4260.9 $\\$
 \text{3D lin+quad} $&$ 298 $&$ 6 $&$ 2.72\times 10^{-4} $&$ -4391.9 $&$ -4366.3 $\\$
 \text{3D lin+mix} $&$ 298 $&$ 7 $&$ 2.91\times 10^{-4} $&$ -4339.3 $&$ -4310.1 $\\$
 \text{3D lin+quad+mix} $&$ 298 $&$ 8 $&$ 2.62\times 10^{-4} $&$ -4417.8 $&$ -4385.0 $\\
\hline\hline\end{tabular}
 \captionof{table}{
  \label{tbl:Erad_summary}
  Summary statistics for the various steps of the hierarchical radiated-energy fit.
  Evidence for spin-difference terms beyond linear order is
  weaker than for the final-spin fits in Table~\ref{tbl:af_summary}.
  \vspace{-\baselineskip}
 }
\end{table}

\begin{figure}[H]
 \includegraphics[width=\columnwidth]{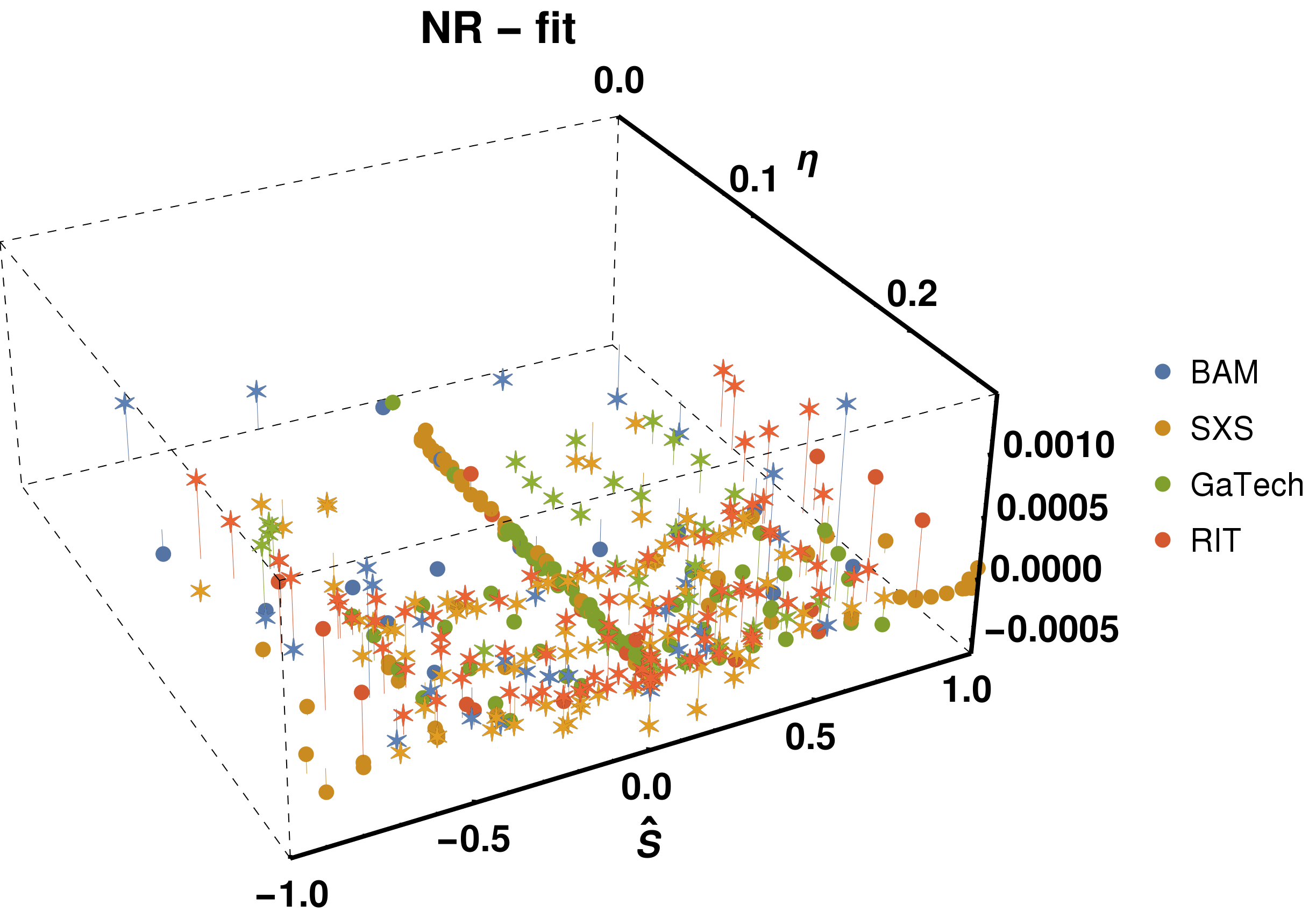}
 \caption{
  \label{fig:Erad_residuals_paramspace}
  Residuals of the radiated-energy fit, projected to the 2D parameter space of $\eta$ and $\Seff$.
  Stars: unequal-spin points.
 }
\end{figure}
\begin{figure}[thbp]
 \includegraphics[width=\columnwidth]{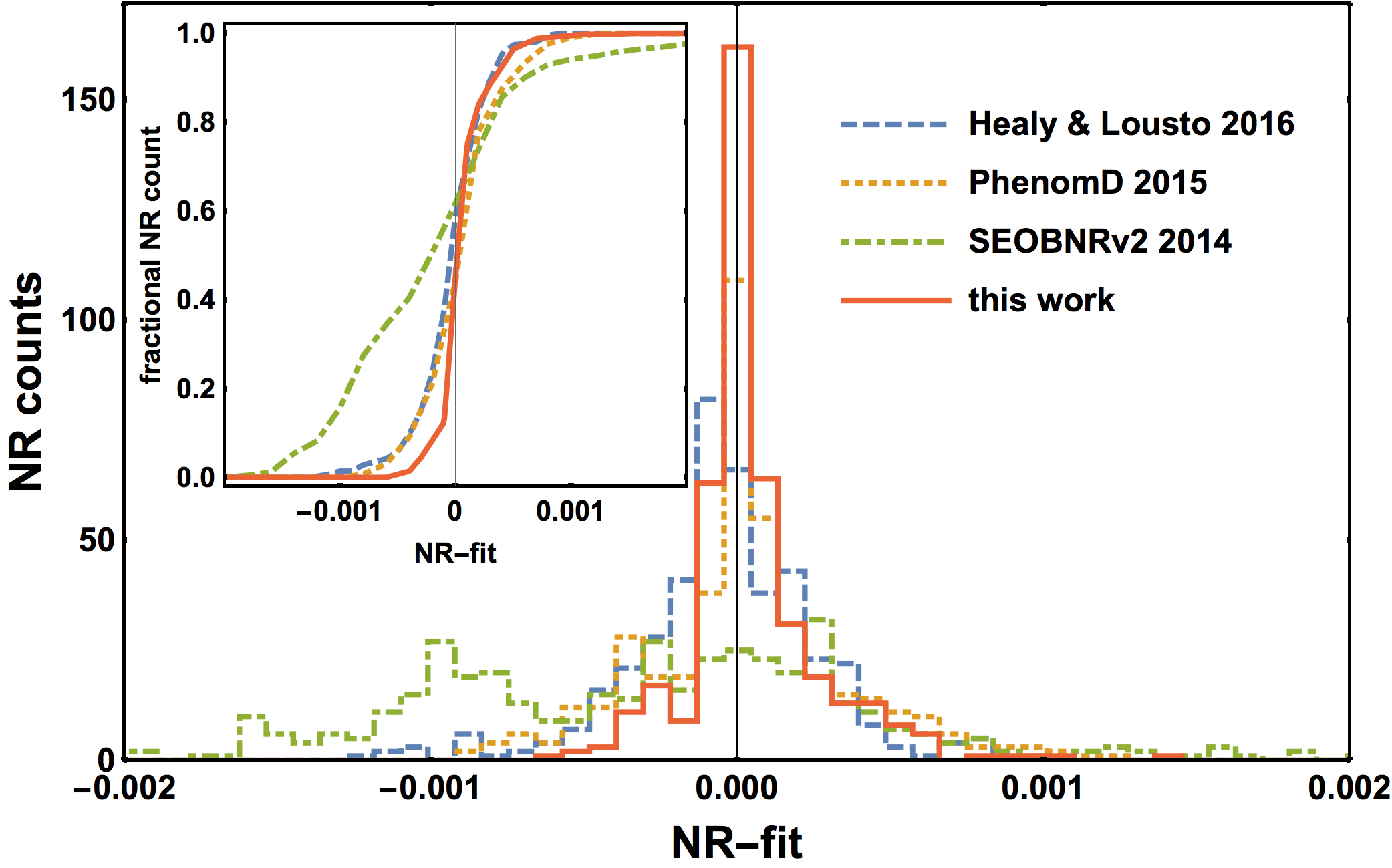}
 \captionof{figure}{
  \label{fig:Erad_resid_hists}
  Fit residuals of the radiated energy $\Erad$, for this work (cf. Table~\ref{tbl:Erad_final_fit_coeffs})
  and for previously published fits (SEOBNRv2 2014~\cite{Taracchini:2013rva}, Healy \& Lousto 2016~\cite{Healy:2016lce}, PhenomD 2015~\cite{Husa:2015iqa}),
  evaluated over the set of \NRcount NR simulations shown in \autoref{fig:eta_chi1_chi2_NR}.
  Main panel: histograms, with 10 outliers for SEOBNRv2 (a recalibration of the fit from~\cite{Barausse:2012qz})
  with \mbox{$|\text{NR}-\text{fit}|>0.002$} outside of the plot range.
  Inset: cumulative distributions over the same range.
 }
 \vspace{0.5\baselineskip}
 \begin{tabular}{lrrccc}\hline\hline
  &$ N_{\text{coef}} $&$ \text{mean} $&$ \text{stdev} $&$ \text{AICc} $&$ \text{BIC} $\\\hline$
 \text{HLZ2014~\cite{Healy:2014yta}} $&$ 19 $&$ -5.4\times 10^{-5} $&$ 3.4\times 10^{-4} $&$ -5802.5 $&$ -5723.2 $\\$
 \text{HL2016~\cite{Healy:2016lce}} $&$ 19 $&$ -4.4\times 10^{-5} $&$ 3.0\times 10^{-4} $&$ -5909.8 $&$ -5830.5 $\\$
 \text{PhenomD} $&$ 10 $&$ 2.5\times 10^{-5} $&$ 3.4\times 10^{-4} $&$ -5914.9 $&$ -5870.8 $\\$
 \text{ (refit)} $&$ 10 $&$ 6.1\times 10^{-5} $&$ 3.3\times 10^{-4} $&$ -5947.7 $&$ -5899.6 $\\$
 \text{SEOBNRv2~\cite{Taracchini:2013rva}} $&$ 2 $&$ -1.7\times 10^{-4} $&$ 1.0\times 10^{-3} $&$ -5036.1 $&$ -5023.9 $\\$
 \text{This work} $&$ 15 $&$ 4.7\times 10^{-5} $&$ 2.2\times 10^{-4} $&$ -6454.8 $&$ -6391.0 $\\$
 \text{ (refit)} $&$ 15 $&$ 6.3\times 10^{-5} $&$ 2.1\times 10^{-4} $&$ -6482.8 $&$ -6419.0 $\\$
 \text{ (uniform} $&$ 15 $&$ -4.0\times 10^{-6} $&$ 2.1\times 10^{-4} $&$ -5987.3 $&$ -5923.5 $\\$
 \text{ (uniform refit)} $&$ 15 $&$ 1.4\times 10^{-6} $&$ 2.0\times 10^{-4} $&$ -6034.2 $&$ -5970.4 $\\
\hline\hline\end{tabular}
 \captionof{table}{
  \label{tbl:Erad_residuals}
  Summary statistics for the new radiated-energy fit
  compared with previously published fits~\cite{Taracchini:2013rva,Healy:2014yta,Husa:2015iqa,Healy:2016lce},
  evaluated over the full set of \NRcount NR simulations shown in \autoref{fig:eta_chi1_chi2_NR}.
  Also listed are a refit of the PhenomD~\cite{Husa:2015iqa} ansatz to the present NR data set,
  a refit of our hierarchically obtained ansatz directly to the full data set,
  and results with the same fitting procedure, but using uniform weights.
  \vspace{-\baselineskip}
 }
\end{figure}

The residuals of the new $\Erad$ fit are shown over the \mbox{$\left(\eta,\Seff\right)$} parameter space in \autoref{fig:Erad_residuals_paramspace}
and as histograms in \autoref{fig:Erad_resid_hists},
compared with several previously published fits~\cite{Taracchini:2013rva,Healy:2014yta,Husa:2015iqa}.
Comparison statistics are also summarized in Table~\ref{tbl:Erad_residuals}.
Again we find significant improvement,
by a factor of 5 in residual standard deviation
over the simple two-coefficient fit of \cite{Taracchini:2013rva}
and about 40\% improvement over the previous PhenomD fit of~\cite{Husa:2015iqa} and the RIT fits of~\cite{Healy:2014yta,Healy:2016lce}.
As the maximum possible emitted energy fraction --
for equal-mass BBHs with extremal positive spins --
the new fit predicts \mbox{$\Erad\approx0.1142$},
consistent with the fit from \cite{Hemberger:2013hsa} which was specifically calibrated to \eqmeqS cases
and with \cite{Healy:2014yta,Husa:2015iqa},
but 15\% higher than \cite{Barausse:2012qz},
underlining the importance of high-order $\Seff$ terms at extremal spins.

\section{Conclusions}
\label{sec:conclusions}

We have developed a hierarchical step-by-step fitting method for results of numerical-relativity simulations of BBH mergers,
assuming nonprecessing spins and negligible eccentricity,
so that the parameter space is given by the mass ratio and two spin components: \mbox{$\left(\eta,\chi_1,\chi_2\right)$}.
We have then applied the method to the spin $\af$ and mass $\Mf$ of the remnant Kerr BH, the latter being equivalent to radiated energy $\Erad$.

An appropriate fit is constructed in simple steps which reduce the dimensionality of the problem, 
with the ansatz choice at each step driven by inspecting the data.
The full higher-dimensional fit is then built in a bottom-up fashion,
modeling each parameter's contribution in order of its importance, 
as illustrated in the flowchart of \autoref{fig:flowchart}.

A key goal of our approach is to avoid overfitting.
To achieve this, and to evaluate the quality of the fits,
we use information criteria (AICc and BIC) described in Appendix~\ref{sec:appendix-stats}.
At least as important, however, is also the hierarchical data-driven nature of our procedure,
e.g. when modeling the subdominant dependence on the difference between the spins.
Through a reduction to one-dimensional problems, and inspecting the data at different mass ratios,
an appropriate model with a small number of parameters is easily identified.

We compare our results with previous fits in the literature,
also refitting these previous models to our data set,
and we find a clear preference for the new fits in terms of residuals and information criteria.

Our emphasis on inspecting the underlying data set,
and comparing its quality with the statistical analysis of fit errors,
highlights that for further improvement of these numerical fits,
it will be essential to understand the uncertainties and systematic effects in a data set,
and to cleanly combine data from different codes,
rather than simply increasing the number of calibration simulations without controlling the error budget.

In addition to the quality of the fit as applied to the available numerical relativity data,
we investigate extrapolation properties beyond the calibration region, in particular for extreme spins,
where we check that our fit does not overshoot the extreme Kerr limit, and avoids pathological oscillations.
The \emrl has previously been incorporated into fits for final mass and spin in different ways,
in particular for the final spin, where the influence of radiated energy is not always accounted for (e.g. in~\cite{Buonanno:2007sv,Healy:2014yta}). 
Doing so is however important to avoid oscillatory features
due to an unphysical value of the derivative with respect to the symmetric mass ratio $\eta$ at \mbox{$\eta=0$}, 
and indeed we achieve a more robust behavior for close-to-extremal final spins in comparison to other recent fits~\cite{Healy:2014yta,Hofmann:2016yih}
-- see \autoref{fig:af_extreme_spin}.
In order to avoid numerical errors in this derivative,
we use an analytical expansion around the \mbox{$(\eta=0, \Seff=1)$} corner.

We have also verified that our fits for final mass and spin are consistent with the Hawking area theorem for black holes~\cite{Hawking:1971tu}:
the area of the final horizon is larger than the sum of the individual horizons of the initial black holes.
The difference in areas is much larger than the fit errors,
thus the area theorem does not provide an additional constraint on the fits.

For both $\af$ and $\Erad$, we can robustly identify the main unequal-spin contribution of the form \mbox{$f(\eta) \, (\chi_1-\chi_2)$}.
The shape of the function $f(\eta)$ is well determined for final spins,
but has larger errors for the smaller unequal-spin contribution to radiated energy.
We also discuss and find statistical evidence for two further unequal-spin terms,
one quadratic in $\chidiff$,
and a mixed  $\Seff\,\chidiff$ term,
which however are comparable in size to errors in the numerical data,
so that their $\eta$-dependent shape cannot be tightly constrained with the current data set.

The same hierarchical fitting procedure can be easily applied to other quantities such as peak luminosity~\cite{T1600018,Keitel:2016krm}.
An important goal of our work is the accurate calibration of unequal-spin effects for full inspiral-merger-ringdown waveform models,
in particular toward extending the PhenomD model \cite{Husa:2015iqa,Khan:2015jqa}.
This will allow investigating under which conditions gravitational-wave observations can reveal the individual spins of a coalescing binary.
While no fundamental modifications are necessary to the method itself,
a careful analysis of the data sets will be required to make appropriate choices
in combining the 1D subspace fits, including \emr information, and adding spin-difference terms.
A more ambitious goal for the future is the extension of our hierarchical fit construction to the higher-dimensional problems of generic precessing BBHs.

\vspace{-\baselineskip}

\section*{Acknowledgments}

\vspace{-0.5\baselineskip}

X.J., D.K. and S.H. were supported by the Spanish Ministry of Economy and Competitiveness grants
FPA2016-76821-P, CSD2009-00064 and FPA2013-41042-P,
the Spanish Agencia Estatal de Investigaci\'on,
European Union FEDER funds,
Vicepresid\`encia i Conselleria d'Innovaci\'o, Recerca i Turisme,
Conselleria d’Educaci\' i Universitats del Govern de les Illes Balears,
and the Fons Social Europeu.
M.H. was supported by the Science and Technology Facilities Council grants
ST/I001085/1 and ST/H008438/1,
and by European Research Council Consolidator Grant 647839,
and M.P. by ST/I001085/1.
S.K. was supported by Science and Technology Facilities Council.
The authors thankfully acknowledge the computer resources at Advanced Research Computing (ARCCA) at Cardiff,
as part of the European PRACE Research Infrastructure on the clusters Hermit, Curie and SuperMUC,
on the U.K. DiRAC Datacentric cluster and on the BSC MareNostrum computer
under PRACE and RES (Red Espa\~nola de Supercomputaci\'on) grants,
2015133131, AECT-2016-1-0015, AECT-2016-2-0009, AECT-2017-1-0017.
We thank the CBC working group of the LIGO Scientific Collaboration,
and especially Nathan Johnson-McDaniel, Ofek Birnholtz, Deirdre Shoemaker and Aaron Zimmerman,
for discussions of the fitting method, results and manuscript.
This paper has been assigned Document No. \mbox{\dcc}.

\appendix

\section{Data sets and NR uncertainties}
\label{sec:appendix-data}

For NR calibration of BBH coalescence models,
it is useful to combine data sets from different research groups and numerical codes,
both to increase robustness against code inaccuracies and errors in the preparation of data products (such as incorrect metadata),
and to benefit from the combined computational resources of different groups.
Computational cost increases significantly with mass ratio and spin,
thus the high-mass-ratio and high-spin regions are still poorly sampled,
and numerical errors are often higher (as we will see below).
Very different spins on the two black holes will typically also increase computational cost
due to the need to resolve different scales in the grids around the two black holes
(higher spin leads to smaller black holes for typical coordinate gauges in numerical relativity),
which is a challenge for accurately capturing small subdominant effects
like the nonlinear spin-difference effects we discuss in this work.

In this Appendix, we discuss our procedures to eliminate data points of poor quality,
to assign fit weights, and to check consistency between assumed error bars and our fit results.
We expect two main avenues to significantly improve over the fits we have presented in this paper:
(a) providing more data points with high spins and unequal masses,
in order to improve the accuracy of the fit near the boundaries of the fitting region
and to reduce the need for extrapolation;
and (b) determining more accurate and robust error bars for NR data,
which would allow one to isolate small subdominant effects.

\begin{figure}[thbp]
 \includegraphics[width=\columnwidth]{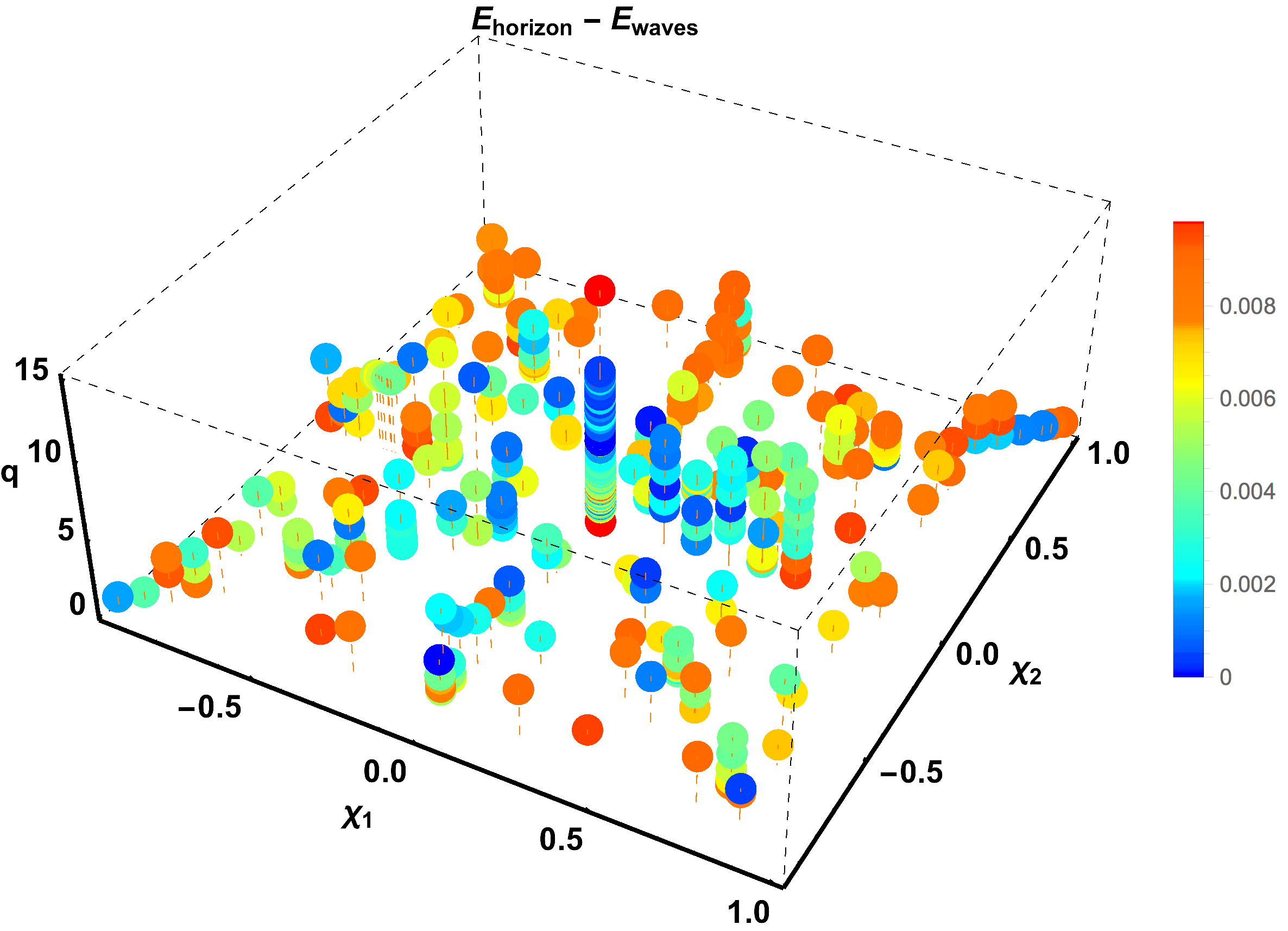}
 \caption{
  \label{fig:NR_HorizonErrorsv2}
  Differences between radiated energy computed from either horizon or waveform data, across the parameter space.
  The color scale quantifies the differences between the two computations.
  Differences are largest for high-mass-ratio and high-spin cases,
  where high NR accuracy is more demanding.
  \vspace{-\baselineskip}
 }
\end{figure}

Final spin and final mass are usually computed as surface integrals over the apparent horizon using the isolated-horizon formalism \cite{Dreyer:2002mx}
(see \cite{Hinder:2013oqa} for a summary of methods and references for different codes),
from surface integrals over spheres at large or infinite radius (as in \cite{Bruegmann:2006at}), 
or from the energy or angular momentum balance computed from initial and radiated quantities (see \autoref{eq:erad_waves} below).
Final spin and final mass can also be obtained from fits to the ringdown \cite{Berti:2007fi}.

Final mass and spin from the apparent horizon (AH) are generally expected to be more accurate
than those based on the evaluation of asymptotic quantities
(such as Bondi mass, angular momentum, or radiated energy and angular momentum) at finite radius,
where errors may arise due to finite radius truncation, insufficient extrapolation to infinity
or numerical inaccuracies in propagating the wave content to large distances at sufficient numerical resolution.
On the other hand, AH quantities may suffer from inaccuracies in finding the apparent horizon and from gauge ambiguities.

For SXS, RIT and GaTech results,
we take the values provided in the catalogs~\cite{Mroue:2013xna,SXScatalog,Healy:2014yta,Jani:2016wkt,GaTechcatalog},
which in general have been computed at the horizon.
For a comparison of horizon vs radiated quantities for the RIT results, see Table~V in~\cite{Healy:2014yta},
where error bars for the horizon quantities are significantly smaller. 
For the \BAM code, for some large-mass-ratio cases the AH finder fails
due to the unfortunate choice of a shift condition,
which results in a coordinate growth in the horizon which is roughly linear in time during the evolution.
After several orbits the horizon of the larger BH is then no longer contained within the fine grid of the mesh refinement,
which may trigger a failure of the horizon finder code.
Due to the high computational cost of the simulations,
we have not rerun these cases with improved parameters for the apparent horizon finder code.
But rather, we compute the final angular momentum from the angular momentum surface integral at large radius,
and the energy from the radiated GWs.
For processing GaTech waveforms we have also followed~\cite{Bustillo:2016gid}.

\begin{figure}[thbp]
 \includegraphics[width=\columnwidth]{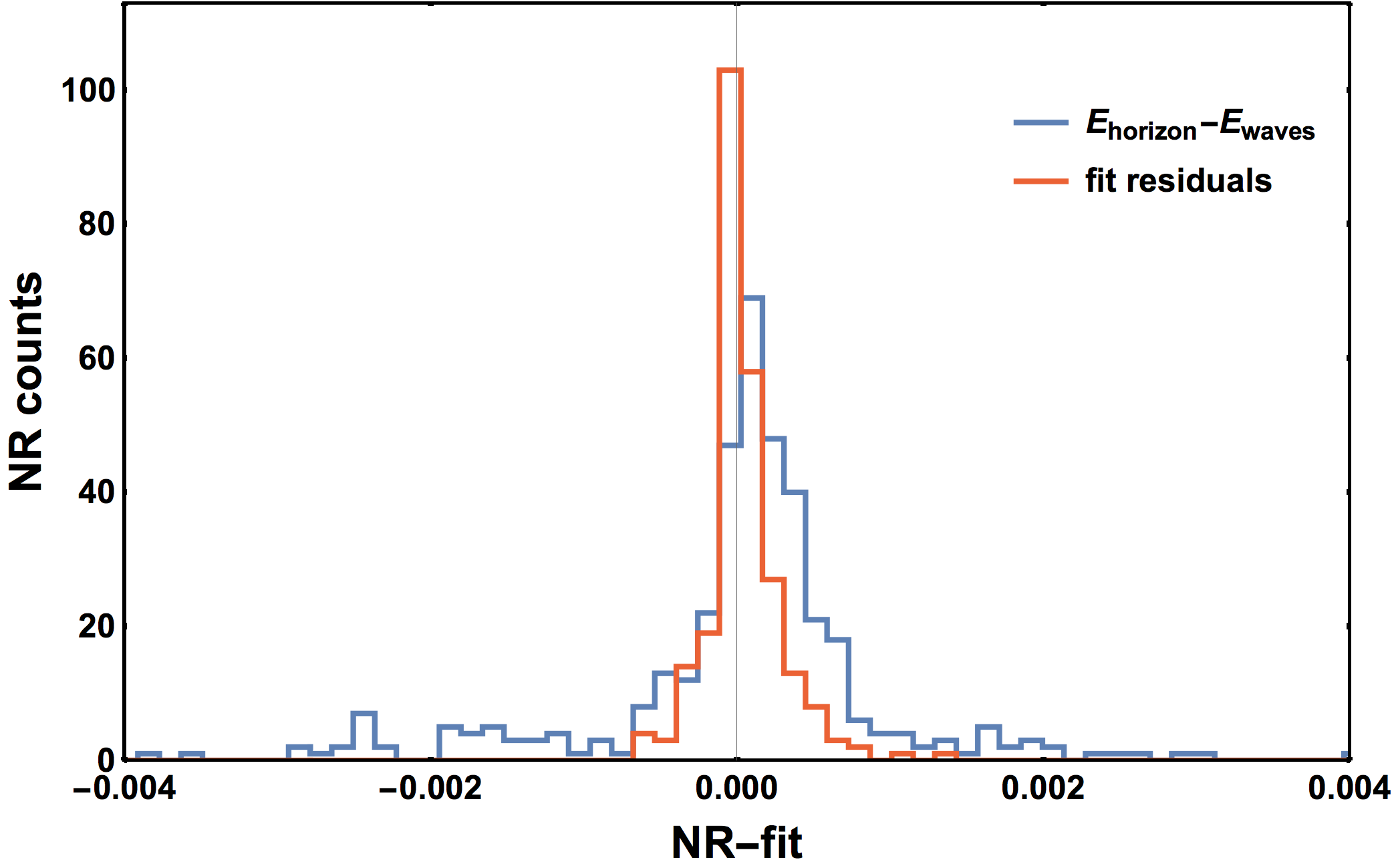}
 \captionof{figure}{
  \label{fig:horizonresidualscomparison}
  Histograms of the differences between radiated energy computed from either horizon or waveform data (as in \autoref{fig:NR_HorizonErrorsv2}),
  compared with the residuals of the new radiated-energy fit (as in \autoref{fig:Erad_resid_hists}).
  \vspace{-2\baselineskip}
 }
\end{figure}

For all 414 cases where we have both the waveform and AH estimate available,
we perform cross-checks between AH quantities and those obtained from integrating radiated energy and angular momentum.
In order to compute the final mass from the radiation quantities, we integrate
\begin{equation}
\label{eq:erad_waves}
\frac{dE}{dt} = \lim_{r\to\infty}\left[
\frac{r^2}{16 \pi} \int_{\Omega}^{} \mathrm{d}\Omega \left| \int_{-\infty}^{t} \psi_4 \mathrm{d}\tilde{t} \; \right|^2\right]
\end{equation}
over time,
starting after the initial burst of ``junk radiation''.

We extrapolate the result from different extraction radii to infinity at linear order in inverse radius.
To account for energy radiated at separations larger than the initial separation of the NR simulation,
we compute the radiated energy at 3.5 PN order~\cite{Iyer:1993xi,Iyer:1995rn,Jaranowski:1996nv,Pati:2002ux,Konigsdorffer:2003ue,Blanchet:2006xj,Nissanke:2004er}
from \mbox{$\omega \in \left[ 0, \omega_0 \right]$},
with $\omega_0$ being the initial orbital frequency of the NR simulation (after junk radiation).
Then we can consider the differences between radiated energy values from the horizon and from the integrated waveforms,
shown in \autoref{fig:NR_HorizonErrorsv2},
as an estimate of NR errors.
However, this will typically be a \textit{pessimistic} estimate because
horizon quantities are in general more reliable and thus big differences are typically caused by inaccuracies in the integrated emission.
In \autoref{fig:horizonresidualscomparison} we show that the distribution of this pessimistic estimate is similar to,
but much wider-tailed than,
the residuals from our radiated-energy fit.

\begin{figure}[t!]
 \includegraphics[width=\columnwidth]{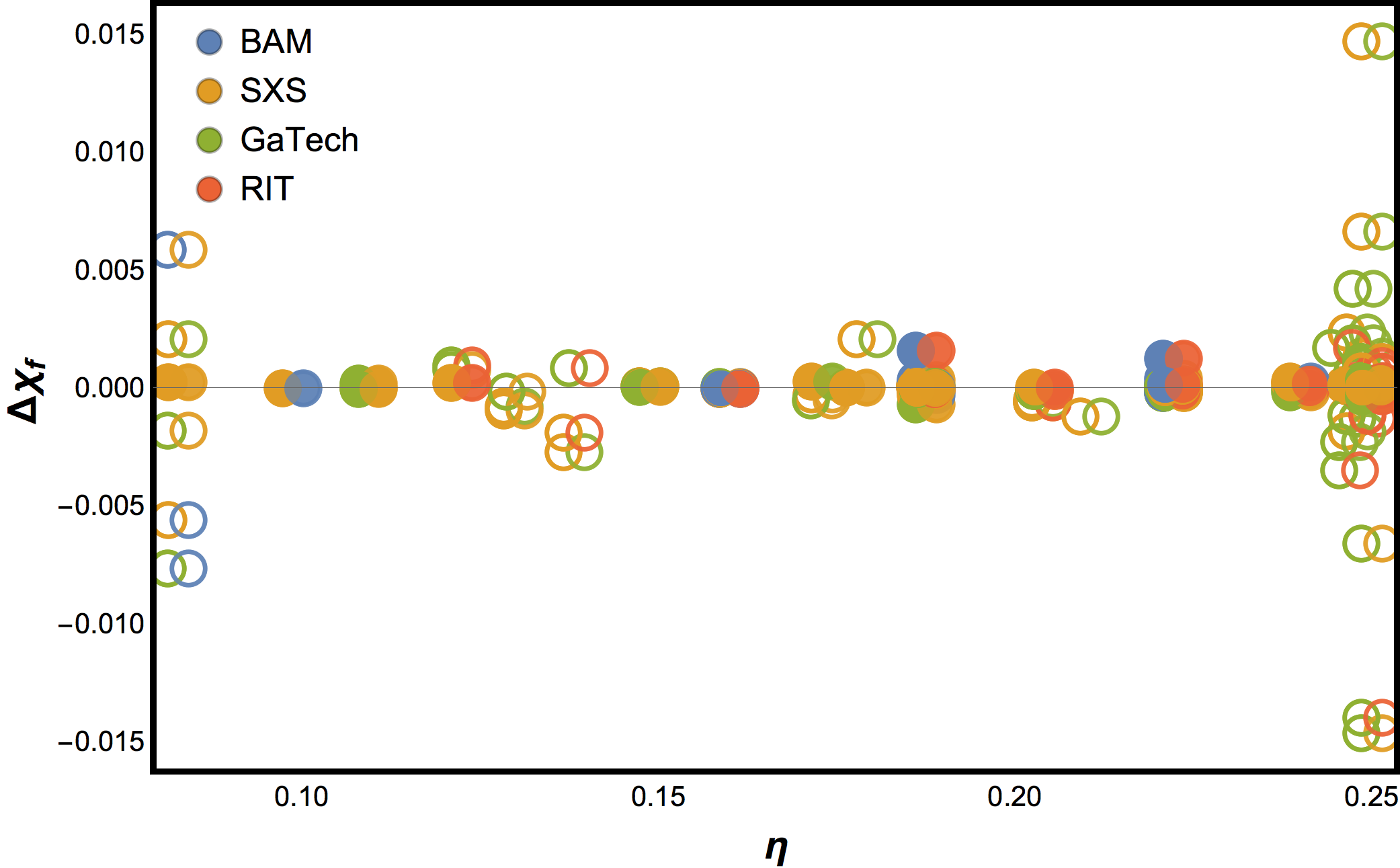} \\
 \includegraphics[width=\columnwidth]{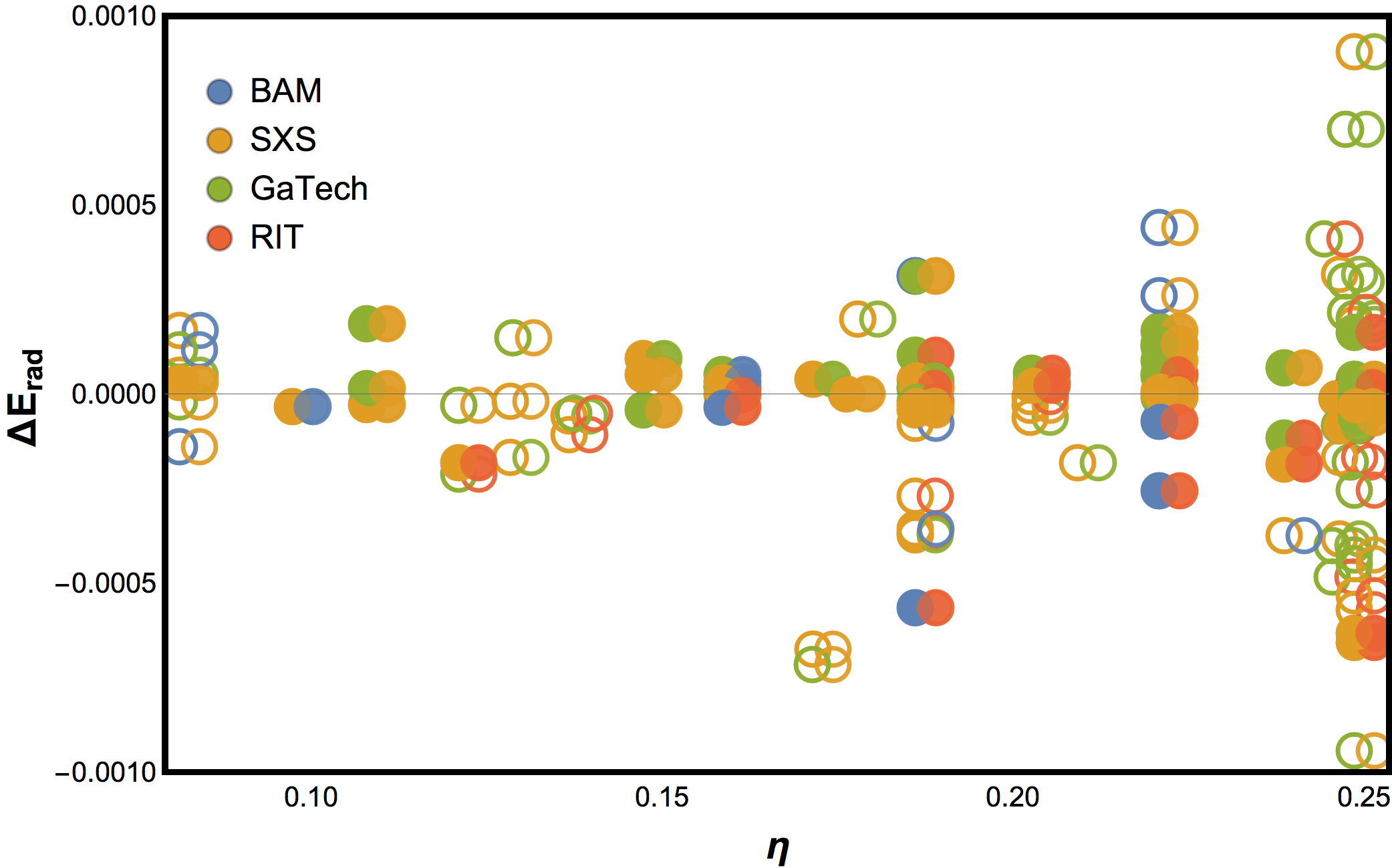}
 \caption{
  \label{fig:eraderrestimateetav2}
  Differences in final spin $\af$ (top panel) and radiated energy $\Erad$ (lower panel)
  for equal-parameter configurations but different NR codes.
  Solid circles: configurations with parameters equal to within numerical accuracy
  (narrow tolerance).
  Open circles: similar configurations but with some deviation in the parameters
  (wider tolerance, e.g. up to \mbox{$\Delta\eta \approx 0.001$}).
  Pairs of simulations are shown with a small horizontal offset for ease of visual identification.
  \vspace{-\baselineskip}
 }
\end{figure}

\begin{figure}[t!]
 \vspace{-\baselineskip}
 \includegraphics[width=\columnwidth]{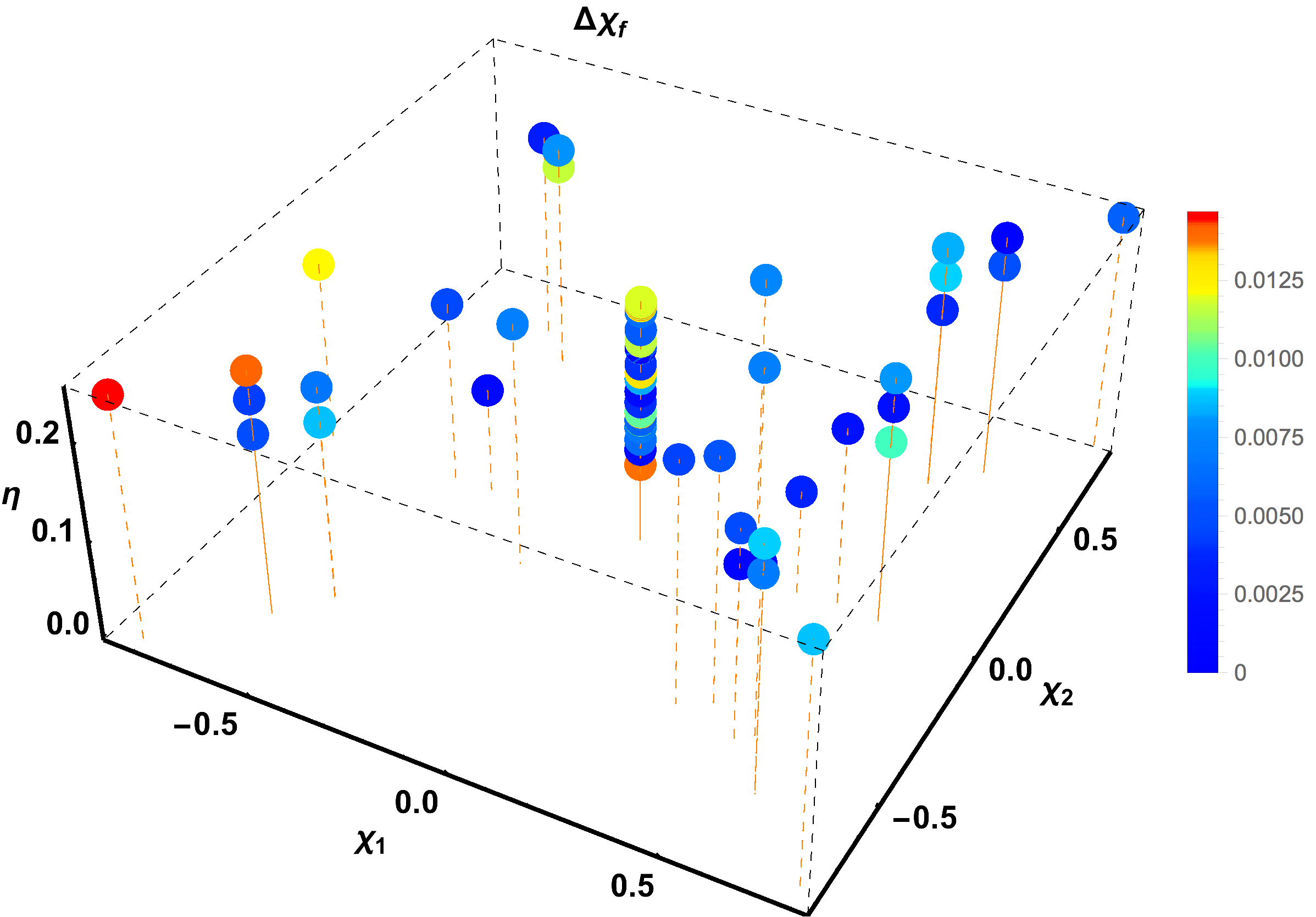}\\
 \includegraphics[width=\columnwidth]{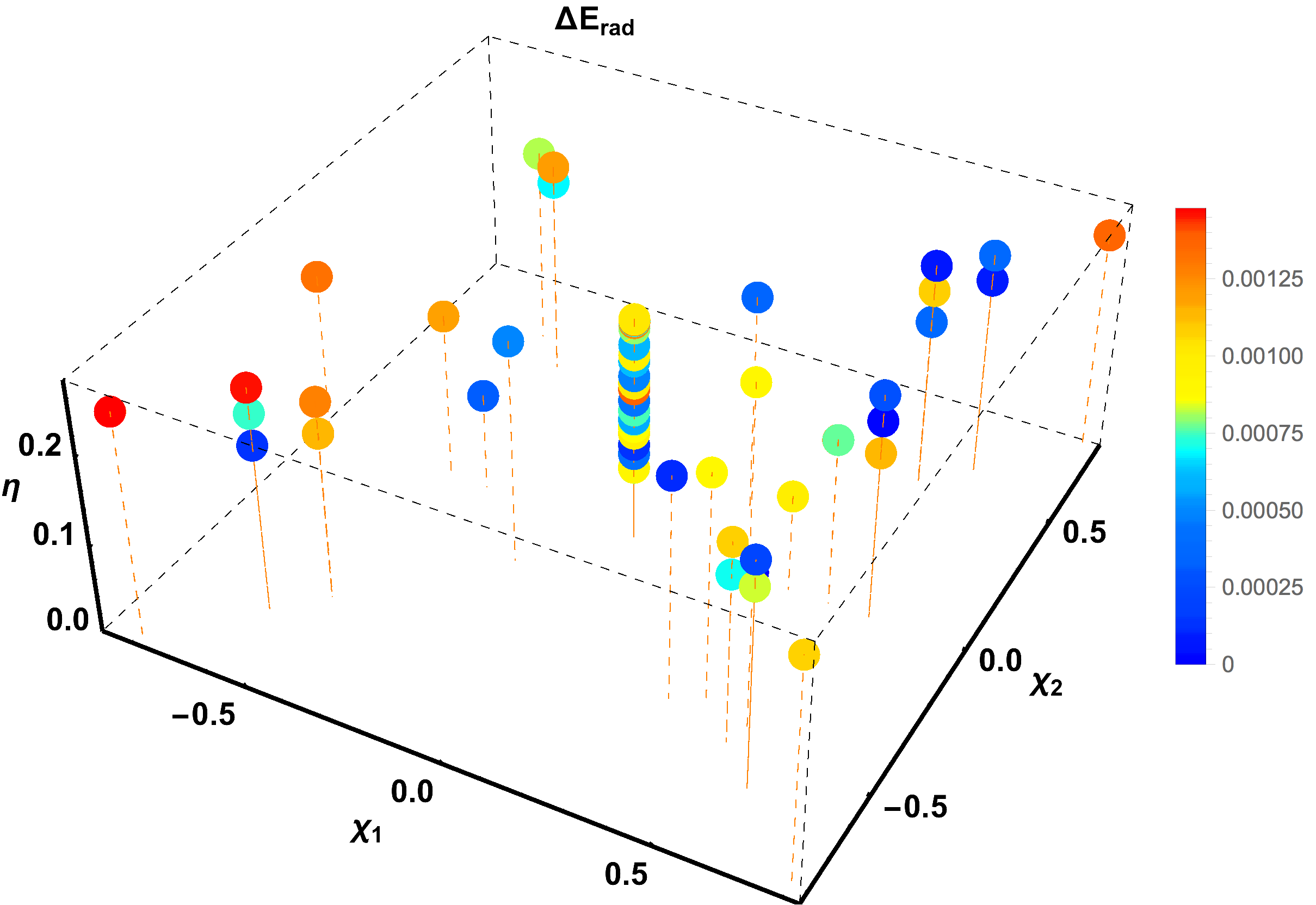}
 \caption{
  \label{fig:duplicated3DrelerrorEn}
  Differences in the final-state quantities for equal-parameter configurations and different NR codes.
  Top panel: final spin $\af$,
  lower panel: radiated energy $\Erad$.
  Points here correspond to both open and solid circles from \autoref{fig:eraderrestimateetav2} (wider tolerance).
  \vspace{-1.5\baselineskip}
 }
\end{figure}

A more realistic measure of NR errors is the difference between results from different codes for equal initial parameters.
With a strict tolerance requiring equal initial parameters to within numerical accuracy,
\begin{equation}
\label{eq:NRtolerance}
\left| \lambda_i-\lambda_j \right| \leq \epsilon = \NRStricttolerance \; \text{with} \; \lambda_i=\left\lbrace \eta_i, \, \chi_{1i}, \, \chi_{2i}  \right\rbrace \,,
\end{equation}
we find \NRduplicated such duplicate configurations out of the total of \NRcount cases.\footnote{With a more relaxed tolerance, 
\mbox{$\epsilon=\NRReltolerance$} in \autoref{eq:NRtolerance},
we find 33 duplicates and 19 sets of two or more configurations with reasonably similar parameters,
corresponding to a total of 131 cases (30\% of the total data set),
compatible with the 71 ``twins" out of a data set of 248 reported in \cite{Hofmann:2016yih}.
The wider-tolerance tuples are shown as open circles in \autoref{fig:eraderrestimateetav2}.}
We evaluate differences between these equal-parameter cases for final spin and radiated energy.
Figure~\ref{fig:eraderrestimateetav2} shows that, with strict tolerance, these error estimates
(standard deviations of $3.1 \times 10^{-4}$ for $\af$ and $1.6 \times 10^{-4}$ for $\Erad$)\footnote{For the relaxed tolerance,
the values are $2.8 \times 10^{-3}$ for $\af$ and $3.5 \times 10^{-4}$ for $\Erad$,
compatible with the $2 \times 10^{-3}$ given in \cite{Hofmann:2016yih} for $\af$.}
are still on the same order but smaller than the respective fit residuals (RMSE of $5.2 \times 10^{-4}$ for $\af$ and $2.2 \times 10^{-4}$ for $\Erad$).
However, the set of true duplicates is small and mostly concentrated in equal-spin-similar-mass regions of the parameter space
(cf. \autoref{fig:duplicated3DrelerrorEn}),
preventing us from naively extrapolating this error estimate to the full parameter space.
Hence we consider it as a somewhat \textit{optimistic} estimate of final-state NR errors.

\begin{table*}[htp!]
 \begin{tabular}{rrrrrrrrrrll}
\hline
 id & $q$ & $\chi_1$ & $\chi_2$ & $\omega_0$ & $D_0$ & $E_{\text{rad}}$ & $\Delta E_{\text{rad}}$ & $\chi_{\text{f}}$ & $\Delta \chi_{\text{f}}$ & tag & code\\
\hline
 1 & 1.00 & -0.80 & -0.80 & 0.060 & 5.88 & 0.0325 & -0.0010 & 0.4122 & -0.0146 & \text{D6.2\_q1\_a-0.8\_m100} & \text{GaTech} \\
 2 & 1.00 & -0.60 & -0.60 & 0.058 & 5.93 & 0.0349 & -0.0013 & 0.4876 & -0.0066 & \text{D6.2\_q1\_a-0.6\_m100} & \text{GaTech} \\
 3 & 1.00 & 0.80 & -0.80 & 0.025 & 10.92 & 0.0491 & 0.0002 & 0.6839 & -0.0000 & \text{D11\_a0.8\_q1.00\_m103\_As} & \text{GaTech} \\
 4 & 1.00 & 0.80 & 0.80 & 0.024 & 11.07 & 0.0883 & -0.0005 & 0.9086 & 0.0010 & \text{D11\_q1.00\_a0.8\_m200} & \text{GaTech} \\
 5 & 2.50 & 0.60 & 0.60 & 0.051 & 6.27 & 0.0528 & 0.0002 & 0.8255 & 0.0004 & \text{Lq\_D6.2\_q2.50\_a0.6\_th000\_m140} & \text{GaTech} \\
 6 & 3.50 & 0.00 & 0.00 & 0.015 & 15.90 & 0.0258 & 0.0007 & 0.5046 & 0.0005 & \text{BBH\_CFMS\_d15.9\_q3.50\_sA\_0\_0\_0\_sB\_0\_0\_0} & \text{SXS} \\
 7 & 5.00 & -0.73 & 0.00 & 0.030 & 9.53 & 0.0129 & 0.0004 & 0.0222 & 0.0460 & \text{D10\_q5.00\_a-0.73\_0.00\_m240} & \text{GaTech} \\
 8 & 5.00 & -0.72 & 0.00 & 0.030 & 9.54 & 0.0129 & 0.0004 & 0.0164 & 0.0340 & \text{D10\_q5.00\_a-0.72\_0.00\_m240} & \text{GaTech} \\
 9 & 5.00 & -0.71 & 0.00 & 0.029 & 9.55 & 0.0130 & 0.0005 & 0.0105 & 0.0220 & \text{D10\_q5.00\_a-0.71\_0.00\_m240} & \text{GaTech} \\
 10 & 5.00 & 0.00 & 0.00 & 0.027 & 10.07 & 0.0176 & -0.0001 & 0.4175 & 0.0009 & \text{D10\_q5.00\_a0.0\_0.0\_m240} & \text{GaTech} \\
 11 & 5.50 & 0.00 & 0.00 & 0.031 & 9.16 & 0.0161 & 0.0001 & 0.3932 & -0.0002 & \text{D9\_q5.5\_a0.0\_Q20} & \text{GaTech} \\
 12 & 6.00 & 0.00 & 0.00 & 0.027 & 10.13 & 0.0145 & -0.0001 & 0.3732 & 0.0007 & \text{D10\_q6.00\_a0.00\_0.00\_m280} & \text{GaTech} \\
 13 & 6.00 & 0.40 & 0.00 & 0.026 & 10.35 & 0.0195 & 0.0000 & 0.6257 & -0.0000 & \text{D10\_q6.00\_a0.40\_0.00\_m280} & \text{GaTech} \\
 14 & 8.00 & 0.85 & 0.85 & 0.048 & 6.50 & 0.0248 & -0.0027 & 0.8948 & -0.0012 & \text{q8++0.85\_T\_80\_200\_-4pc} & \text{BAM} \\
 15 & 10.00 & 0.00 & 0.00 & 0.035 & 8.39 & 0.0082 & -0.0000 & 0.2588 & -0.0019 & \text{D8.4\_q10.00\_a0.0\_m400} & \text{GaTech} \\
 16 & 10.00 & 0.00 & 0.00 & 0.035 & 8.39 & 0.0081 & -0.0001 & 0.2665 & 0.0058 & \text{q10c25e\_T\_112\_448} & \text{BAM} \\
\end{tabular}
 \caption{
  \label{tbl:outliers}
  NR cases from the source catalogs not included in the fit calibration,
  for reasons detailed below.
  }
\end{table*}

\begin{figure}[tbhp]
 \includegraphics[width=\columnwidth]{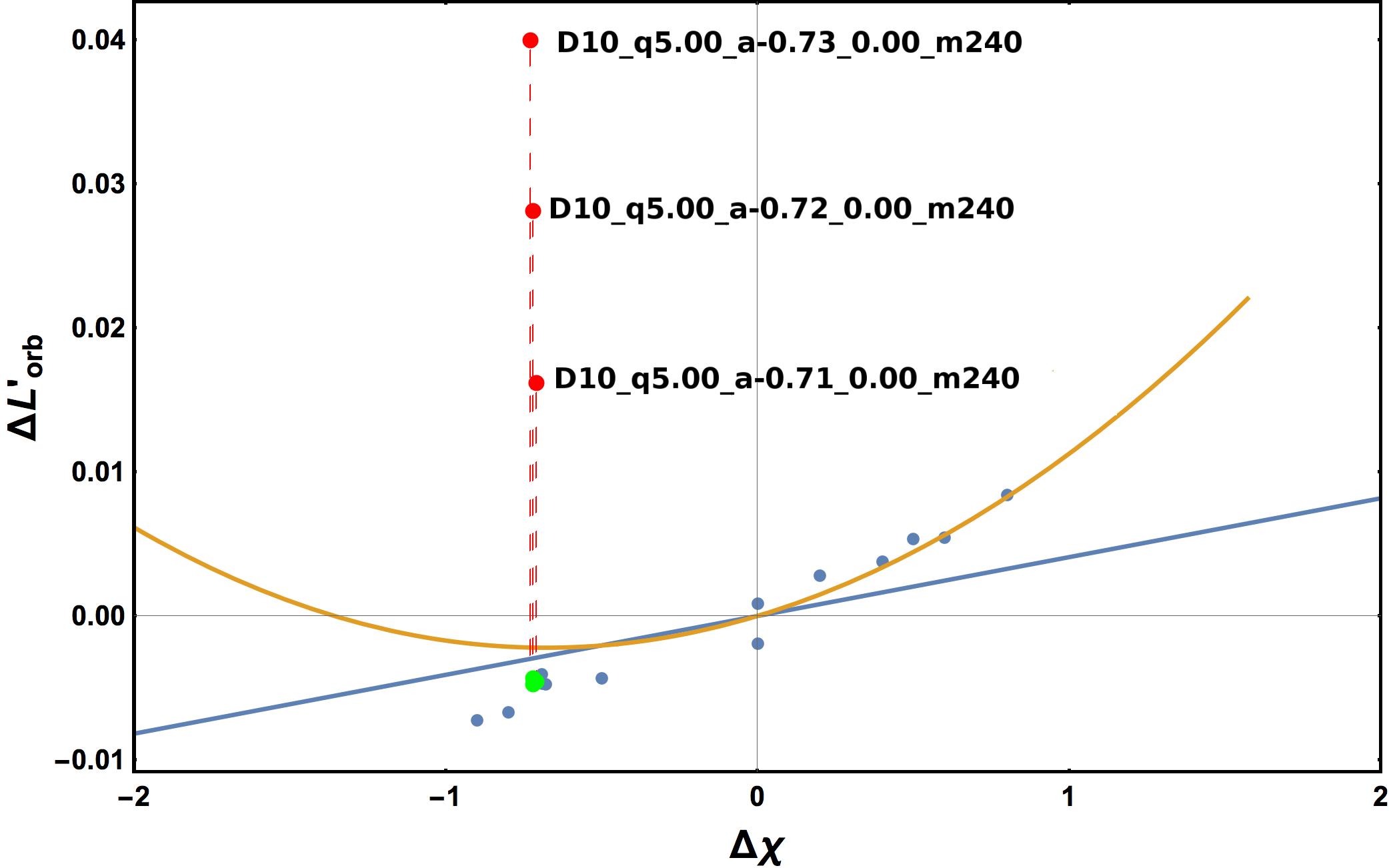}
 \caption{
  \label{fig:spinbyetaq5plot}
  Unequal-spin effects for final spin $\af$ at \mbox{$q=5$},
  shown as residuals against the 2D equal-spin fit (cf. \autoref{fig:spin_diff_per_q}).
  The three points highlighted in red
  are similar configurations from the GaTech catalog,
  for which it has since been confirmed~\cite{Shoemaker:2016priv} that the sign of $\af$ should be negative instead,
  making $\ourLorb$ fit with the trend -- corrected values are shown in green.
  \vspace{-\baselineskip}
  }
\end{figure}

We therefore have a rough expectation for the range of possible NR errors bracketed by these pessimistic and optimistic estimates,
but no detailed information for each case over the whole parameter space.
Instead, we use simple heuristic fit weights.
The overall scale of the NR error is not relevant for determining fit weights,
so we only need to assign relative weights between the cases,
emulating the usual quadratic scaling with data errors which can also be deduced from \autoref{fig:NR_HorizonErrorsv2}.
For SXS data we down-weight cases with \mbox{$\eta < 0.1$} by a factor of $2^2$;
while for the puncture codes (\BAM, GaTech, RIT) we expect larger inaccuracies especially at low $\eta$,
and so we down-weight by a factor of $2^2$ above \mbox{$\eta=0.223$} and $3^2$ below that mass ratio,
and a factor of $5^2$ below \mbox{$\eta=0.05$}
(including the computationally challenging \mbox{$q=18$} cases).
As mentioned before, a more detailed NR error study,
leading to better-determined weights,
would be a clear avenue to further improve fit results.

From the original set of NR simulations we have removed \NRcountOutliers cases as outliers,
which are listed in Table~\ref{tbl:outliers}.
For this decision, we have considered three main sources of outliers:
cases whose NR setup is not appropriate for the purpose of this study,
duplicated configurations for which the variations in the final quantities are much larger than the RMSE,
and cases that are found to be drastically off the trend of otherwise smooth data sets in any of the one-dimensional plots in our hierarchical fitting procedure.
Outliers 1,2,5,16 have rather short orbital evolutions,
so that they can be used for ringdown-only studies, but not for our purpose of predicting the final state from initial parameters.
For outliers 10--12 and 15--16 we have found large variations in the final-state values for different codes (see \autoref{fig:eraderrestimateetav2}).
Here we have used only the equivalent SXS configuration, 
in each case corresponding to longer and presumably more accurate evolutions.
The remaining seven outliers have been identified after performing the step-by-step one-dimensional analysis of the data,
each deviating so clearly that there must be an underlying systematic problem and not just a statistical fluctuation
(in which case they could not be excised from the data set).
As an example, we highlight in \autoref{fig:spinbyetaq5plot} three clear GaTech outliers found in the unequal-spin calibration step;
however, it was recently confirmed~\cite{Shoemaker:2016priv} that these three cases should have a negative sign of their final spin,
and with this change they are fully consistent with our fits.
We note that the overall data quality of the omitted cases may be perfectly adequate for other studies;
while for this final-state study,
due to good data coverage in the corresponding parameter-space regions
and clear global trends in the full data set,
the consistency requirements are quite narrow.

\begin{table}
 \begin{tabular}{rrrcrrcr}
\hline
 $q$ & $\chi_1$ & $\chi_2$ & $\omega_0$ & $D_0$ & $e\left[\times 10^{-3}\right]$ & $E_{\text{rad}}$ & $\chi_\text{f}$ \\
\hline
 1.00 & 0.00 & -0.85 & 0.022 & 11.97 & 2.25 & 0.0392 & 0.5514 \\
 1.00 & 0.85 & -0.85 & 0.023 & 11.61 & 2.61 & 0.0491 & 0.6854 \\
 1.00 & 0.50 & -0.50 & 0.023 & 11.58 & 1.59 & 0.0485 & 0.6858 \\
 1.20 & 0.00 & -0.85 & 0.020 & 12.79 & 0.74 & 0.0401 & 0.5747 \\
 1.20 & 0.50 & -0.50 & 0.028 & 10.00 & 1.76 & 0.0503 & 0.7142 \\
 1.20 & 0.85 & -0.85 & 0.028 & 10.00 & 5.16 & 0.0527 & 0.7359 \\
 1.50 & -0.50 & 0.50 & 0.024 & 11.00 & 1.80 & 0.0408 & 0.5865 \\
 1.75 & 0.00 & 0.85 & 0.022 & 11.69 & 2.35 & 0.0484 & 0.7033 \\
 1.75 & 0.00 & -0.85 & 0.021 & 12.18 & 1.93 & 0.0369 & 0.5810 \\
 1.75 & 0.85 & -0.85 & 0.023 & 11.59 & 4.95 & 0.0567 & 0.8167 \\
 1.75 & -0.85 & 0.85 & 0.021 & 12.35 & 2.66 & 0.0343 & 0.4607 \\
 1.75 & 0.85 & 0.00 & 0.021 & 12.35 & 1.00 & 0.0682 & 0.8724 \\
 1.75 & -0.85 & 0.00 & 0.020 & 12.69 & 0.58 & 0.0307 & 0.3934 \\
 2.00 & 0.50 & -0.50 & 0.024 & 11.10 & 1.76 & 0.0464 & 0.7510 \\
 2.00 & 0.00 & -0.85 & 0.023 & 11.47 & 2.85 & 0.0347 & 0.5693 \\
 2.00 & 0.00 & 0.85 & 0.024 & 11.16 & 2.52 & 0.0436 & 0.6732 \\
 2.00 & 0.85 & -0.85 & 0.024 & 11.00 & 1.78 & 0.0556 & 0.8344 \\
 2.00 & -0.85 & 0.85 & 0.022 & 11.73 & 3.07 & 0.0310 & 0.4002 \\
 2.00 & -0.50 & 0.50 & 0.023 & 11.53 & 2.60 & 0.0336 & 0.4925 \\
 2.00 & -0.85 & 0.00 & 0.022 & 11.97 & 2.70 & 0.0292 & 0.3425 \\
 2.00 & 0.85 & 0.00 & 0.023 & 11.36 & 4.02 & 0.0646 & 0.8782 \\
 3.00 & -0.50 & 0.50 & 0.024 & 11.26 & 1.69 & 0.0237 & 0.3339 \\
 3.00 & 0.50 & -0.50 & 0.025 & 10.63 & 1.41 & 0.0373 & 0.7410 \\
 3.00 & -0.85 & 0.00 & 0.023 & 11.74 & 3.25 & 0.0201 & 0.1562 \\
 4.00 & 0.00 & 0.85 & 0.026 & 10.52 & 14.79 & 0.0230 & 0.4900 \\
 4.00 & -0.85 & 0.85 & 0.023 & 11.37 & 5.04 & 0.0158 & 0.0323 \\
 4.00 & -0.50 & 0.50 & 0.024 & 10.99 & 1.68 & 0.0177 & 0.2152 \\
 18.00 & 0.80 & 0.00 & 0.042 & 9.00 & 5.00 & 0.0087 & 0.8308 \\
 18.00 & -0.80 & 0.00 & 0.027 & 10.00 & 2.50 & 0.0031 & -0.5270 \\
 18.00 & -0.40 & 0.00 & 0.031 & 9.00 & 1.30 & 0.0035 & -0.1813 \\
 18.00 & 0.00 & 0.00 & 0.040 & 7.58 & 1.30 & 0.0041 & 0.1623 \\
 18.00 & 0.40 & 0.00 & 0.040 & 7.43 & 5.00 & 0.0057 & 0.5030 \\
\end{tabular}
 \caption{
  \label{tbl:newBAM}
  New \BAM simulations used in this work,
  with a focus on high unequal spins;
  and recalculated values for \mbox{$q=18$} cases from~\cite{Husa:2015iqa}.
  For each simulation, we list mass ratio \mbox{$q=m_1/m_2$},
  initial spins $\chi_1$ and $\chi_2$,
  reference orbital frequency $\Omega_0$,
  initial separation $D_0$ (after junk radiation),
  eccentricity $e$,
  radiated energy $\Erad$ (scaled to unit initial mass)
  and dimensionless final spin $\af$.
  \vspace{-\baselineskip}
  }
\end{table}

We also summarize in Table~\ref{tbl:newBAM} \NRcountBAMnew new \BAM simulations first used in this paper,
including recalculated values for previously published~\cite{Husa:2015iqa} mass ratio \mbox{$q=18$} simulations.

\section{Fit assessment and model selection criteria}
\label{sec:appendix-stats}

Both while constructing the full ansatz in our hierarchical process,
and when selecting the final fit,
we rank fits by several standard statistical quantities,
which are briefly summarized here for the benefit of the reader.

A basic figure of merit is the root-mean-square-error,
\begin{equation}
 \RMSE = \sqrt{ \tfrac{1}{\Ndata} \sum\limits_{n=1}^{\Ndata} \left[
                X_{\mathrm{NR}}(\eta_n,\chi_{1,n},\chi_{2,n}) - \mathrm{fit}(\eta_n,\chi_{1,n},\chi_{2,n})
                \right]^2 } \,,
\end{equation}
which just checks the overall goodness of fit.
One caveat here is that down-weighted NR cases are fully counted in the RMSE,
so that a generalized variance estimator using weights can be more useful.

Furthermore, it is important in model selection to penalize models with too many free coefficients,
as in principle the RMSE can be made arbitrarily small when the number of coefficients approaches
the number of data points.
A popular figure of merit for model selection considering the number of coefficients is the
Akaike information criterion~\cite{Akaike:1974},
\begin{equation}
 \AIC = -2 \ln \maxlnL + 2\Ncoef \,,
\end{equation}
which intuitively can be understood as weighing up goodness of fit
(measured by the maximum log-likelihood $\maxlnL$)
against parsimony.
Standard implementations, as the one from Wolfram Mathematica,
assume Gaussian likelihoods.

A generalization that corrects the AIC for low numbers of observations and reproduces it for large
data sets is the AICc:
\begin{equation}
 \AICc = \AIC + \frac{2\Ncoef(\Ncoef+1)}{\Ndata-\Ncoef-1} \,.
\end{equation}
In this work, we always use AICc instead of AIC.

A related quantity, similar in form but with a completely different theoretical justification
and with subtle differences in practice,
is the Bayesian information criterion or Schwarz information criterion~\cite{Schwarz:1978}:
\begin{equation}
 \BIC = -2 \ln \maxlnL + \Ncoef \ln(\Ndata) \,.
\end{equation}
Though based on an approximation to full Bayesian model selection
(while the AIC is derived from information theory),
the BIC in general cannot be interpreted as a direct measure of Bayesian evidence between models.

There is much literature on advantages and disadvantages of these two criteria,
and several other alternatives exist --
see e.g.~\cite{Liddle:2007fy} and references therein.
In practice, the BIC tends to impose a slightly stronger penalty on extra parameters than the AIC(c).
Both criteria have been criticized~\cite{Liddle:2007fy} for not only penalizing completely extraneous parameters,
but also the existence of degeneracies between parameters.
However, this is a virtue rather than a problem for our purpose of selecting parsimonious model \ansaetze.

For all of AIC, AICc and BIC, the model with the \emph{lowest} value is preferred.
Higher than unit differences between two models are generally required to count as significant evidence;
\cite{Liddle:2007fy} quotes $\pm5$ as ``strong'' and $\pm10$ as ``decisive'' evidence.

By default, we rank the one-dimensional $\eta$ and $\Seff$ \ansaetze by BIC,
and apply the same criterion to judge how many $\chidiff$ terms to include in the final 3D model.
In general these are not guaranteed to be the best by AICc or RMSE as well,
but in practice we check that the three criteria give a very similar ranking of models.

Still, we find that sometimes the best fit by any of these three criteria (RMSE, AICc, BIC)
can suffer from one or several parameters being not well constrained.
In parameter estimation over large measured data sets, such as the CMB example in~\cite{Liddle:2007fy},
this is a desirable feature of model selection, as it allows to assess which physics is actually constrained by the data.
However, in our case we are well aware that our data set does not constrain all possible functions over the parameter space,
and we are more interested in reporting a well-constrained model
that captures the information in the data set than to dig out weak constraints on possible extensions of that model.
So we augment our model selection criteria by considering also the well-constrainedness of each individual fit coefficient,
and allow for picking a fit with slightly worse summary statistics if it has better-constrained coefficients;
or we drop individual coefficients from a high-order ansatz and reassess the quantitative criteria for that reduced model.

\begin{figure}[t!]
 \includegraphics[width=\columnwidth]{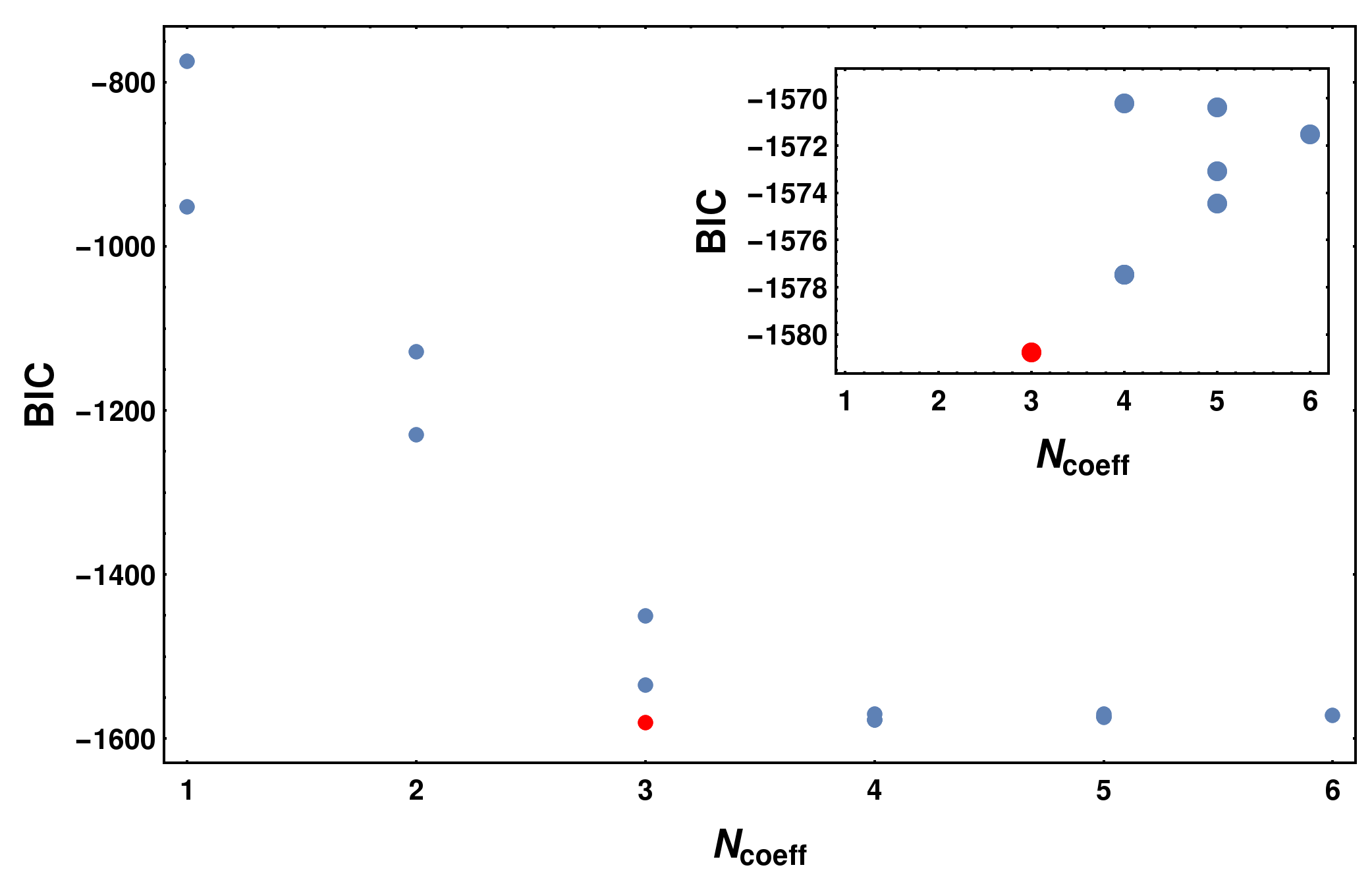}
 \caption{
  \label{fig:af_eta_fit_BIC}
  BIC for the one-dimensional \mbox{$\LorboneDeta$} fits from \autoref{sec:spinfits-1d}.
  The inset panel is a zoom-up of the top-ranked fits.
  The tested set of \ansaetze includes all polynomials from second to seventh order in $\eta$
  and all rational functions of order $(i,j)$, \mbox{$j \leq i$}, up to \mbox{$i+j=6$}.
  The preferred ansatz, a rational function of order (3,1) with three free coefficients, is highlighted.
  \vspace{-\baselineskip}
 }
\end{figure}

As an example of how e.g. the BIC can guide model selection,
we show in \autoref{fig:af_eta_fit_BIC} the BIC ranking for the one-dimensional \mbox{$\LorboneDeta$} fits from \autoref{sec:spinfits-1d}.
A plateau of almost constant BIC is made up of several fits with $\Ncoef \geq 3$,
with the more complex fits yielding no additional improvement,
so that we choose the simplest fit among this group.
Still, even if it had not come up actually top-ranked, as in this case,
choosing a low-$\Ncoef$ fit from within the high-ranked group would be preferable over some slightly higher-ranked,
but less-well-constrained fit.

\section{Spin parameter selection}
\label{sec:appendix-spinpar}

The results of the main text are given in terms of the spin parameter $\Seff$.
However, there is no unique definition of an ``effective spin'',
and alternative parametrizations have been used in the literature~\cite{Husa:2015iqa,Hofmann:2016yih}.
We have tested the robustness of our hierarchical approach for two additional spin parameters:
\begin{equation*}
 \Seff=\frac{S}{m_1^2+m_2^2}, \qquad \Stot=m_1^2\,\chi_1+m_2^2\,\chi_2 , \qquad \chieff = m_1 \chi_1 + \chi_2 m_2 \,.
\end{equation*}

\begin{figure}[htp!]
 \includegraphics[width=\columnwidth]{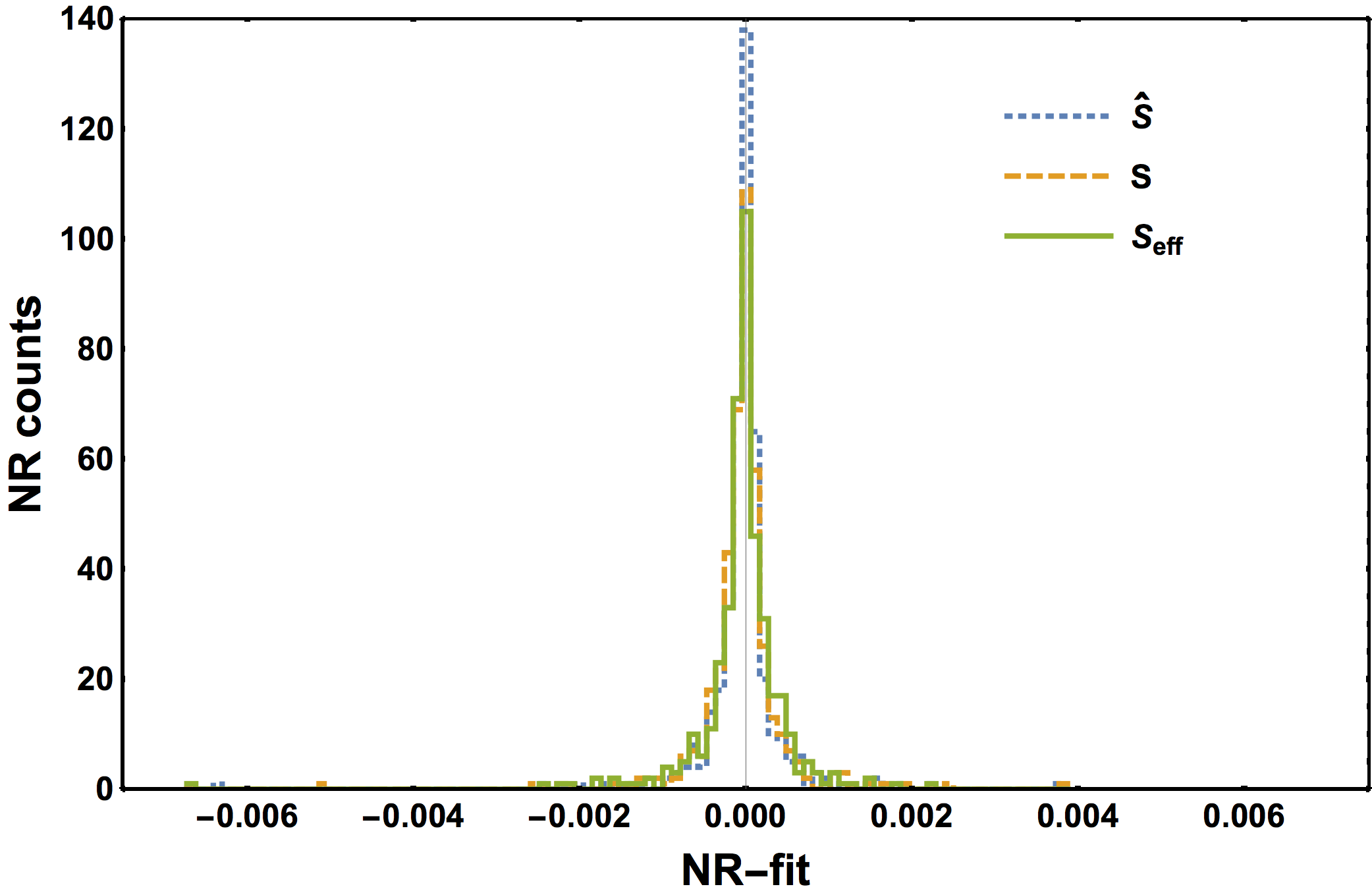} \\[0.5\baselineskip]
 \includegraphics[width=\columnwidth]{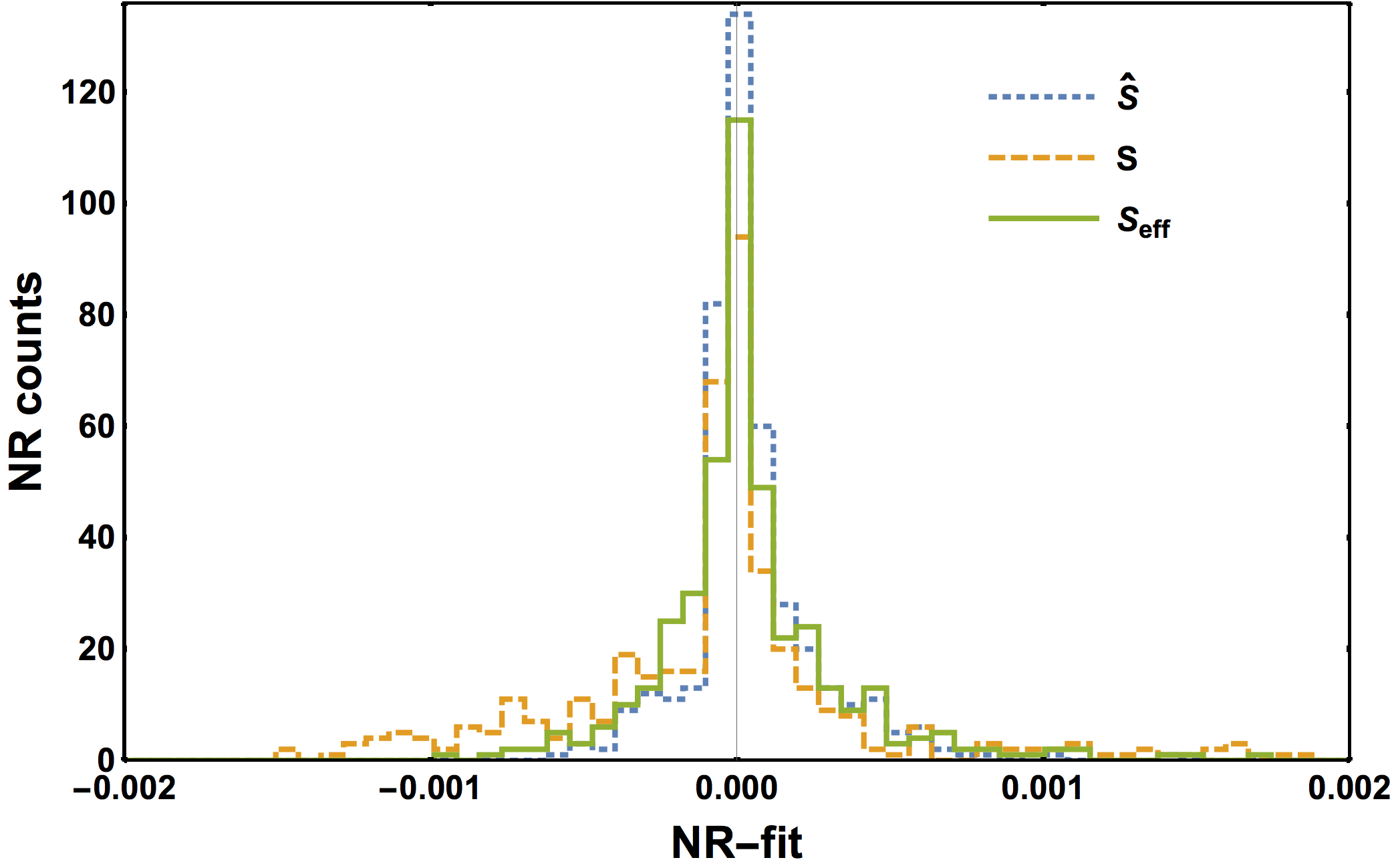}
 \caption{
  \label{fig:af_spinPar_hist}
  Fit residuals for three different choices of effective spin parameter.
  Top panel: final spin;
  lower panel: radiated energy.
 }
\end{figure}

We redid the hierarchical ansatz construction and fitting for $\Stot$ and $\chieff$,
making the same ansatz choices for $\chieff$ as we did for $\Seff$ in the main text,
but changing the 1D spin ansatz to a polynomial P(7) for $\Stot$
(instead of R(3,1) for $\Seff$ and $\chieff$)
because rational functions in $\Stot$ tend to yield singularities.
Checking other possible choices,
we have not found any ansatz combination that makes these alternatives match or exceed the performance of the $\Seff$-based fits presented in the main part of this paper.
Results in terms of the RMSE, AICc and BIC are listed in Table~\ref{tbl:af_spin_parameters},
and residual histograms shown in \autoref{fig:af_spinPar_hist}.
We still obtain better results than most previous fits (see Tables~\ref{tbl:af_residuals} and \ref{tbl:Erad_residuals}) with any parametrization,
thus demonstrating the robustness of our method.

\begin{table}[thbp]
 \begin{tabular}{lccc}\hline\hline
  &$ \text{RMSE} $&$ \text{AICc} $&$ \text{BIC} $\\\hline$
 \hat{S} $&$ 5.15\times 10^{-4} $&$ -5991.5 $&$ -5923.9 $\\$
 \text{S} $&$ 5.24\times 10^{-4} $&$ -5930.9 $&$ -5863.3 $\\$
 \chi _{\text{eff}} $&$ 5.97\times 10^{-4} $&$ -5799.6 $&$ -5731.9 $\\
\hline\hline\end{tabular} \\[\baselineskip]
 \begin{tabular}{lccc}\hline\hline
  &$ \text{RMSE} $&$ \text{AICc} $&$ \text{BIC} $\\\hline$
 \hat{S} $&$ 2.24\times 10^{-4} $&$ -6454.8 $&$ -6391.0 $\\$
 \text{S} $&$ 6.45\times 10^{-4} $&$ -5526.1 $&$ -5439.1 $\\$
 \chi _{\text{eff}} $&$ 4.23\times 10^{-4} $&$ -5962.7 $&$ -5898.9 $\\
\hline\hline\end{tabular}
 \caption{
  \label{tbl:af_spin_parameters}
  Summary statistics for fits with three different choices of effective spin parameter
  and ansatz choices as discussed below,
  evaluated over the full \NRcount point NR data set.
  Top table: Final spin, lower table: radiated energy.
  \vspace{-\baselineskip}
 }
\end{table}

\begin{figure}[thbp]
 \includegraphics[width=\columnwidth]{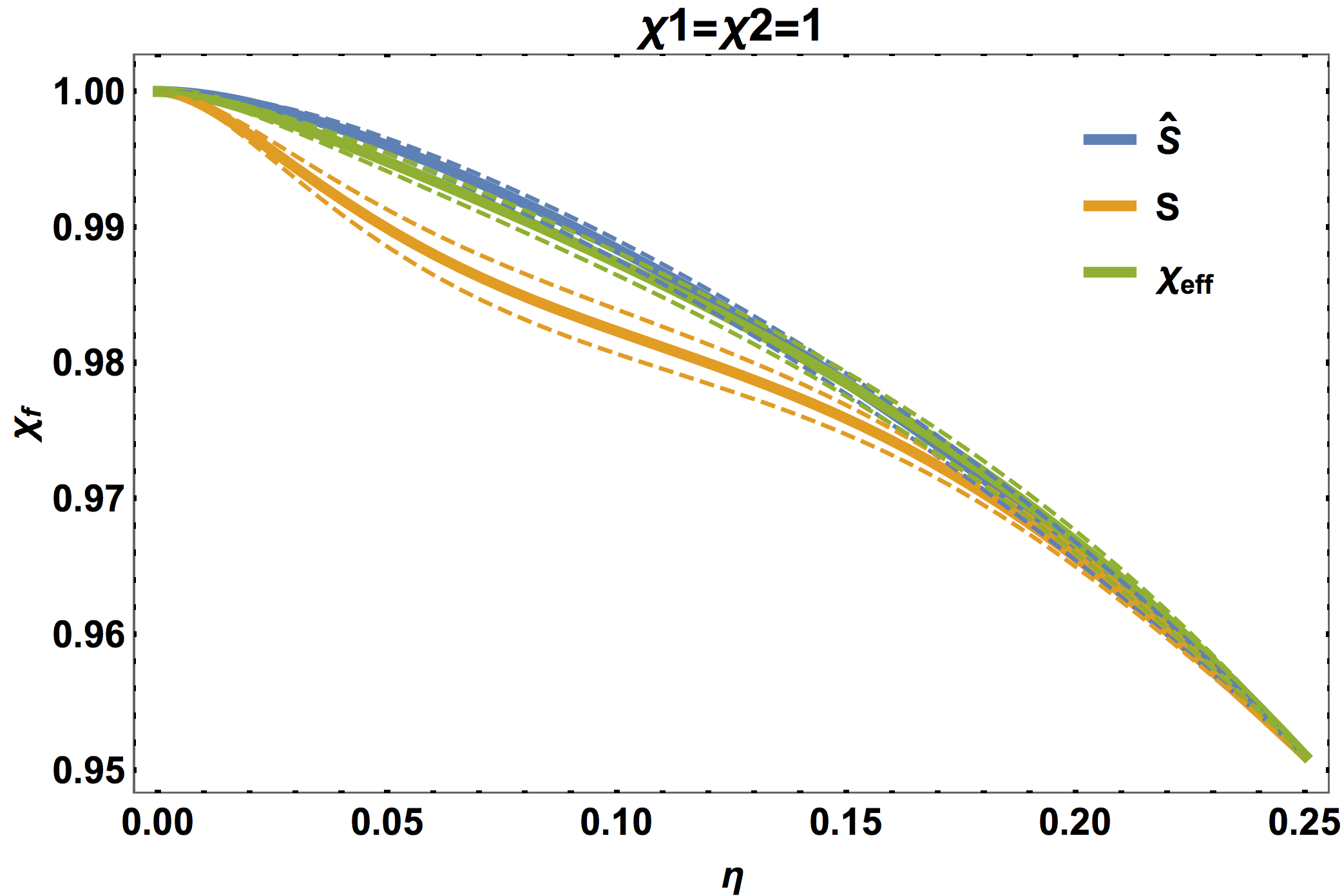} \\[0.5\baselineskip]
 \includegraphics[width=\columnwidth]{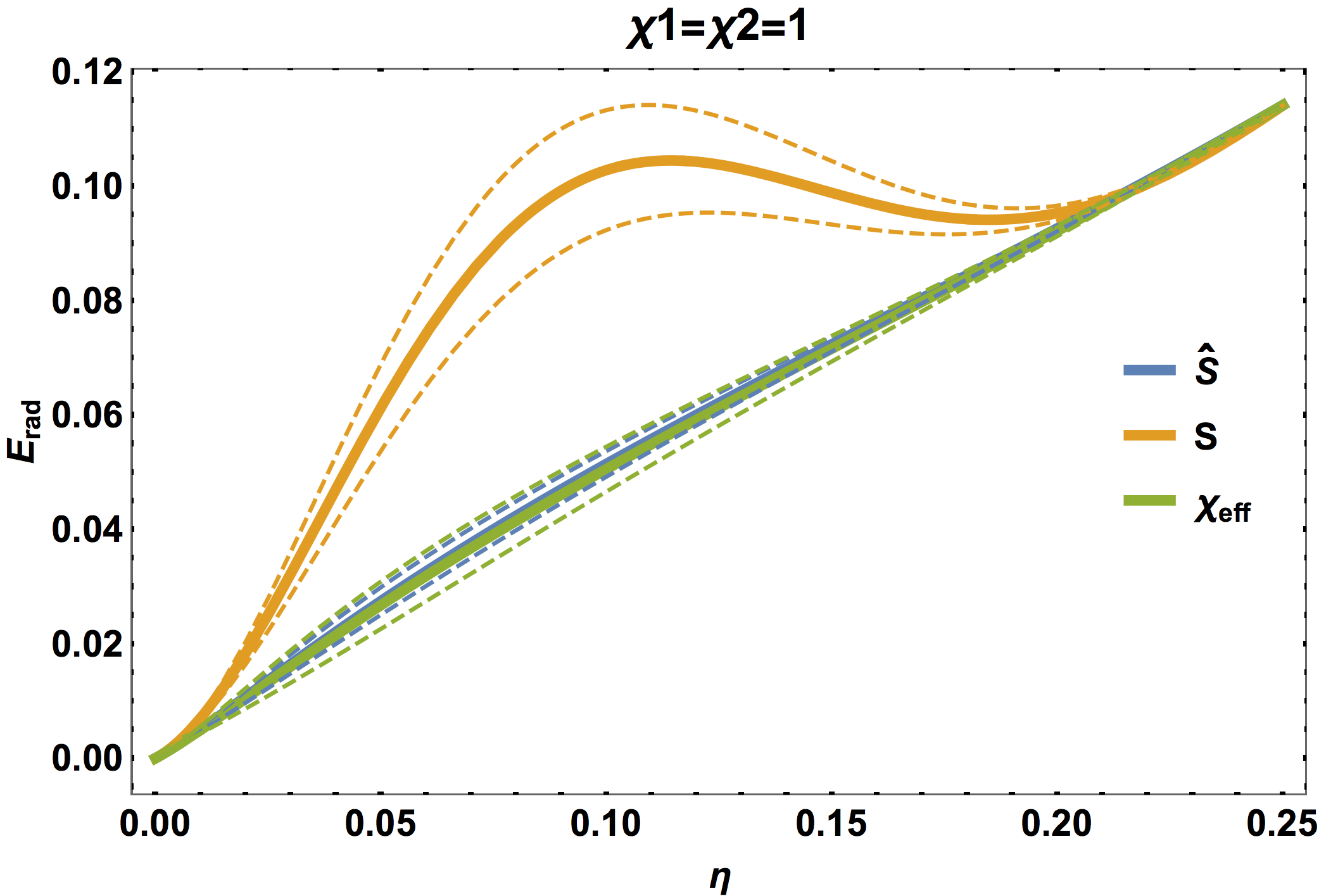}
 \captionof{figure}{
  \label{fig:af_spinPar_extremal}
  Final-state quantities in the extremal \mbox{$\chi_1=\chi_2=1$} limit for three different choices of effective spin parameter.
  Top panel: final spin, lower panel: radiated energy.
 }
\end{figure}

Again, we have also analyzed the fit in the extrapolation regions to detect any artifacts not reflected by the statistical criteria
(which are meaningful only in the calibrated region).
In \autoref{fig:af_spinPar_extremal} we check the extrapolation behavior of fits with the alternative parametrizations
in the notoriously difficult \mbox{$\chi_1=\chi_2=1$} limit.
The approach to this limit is smoother for the fits using $\Seff$ and $\chieff$ than for that using $S$,
which shows some certainly nonphysical oscillations.

The conclusion is that the hierarchical fitting method is quite robust under a change of effective-spin parametrization,
and indeed we would expect full equivalence in the limit of a huge data set with small, completely known NR errors
(using appropriately adapted \ansaetze for each parametrization).
With the current data set, $\Seff$ and $\chieff$ perform similarly,
while when using $\Stot$ additional high-spin data would be even more important to ensure smooth extrapolation.

\section{Fit uncertainties}
\label{sec:appendix-uncertainties}

The uncertainty of evaluating a fitted quantity $Q$ at a point \mbox{$\left(\eta,\chi_1,\chi_2\right)$}
can be expressed through \textit{prediction intervals}~\cite{faraway:2005lmr}
\begin{equation}
 \label{eq:fit-pred-int}
 Q\left(\eta,\chi_1,\chi_2\right) \pm q_t\left(x,\Ndata-\Ncoef\right)
 \sqrt{ \, \estVar + \sigma^2_{\mathrm{fit}} } \,,
\end{equation}
where $q_t$ is the student-t quantile for a confidence level $x$,
\;$\estVar$ is the error variance estimator
from the (weighted) mean-square error of the calibration data under the fit,
and $\sigma^2_{\mathrm{fit}}$ is the standard error estimate of the fitted model,
which for a single-stage fit is
\begin{equation}
 \sigma^2_{\mathrm{fit}} =             \mathrm{grad}^t\left(\eta,\chi_1,\chi_2\right)
                           \, \cdot \, \covfit
                           \, \cdot \, \mathrm{grad}\left(\eta,\chi_1,\chi_2\right)
\end{equation}
with the gradient vector
\mbox{$\mathrm{grad}\left(\eta,\chi_1,\chi_2\right)$} of the fit ansatz in the coefficients,
evaluated at this point,
and the covariance matrix $\covfit$ of the fit.
Note that \autoref{eq:fit-pred-int} gives the uncertainty for a single additional observation,
as opposed to the narrower \textit{confidence interval} of the mean prediction,
which lacks the $\estVar$ term.

In our hierarchical fitting approach,
to propagate the uncertainties from the nonspinning, equal-mass and \emr limits,
we have to assume that the uncertainties in these regimes and that of the final fit are independent,
so that we can take the full covariance matrix as a block-diagonal composition of these four contributions.
The half width of a prediction interval at confidence $x$ is then
\begin{equation}
 q_t\left(x,\Ndata-\Ncoef\right)
 \sqrt{ \,
          \estVar
          + \sigma^2_{\mathrm{final}}
          + \sigma^2_{\eta}
          + \sigma^2_{\Seff}
          + \sigma^2_{\eta=0}
      } \,.
\end{equation}
As these three particular regimes are significantly better constrained than the bulk of the parameter space
(which is the main motivation for the hierarchical approach, in the first place),
their uncertainty contribution is small,
so that the accuracy of this approximation is not critical.

The covariance matrices
for the fits to both final spin and radiated energy are
provided in ASCII format as supplementary material.
The estimated variances are
$\estVar_{\af}\approx6.225\times10^{-8}$ for final spin
and
$\estVar_{\Erad}\approx2.061\times10^{-8}$ for radiated energy.

\vfill

\bibliography{../biblio.bib}

\end{document}